\documentclass[namedreferences]{solarphysics}
\usepackage[optionalrh,solaromanenum]{spr-sola-addons} 
\usepackage{graphicx}                    
\usepackage{color}                       
\usepackage{url}                         
\usepackage{mathrsfs}
\bibpunct{(}{)}{;}{}{,}{,}
\usepackage[margin=1.25in]{geometry}
\usepackage[pdfborder={0 0 0 },urlcolor=blue,breaklinks]{hyperref}
\usepackage{listings}

\ifx \urlurl    \undefined \def \urlurl#1{\href{http://#1}{\textsf{#1}}}\fi
\ifx \doiurl    \undefined \def \doiurl#1{\href{http://dx.doi.org/#1}{\textsf{\textsf{DOI}}}}\fi
\ifx \adsurl    \undefined \def \adsurl#1{\href{http://adsabs.harvard.edu/abs/#1}{\textsf{\textsf{ADS}}}}\fi
\ifx \arxivurl  \undefined \def \arxivurl#1{\href{http://arxiv.org/abs/#1}{\textsf{\textsf{arXiv}}}}\fi

\begin{document}

\begin{article}
\begin{opening}

\title{Observables Processing for the {\it Helioseismic and Magnetic Imager} 
	Instrument on the Solar Dynamics Observatory}
\author{S.~\surname{Couvidat}$^{1}$\sep 
J.~\surname{Schou}$^{2,1}$\sep 
J.T.~\surname{Hoeksema}$^{1}$\sep 
R.S.~\surname{Bogart}$^{1}$\sep 
R.I.~\surname{Bush}$^{1}$\sep
T.L.~\surname{Duvall, Jr.}$^{2}$\sep 
Y.~\surname{Liu}$^{1}$\sep 
A.A.~\surname{Norton}$^{1}$\sep 
P.H.~\surname{Scherrer}$^{1}$
}

\institute{$^{1}$ W.W. Hansen Experimental Physics Laboratory, Stanford University, Stanford, CA 94305 USA 
	Email: \href{mailto:jthoeksema@sun.stanford.edu}{jthoeksema@sun.stanford.edu} \\
	$^{2}$ Max-Planck-Institut f\"ur Sonnensystemforschung, Justus-von-Liebig-Weg 3, 37077 G\"ottingen, Germany\\ }

\date{Received: 12 November, 2015; Accepted: 26 March, 2016}
\runningauthor{Couvidat, Schou, Hoeksema, {\textit {et al.}}}
\runningtitle{HMI Observables Processing}

\begin{abstract}

NASA's Solar Dynamics Observatory (SDO) satellite was launched 11 February 2010 with three instruments onboard, including the Helioseismic and Magnetic Imager (HMI). After commissioning, HMI began normal operations on 1 May 2010 and has subsequently observed the Sun's entire visible disk almost continuously. HMI collects sequences of polarized filtergrams taken at a fixed cadence with two $4096 \times 4096$ cameras from which are computed arcsecond-resolution maps of photospheric observables that include line-of-sight velocity and magnetic field, continuum intensity, line width, line depth, and the Stokes polarization parameters, {\sf [I\,Q\,U\,V]}. Two processing pipelines have been implemented at the SDO Joint Science Operations Center (JSOC) at Stanford University to compute these observables from calibrated Level-1 filtergrams, one that computes line-of-sight quantities every 45 seconds and the other, primarily for the vector magnetic field, that computes averages on a 720-second cadence. Corrections are made for static and temporally changing CCD characteristics, bad pixels, image alignment and distortion, polarization irregularities, filter-element uncertainty and non-uniformity, as well as Sun-spacecraft velocity.  This report details the functioning of these two pipelines, explains known issues affecting the measurements of the resulting physical quantities, and describes how regular updates to the instrument calibration impact them. We also describe how the scheme for computing the observables is optimized for actual HMI observations.  Initial calibration of HMI was performed on the ground using a variety of light sources and calibration sequences. During the five years of the SDO prime mission, regular calibration sequences have been taken on orbit in order to improve and regularly update the instrument calibration, and to monitor changes in the HMI instrument. This has resulted in several changes in the observables processing that are detailed here. The instrument more than satisfies all of the original specifications for data quality and continuity. The procedures described here still have significant room for improvement. The most significant remaining systematic errors are associated with the spacecraft orbital velocity.

\end{abstract}


\end{opening}
\pagebreak

\section{Introduction}

The Helioseismic and Magnetic Imager (HMI) investigation \citep{Scherrer2012} provides continuous observations of the full solar disk from the {\it Solar Dynamics Observatory} (SDO, \opencite{Pesnell2012}). The HMI instrument \citep{Schou2012} obtains narrow-band filtergrams at six wavelengths centered on the Fe\,{\sc i} spectral line at 6173~\AA. Sequences of filtergrams in different polarizations are obtained every 45 or 135 seconds in order to determine the photospheric velocity, magnetic field, intensity, and
spectral line parameters. This report describes the processing pipelines that produce these  ``observables" from the calibrated Level-1 filtergrams (see Table \ref{tab:DataSeries}).

The HMI prime mission began on 1 May 2010 and was completed on 30 April 2015.
During those five years of nearly continuous operation, HMI recorded more
than 84 million filtergrams with its two $4096\,\times\,4096$ pixel CCD
cameras. That number is 99.86\% of the expected number of exposures. The
extended mission is expected to provide the same level of high-quality data
to the scientific community.

More than a thousand articles using HMI data were listed on the NASA
Astrophysics Data System website as of spring 2015. The success, scope,
and breadth of use of the HMI data make it necessary to provide the solar
physics community with up-to-date information regarding their processing and
the issues affecting these observables. This is required to ensure a better
understanding of their limitations and what can be accomplished with them.

Elements of the line-of-sight and vector-magnetic-field pipelines have been
described in varying levels of detail in other publications. This report
provides a more comprehensive description and details any available updates
at the time of writing. This analysis draws on published articles based on
ground calibrations, {\it e.g.}, \citet{SchouPolarization2012, Couvidat2012,
Liu2012}, and \citet{Hoeksema2014}. The on-orbit performance of the
HMI instrument and the processing of the HMI science data up to Level-1
filtergrams are described in another report \citep{Bush2015}.

A successor to the Michelson Doppler Imager (MDI; \opencite{Scherrer1995})
on board the {\it Solar and Heliospheric Observatory} (SOHO;
\opencite{Domingo1995}), HMI benefits from a more detailed ground calibration
program that provides better characterization of many essential properties
of the instrument and thus an improved understanding of the data. However,
unlike SOHO's relatively benign halo orbit around the Earth-Sun L1 Lagrange
point, large changes in the Sun-SDO radial velocity associated with the
geosynchronous orbit of SDO produce significant daily variations in the
measurements of some physical quantities that have proven difficult to
eliminate from the data. This results in daily artifacts at the level of
a few percent in most observables.

This article reviews these HMI observables: how they are computed, how
on-orbit calibration sequences and instrument monitoring steps are used to
ensure that they are produced with up-to-date information, what the known
issues are, and some of the future plans we have to improve them.
Section 2 reminds the reader how the observables are computed from the
Level-1 filtergrams, both for line-of-sight (LoS) and vector-field quantities.
That section provides more information than previously available regarding
the observables processing and the several calibration steps performed to
improve the quality of the data. It also describes updates to the processing
pipelines that are based on on-orbit calibration results. Section 3 details
some of the known errors and uncertainties affecting the observables, reviews
known instrumental issues, and outlines improvements planned for implementation
in the observables pipelines. Section 4 provides a summary of the report.

\section{Observables Computation}

The HMI observables are also known as Level-1.5 data, in contrast to Level-0
data (raw HMI images) and the Level-1 filtergrams (Level-0 images at a
particular wavelength and polarization that have
been corrected for various effects). These observables are separated into
two pipelines called LoS and vector magnetic field. The HMI observables are
also used to routinely calculate higher-level HMI data-pipeline products
that are not described in this report, such as vector magnetic field maps
\citep{Hoeksema2014} HMI active region patches, \citep{Bobra2014}, 
synoptic charts and frames,
and sub-surface flow maps \citep{jzhao2012}.

The LoS observables are computed from filtergrams taken using the HMI front
camera (also called the LoS camera). 
The LoS observables are images of velocity (Dopplergrams), LoS magnetic field
(magnetograms), continuum intensity, and the Fe\,{\sc i} line width and
line depth. They are produced in two modes: definitive and near-real-time
(NRT). The top section of Table \ref{tab:DataSeries} lists the DRMS (Data
Record Management System) series names computed in the LoS pipeline. All
are produced with a 45-second cadence using the front-camera filtergrams,
which are observed only in left or right circular polarization. 

\begin{table}
\begin{center}
\caption{SDO JSOC Data Series with HMI Observables}
\label{tab:DataSeries}
\begin{tabular}{r|r|r|l}
\hline
Definitive & NRT Series Name & Photon Noise & Description \\
Series Name & & (Disk Center) & \\
\hline
{\bf LoS Pipeline}~~~ & & \\
{\sf hmi.V\_45s} & {\sf hmi.V\_45s\_nrt} & 17 m\,s$^{-1}$& Line-of-sight Velocity \\
{\sf hmi.M\_45s} & {\sf hmi.M\_45s\_nrt} & 7 G$^\ddagger$ & Line-of-sight Magnetic Field$^\dagger$ \\
{\sf hmi.Ic\_45s} & {\sf hmi.Ic\_45s\_nrt} & 0.03\% & Computed Continuum Intensity \\
{\sf hmi.Lw\_45s} & {\sf hmi.Lw\_45s\_nrt} & 1 m\AA & Fe\,{\sc i} Line Width \\
{\sf hmi.Ld\_45s} & {\sf hmi.ld\_45s\_nrt} & 0.05\% & Fe\,{\sc i} Line Depth \\
\hline
{\bf Vector Pipeline}~~~ & & & \\
{\sf hmi.S\_720s} & {\sf hmi.S\_720s\_nrt} & 0.05\% for I$^\star$ & Stokes Polarization Parameters, {\sc I\,Q\,U\,V} \\
 & & 0.09\% of I for Q\,U\,V & \\
{\sf hmi.V\_720s} & {\sf hmi.V\_720s\_nrt} & 7 m\,s$^{-1}$ & Line-of-sight Velocity \\
{\sf hmi.M\_720s} & {\sf hmi.M\_720s\_nrt} & 3 G$^\diamond$ & Line-of-sight Magnetic Field$^\dagger$ \\
{\sf hmi.Ic\_720s} & {\sf hmi.Ic\_720s\_nrt} & 0.01\% & Computed Continuum Intensity \\
{\sf hmi.Lw\_720s} & {\sf hmi.Lw\_720s\_nrt} & 0.4 m\AA & Fe\,{\sc i} Line Width \\
{\sf hmi.Ld\_720s} & {\sf hmi.Ld\_720s\_nrt} & 0.02\% & Fe\,{\sc i} Line Depth \\
\hline
\end{tabular}
\end{center}

\noindent
$^\dagger$\,HMI measures flux density in each pixel. Because a filling factor
of 1 is assumed, a flux density of 1~Mx\,cm$^{-2}$ \\ 
is equivalent to a field strength of 1~G, and we use the two interchangeably in this report.

\vspace{1ex}
$^\ddagger$\,Compared with 8.5 G observed in a near disk-center weak-field histogram \citep[Fig. 3 of][]{Liu2012}.

\vspace{0.5ex}
$^\star$\,Photon noise is 26 DN for a typical Intensity of 50K DN/s. Larger noise of 43 DN is expected for \\
Q\,U\,\&\,V, which are ten times smaller than I, even in strong-field regions.

\vspace{0.5ex}
$^\diamond$\,Compared with 4 G observed in a near disk-center weak-field histogram \citep[Fig. 3 of][]{Liu2012}.

\end{table}

The vector-field pipeline computes observables using both
linearly and circularly polarized filtergrams obtained with the side camera
(also called the vector camera).
The primary vector-field observable is the set of 24 images comprising the four Stokes-vector 
elements at each of six wavelengths. The vector-field pipeline runs on a
12-minute cadence and combines filtergrams from ten 135-second sequences. The
pipeline also applies the standard LoS observables algorithms to the 720-s Stokes
I+V (RCP) and I-V (LCP) components to determine averaged LoS quantities. The data
series computed in the vector-field pipeline are listed in the lower section
of Table \ref{tab:DataSeries}.

Random uncertainties in the observables are determined largely by the photon
noise, which is a feature and consequence of the instrument design. HMI
observables meet or exceed the original performance specifications, and
estimates of disk-center per-pixel uncertainties due to photon noise are
given in Table \ref{tab:DataSeries} for each observable. \citet{Liu2012}
measured magnetic field variations in quiet-Sun regions to be no more than
10\%--20\% larger than the computed photon noise; those observations include
unresolved solar signals as well as the effects of other smaller errors
due to uncertainties in instrument parameters, such as flat-fielding and
shutter noise.

Systematic errors in the observables are larger, more difficult to quantify,
and potentially more impactful. They derive from uncertain, irregular and
evolving characteristics of the instrument (e.g. temperature-dependent filter
transmission), inherent limitations of the instrument design (e.g. wavelength
sampling and resolution), complications of observing a moving, changing
Sun (e.g. photospheric evolution during an observation), and environmental
conditions (e.g. the effects of the spacecraft orbit). Sources of such errors
are a primary subject of this report.

\begin{table}
\begin{center}
\caption{JSOC, DRMS and Other Terminology}
\label{tab:DRMS}
\begin{tabular}{r|l}
\hline
JSOC & The Joint Science and Operation Center -- in particular the facility for Science \\
 & Data Processing (JSOC-SDP) for SDO's HMI and AIA instrument teams at Stanford. \\
\hline
DRMS & Data Record Management System -- the software system that keeps track of the \\
 & components of {\sf Data Series}: {\sc records, keywords, segments,} etc.\\
\hline
{\sf Data Series} & Set of data records that contain HMI data and metadata, often a time series. \\
 & The elements of the set are identified by {\sc prime\_keys}.\\
\hline
{\sc prime\_keys} & Special keywords that identify unique records in a data series, e.g. time and camera. \\
\hline
{\sc record} & Set of keywords, segments and links that contain HMI data, often a time step. \\
 & Each record is associated with a specific set of valid {\sc prime\_keys}.\\
\hline
{\sc keywords} & Information about a data record or series, e.g. the Sun-spacecraft radial velocity, {\sc obs\_vr}.\\
 & Can be a string or numerical value and be specific to a {\sf Data Series}, {\sc record}, or {\sc segment}.\\
\hline
{\sc segment} & Data in SUMS directory for a {\sc record} described by {\sc keywords}. \\
 & Often a multi-dimensional array, such as a Dopplergram.\\
\hline
SUMS & Storage Unit Managagement System -- File management system that keeps track of \\
 & directories containing data files associated with {\sf Data Series} {\sc records}.\\
\hline
Pipeline & Set of programs applied to the HMI data stream.\\
 & The pipeline creates or extends standard {\sf Data Series} at a regular cadence.\\
\hline
{\sf Module} & Program in the JSOC data reduction pipeline.\\
\hline
Level 0 & Uncalibrated filtergram.\\
\hline
Level 1 & Calibrated filtergram.\\
\hline
Observable & {\sf Data Series} containing a calibrated solar quantity derived from a set of \\
 & HMI filtergrams. See Table \ref{tab:DataSeries}.\\
\hline
\end{tabular}
\end{center}

See Appendix \ref{sec:DRMSApp} for more information about JSOC and DRMS.

Note that different fonts are used throughout for names of {\sf Modules} and 
{\sf Data Series} and for data series {\sc keywords} and {\sc segments}.
\end{table}

\subsection{Production of Level-1 Filtergrams \label{Level1Filtergrams}} 

All of the observables are computed from corrected HMI Level-1 images produced as described in \citet{Bush2015}. These are stored in the {\sf hmi.lev1} (definitive) and {\sf hmi.lev1\_nrt} (near real time, NRT) DRMS {\em data series} (See Table \ref{tab:DRMS} and the Appendix for brief definitions of terms). Each {\em record} in these series contains two data {\em segments}: an image taken by the instrument and a list of bad pixels.

Images, referred to as filtergrams, are ordinarily taken at a specific wavelength, {\it i.e.} with the instrument filter elements co-tuned. Although not part of the observables processing, it is useful to briefly remind the reader of how Level-1 data are obtained from raw images.

The Level-1 processing makes per-pixel adjustments to the Level-0 data to provide a uniform image for the observable computations. These corrections include removing the CCD overscan rows and columns from the raw Level-0 images, subtracting an offset image to remove the CCD-detector dark current and pedestal, multiplying by a flat-field image to correct gain variations across the detector, and normalizing for exposure time.

The flat fields are monitored and updated weekly. As part of the standard Level-1 flat fielding, pixels with significantly high or low values are flagged and added to the bad-pixel list that accompanies each Level-1 filtergram. In the definitive data, cosmic-ray pixels in each image are identified based on time-dependent variations from exposure to exposure. 

A limb finder determines the nominal coordinates of the solar disk center on the CCD and the observed solar radius. However, because the formation height of the signal changes with wavelength,
the radius determined by the 
limb finder varies as a function of the difference between the target
wavelength and the wavelength at the solar limb.
At each point, the wavelength shift depends on the changing spacecraft velocity, 
fixed and variable solar motions and features, and the limb shift.
Corrected values are recorded in the {\sc crpix1, crpix2}, and {\sc r\_sun} 
keywords, and the reported plate scale 
is made consistent with the corrected values.
The {\sc t\_obs} keyword indicates the center of the time the shutter was 
open for each filtergram. {\sc exptime} is the duration of the exposure.
The Level-1 filtergrams are normalized to units of DN\,s$^{-1}$ by
dividing the raw pixel values by {\sc exptime}.

The production of definitive Level-1 images from Level-0 data may take some
time, depending on a variety of factors. For that reason the Level-1 data
are produced in two modes: NRT and definitive. NRT observations are intended
to be used only for time-sensitive applications, {\it e.g.} space-weather
forecasting. There are generally only minor differences between the NRT Level-1 images
and their definitive counterparts. NRT Level-1 bad-pixel records do
not identify cosmic-ray hits because
the code computing them requires analysis of a time series. Moreover, the flat
fields, daily calibration parameters, drift coefficients, and flight dynamics
data applied to the NRT Level-1 images may not be quite up-to-date. If NRT
science data are not retrieved in a timely fashion, records may be skipped.
Before definitive Level-1 pipeline processing is completed, the operator
manually verifies that all possible data files have been received from the
SDO ground system. The complete set of definitive Level-1 records (usually
available with a 3- to 4-day delay) are used for processing the definitive 
observables. 
NRT filtergrams and downstream observables are produced within 
minutes of receiving the raw HMI images.

The website
\urlurl{jsoc.stanford.edu/cvs/JSOC/proj/lev0/apps/build\_lev1\_hmi.c}
provides the source code for the Level-1 pipeline.

\subsection{The Observables Processing Pipelines} \label{sec:ObservablesPipelines}

Two distinct software pipelines produce the HMI observables from Level-1 filtergrams.
One computes the LoS observables on a 45-s cadence. 
The other computes time-averaged vector and LoS observables every 720\,s.
Observables are computed from sets of filtergrams taken in a fixed,
repeating sequence called a framelist. 
The framelist specifies the wavelength and polarization state for each exposure. 
Throughout the prime mission and until 13 April 2016 each camera
produced a complete and independent series of filtergrams from which observables
have been computed with the corresponding pipeline. 
The standard 45-second 12-frame sequence
used for the LoS camera collects two filtergrams in each of six wavelengths,
one in right and the other in left circular polarization. The 36-frame
vector-camera sequence requires 135 seconds; it collects four additional
linear polarization states in each of the six wavelengths in order to
determine the full Stokes vector.

The pipelines consist of modules written in C and the
JSOC code is available at 

\noindent \urlurl{jsoc.stanford.edu/cvs/JSOC/proj}.

\subsubsection{The LoS Pipeline -- 45-s Data}
\label{sec:LoSObservables}
The LoS observables calculation is implemented in the  
{\sf HMI\_observables} module. For each 45-second time step the module 
identifies and retrieves the proper Level-1 filtergrams, applies a series of 
corrections to the individual images, interpolates the filtergrams to the 
specified time, and combines the calibrated filtergrams to compute 
the observable quantities. 
Details are provided in the indicated subsections.

Section \ref{sec:45sSelection} describes how the individual filtergrams
are selected and summarizes the calibration processing applied to them.
Sections \ref{nonlinearity} and \ref{sec:Alignment} give more details about
CCD linearity and the corrections for optical, spatial, and temporal alignment.
Section \ref{sec:LoSPolarization} briefly describes the polarization correction
made for the LoS magnetic field computation.
Once the filtergrams are fully calibrated, the observables are computed
using the MDI-like algorithm described in Section \ref{sec:MDIIntro}. 

Note that the HMI observables are not cropped right at the limb, 
because some provide valid measurements above the limb, e.g. \cite{HMIOffLimb2014}.
The crop radius increased from 50 to 90 pixels off the solar limb on 15 January 2014. 
While the magnetic field products are just noise off the limb, the project
leaves it to the user to choose where to crop depending on the purpose of
the investigation.

Section \ref{polcor} describes the current approach implemented to correct the reported
velocities for some of the deficiencies caused by SDO's
large orbital velocity using a comparison of the median full-disk velocity to
the accurately known spacecraft velocity.

\subsubsection{The Vector Observables Pipeline --- 720-s IQUV Generation}
\label{sec:IQUV}

The vector pipeline module {\sf HMI\_IQUV\_averaging} produces averaged 
I, Q, U, and V images at six wavelengths on a regular 12-minute cadence.
The vector-field observing sequence, run on the HMI side camera, captures six
polarizations at each wavelength according to a repeating 135-second framelist.
The two pipelines share many steps: the gap filling, the de-rotation, the re-centering of the images,
and the polarization calibration.  However, rather than performing a
temporal interpolation,
it computes a temporal average.
Conceptually, this averaging is executed in two steps. 
First, like the spatial interpolation described in Section \ref{sec:Alignment}, 
a temporal Wiener interpolation of the observed filtergrams onto a regular
temporal grid with a cadence of 45\,s is performed and short temporal gaps
are filled. In this case the assumed covariance is derived
from the observed average power spectrum of a single pixel in an image. 
This results in a set of 25 frames for each
wavelength/polarization state constructed using the ten original 135-s
framelists. The full time window over which the interpolation is performed
is 1350\,s, which is wider than the averaging window; a wider window
is required because the interpolation needs filtergrams before and after the
interpolated times. The 25 interpolated frames are then averaged using an
apodized window with a FWHM of 720\,s; the window is a boxcar with cos$^2$
apodized edges that nominally has 23 nonzero weights, of which the central
nine have weight $1.0$ \citep{Hoeksema2014}. In reality the interpolation 
and averaging are done in a single step for computational efficiency.

Following the temporal averaging
the six polarized filtergram are converted
into a Stokes [I\,Q\,U\,V] vector. The details of the
polarization calibration are given in Section \ref{sec:Polarization}. 
The final results are stored in {\sf hmi.S\_720s}.
The other 720-s observables are computed from I$\pm$V using the same
algorithm as the LoS pipeline (Section \ref{sec:MDIIntro}).

\subsection{Filtergram Selection, Mapping, and Image Processing}\label{sec:SpaceTime}

This section describes how filtergrams for the LoS pipeline are 
selected and reviews how they are processed individually to correct for
various problems. 
Subsequent sections detail how the pipeline deals with nonlinearity in the 
CCD cameras (Section \ref{nonlinearity}) and with spatial alignment and 
distortion corrections (Section \ref{sec:Alignment}). 
Section \ref{sec:Venus} describes the relative and absolute roll 
angles of the CCD cameras and validation of the distortion and 
roll determination using the Venus transit.
Later sections describe issues with polarization (Section \ref{sec:Polarization})
and the wavelength filters (Section \ref{sec:Filters}).

\label{sec:45sSelection}

The filtergram selection module 
gathers selected keywords of all the Level-1 records in a time interval
around the target time {\sc t\_obs} at which an observables computation is
requested by the user.  
The Level-1 image at the proper wavelength nearest to the target time {\sc t\_obs} 
for which the observables record is to be produced is identified as the {\em target filtergram}.
Since launch the HMI filtergram closest to and slightly blueward of the rest wavelength 
of the core of the Fe\,{\sc i} line, has been used. 
Certain keywords of the target filtergram are used as reference values for the
final observable data record. For instance, the focus block used to take the
target filtergram is the reference focus block for the computation.
If another filtergram is taken with a different focus, this is an error 
and no observable will be created.

The code then locates another filtergram taken with the same wavelength
and polarization settings as the target to linearly interpolate values of
the components of the spacecraft velocity ({\sc obs\_vr, obs\_vw}, and {\sc
obs\_vn}), the solar distance ({\sc dsun\_obs}), the Carrington coordinates
of the disk center ({\sc crlt\_obs} and {\sc crln\_obs}), and the position
angle of the Carrington rotation axis ({\sc crota2}) for the target time,
{\sc t\_obs}. 
Spacecraft ranging data regularly confirm that the location of the 
spacecraft is known to [much] better than 120-m accuracy and the velocity to better 
than 0.01 m~s$^{-1}$.
Note that {\sc crota2} is the negative of the classic ``p
angle.'' As SDO has been operated since launch, the HMI {\sc crota2} has
remained close to -180 degrees except during roll maneuvers (see Section
\ref{RollCalibration}). 

A gap-filling routine is called 
to replace the pixels identified in the bad-pixel list. 
Similar to the spatial interpolation (see Section \ref{sec:Alignment} for
details), the gap-filling uses a Wiener interpolation assuming the covariance
function obtained from the perfect modulation transfer function (MTF).
But unlike that interpolation, this is done as a general 2D interpolation, with
weights calculated based on the exact pattern of missing pixels surrounding
the target pixel. Also, a term is added to the optimization to minimize both
the sum of the variance from the inaccurate interpolation and the variance
from the photon noise, rather than just the former.

The HMI cameras are affected by a small non-linearity in their response to
light exposure that 
is corrected separately for the LoS and vector cameras
(see Section \ref{nonlinearity}). 

The code then retrieves and corrects each of the Level-1 filtergrams
needed to produce an image at time {\sc t\_obs} with the required wavelength
and polarization setting. 
When all of the necessary filtergrams have been prepared, an interpolation
module performs several tasks. First it corrects each
image for instrumental distortion. The distortion as a function of
position is reconstructed from Zernike polynomials determined during
pre-launch calibration that employed a random-dot target mounted in the
stimulus telescope \citep{Wachter2012}. Section \ref{sec:Alignment} describes an
evaluation of the measured
instrumental distortion. The routine also corrects the Sun-center coordinates
and solar radius keywords to account for modifications due to the distortion
correction. 
Because each
Level-1 record used to compute an observable is taken at a slightly different
time, features on the solar disk move a small fraction of a CCD
pixel, so that a given pixel in two filtergrams does not sense exactly the same 
location on the solar surface. This rotation is corrected to subpixel accuracy using a
Wiener spatial-interpolation scheme. The time difference used to calculate
the pixel shift is the precise observation time of the Level-1-filtergram
{\sc t\_obs}. The filtergrams are also re-centered and re-sized 
to the same values. These values are obtained by averaging the
characteristics of all the Level-1 images used to produce the
observable. Next the filtergrams are temporally interpolated
to the target observable time, {\sc t\_obs}.
For a given wavelength/polarization pair,
the temporal interpolation ordinarily requires six Level-1 filtergrams in
definitive mode and two in NRT mode to interpolate to the requested time. 
The two-point NRT temporal interpolation is a basic linear scheme,
while the definitive six-point method uses the specific weighting scheme described in
\citet{MartinezOliveros2011}.
When the loop over all wavelengths and polarizations is complete, the result
is a set of filtergrams with the same solar radius and Sun-center position
all interpolated to the proper observable time, {\sc t\_obs} at the spacecraft.

Note that the observables have at least two time keywords: {\sc t\_rec}
and {\sc t\_obs}. {\sc t\_rec} is a prime keyword and is the time the data
would have been observed at precisely 1 AU, whereas {\sc t\_obs} is the
clock time of the midpoint of the observation at the SDO spacecraft: 
{\sc t\_obs = t\_rec + (dsun\_obs}\,-\,1 AU)/$c$), where $c$ is the speed of light. 
The {\sc t\_rec} times are by design uniformly spaced in (TAI) time and
therefore convenient to use. However, the time of observation for which the orbit 
keywords best describe the observable is the time at SDO, {\sc t\_obs}.

The final processing step is to correct for polarization issues 
and create a set of filtergrams that better represent the true I$\pm$V
polarizations with less cross-contamination; those images are called
Level-1p data. See Section \ref{sec:LoSPolarization} for further discussion.
The Level-1p records can optionally be saved.

\subsection{CCD Non-linearity \label{nonlinearity}}

The signal (in DN) measured in a given pixel does not vary exactly linearly
with the number of incident photons. \citet{Wachter2012} determined that
the non-linearity of the HMI CCDs is on the order of 1\% for intensitites
less than 12,000 DN from ground-calibration data. Typical 140-ms exposures
are about 4200 DN, giving median normalized Level-1 filtergram values
of 30,000--50,000 DN/s, depending on wavelength. Even apart from solar
variations, the number of photons received by each CCD pixel is not constant
for a given exposure time for a variety of reasons, including the daily
change in the Sun-SDO distance. A non-linearity correction is therefore
implemented in the observables pipelines for each Level-1 image. 

The initial correction was based on the results of \citet{Wachter2012}. The
difference between the actual intensity and a linear response was fit as a
function of intensity using a 3rd-order polynomial. The coefficients of this
polynomial were: $-11.08$, $0.0174$, $-2.716 \cdot 10^{-6}$, and $6.923 \cdot
10^{-11}$ for the front camera and $-8.28$, $0.0177$, $-3.716 \cdot 10^{-6}$,
and $9.014 \cdot 10^{-11}$ for the side camera.

Since 15 January 2014 different coefficients have been used. The main reason
for the change is that the negative value for the zeroth-order coefficient
(intercept term) in the original fits means that a few pixels ended up with
a negative (albeit small) intensity, which does not make physical sense. The
coefficients used after 15 January 2014 are: $0$, $0.0207$, $-3.187 \cdot
10^{-6}$, and $8.754 \cdot 10^{-11}$ for the front camera and $0$, $0.0254$,
$-4.009 \cdot 10^{-6}$, and $1.061 \cdot 10^{-10}$ for the side camera.

The version of the non-linearity calibration is given in the keyword {\sc calver64},
see Section \ref{sec:Fringe}.

\begin{figure}[!htb]
\centering
\includegraphics[width=0.9\textwidth]{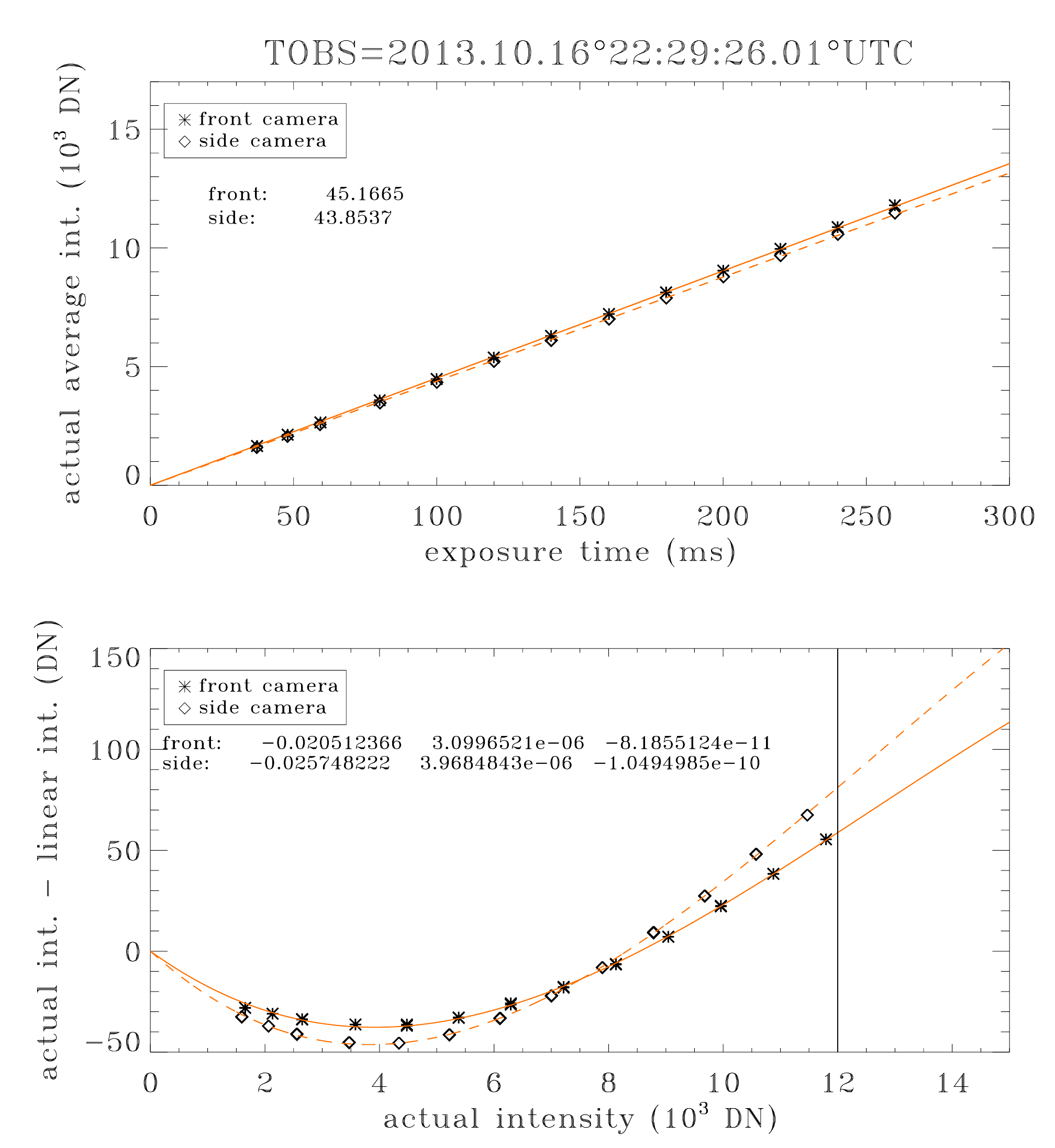}
\caption{Results of the non-linearity analysis for the front and side cameras 
for 16 October 2013. The top panel shows the measured intensity vs. exposure 
time. The bottom panel shows 3rd-order polynomial fits to the residual as a 
function of intensity. A typical HMI exposure is about 140 ms.}
\label{fig9}
\label{fig:Nonlinearity}
\end{figure}

Calibration sequences are taken regularly on orbit to monitor the non-linearity of 
the CCDs. Though a slightly different non-linear response can be detected in each 
quadrant \citep[Figure 19 of][shows a total spread of no more than 5-10 DN]{Wachter2012}, 
we measure the spatial average over the entire CCD. So far, the non-linearity has 
proven constant. Figure \ref{fig:Nonlinearity} shows the result for a typical 
non-linearity sequence recorded on 16 October 2013.

\subsection{Distortion Correction and Image Alignment} \label{sec:Alignment}

Image distortion arises because
of small imperfections in the optics, including the optics that move
to tune the instrument. The correction is based on Zernike
polynomial coefficients measured with ground data taken prior to launch
\citep[Fig. 7 in][]{Wachter2012}. The mean residual distortion is $0.043 \pm 0.005$ pixels,
with a maximum less than 2 pixels near the top and bottom of the CCD camera.
Differences between the cameras are of order 0.2 pixels.
The instrumental distortion correction is applied to each Level-1 filtergram. 
Production of definitive LoS observables typically involves 72 
filtergrams, while a definitive 12-minute averaged Stokes
vector requires 360 Level-1 filtergrams.

Each filtergram has slightly different Sun-center location coodinates
and p~angle.
Therefore, before performing a temporal interpolation to {\sc t\_obs},
each Level-1 filtergram must be registered and aligned. 
Conceptually each image is first rotated to a common 
p angle and adjusted to a common solar inclination 
(B0-angle). Then the effect of solar differential rotation is removed by
spatially interpolating to the proper spatial coordinates at the target time. (Near disk center solar rotation carries features across a pixel in about 3 minutes.)
The de-rotated images are then re-centered around a common Sun-center point 
that is the average of all the input Level-1 filtergrams. In practice all
of these operations are performed in a single interpolation step.

The spatial interpolation is done using a separable (in x and y) Wiener
interpolation scheme of order 10. Such a scheme 
minimizes the rms error of the interpolation for a specified
covariance. Here the covariance function corresponds
to the ideal diffraction-limited MTF of the instrument, i.e. the covariance
expected if a white noise signal were observed by an ideal instrument. Of
course, the instrument is not ideal and the inherent solar signal is not
white, but this only affects the relative convergence at different spatial
frequencies. Because the HMI pixels undersample
the point spread function (PSF) by about 10\%, it is impossible to perfectly
interpolate the data. However, this only affects the spatial frequencies
to which the super-Nyquist signal folds. Consequently the interpolation
(and the resulting power spectrum) is very good up to $\sim$0.9 times the
Nyquist frequency and imperfect above that.

After the image distortion has been corrected an analytical correction
is applied to the original Level-1 Sun-center position and
solar radius determined by the limb finder. Unfortunately, tests
show that the analytical correction is not precisely correct,
particularly for the y-direction Sun-center location reported in keyword {\sc
crpix2} (nominally the solar north-south location). The {\sc crpix2} value
is systematically greater than the more accurate center position determined
when the limb finder is applied to the distortion-corrected image by $0.13$ to
$0.17$ pixels, depending on the filtergram wavelength. The systematic y-axis
difference also depends predictably on image location. In the x direction
(reported in {\sc crpix1}) the center-position difference is roughly ten
times smaller, and for the solar radius it is less than a hundredth of a
pixel. Consequently, users are cautioned that currently the {\sc crpix2}
keyword in the observables records is systematically off by $\sim 0.15$
pixels. Random variations in measured center position from one image to the
next due to all noise sources are $< 0.15$ pixels for the 45-s observables 
and $< 0.05$ pixels for the 720-s observables.

\subsection{Roll, Absolute Roll Calibration, Distortion, and the Venus Transit}
\label{sec:Venus}
\label{distortion}

This section describes the determination of the relative roll angle
of the two HMI cameras from daily calibration measurements and how the roll 
difference varies with time. Analysis of Venus-transit observations provides 
an accurate determination of the absolute roll angle of the HMI instrument,
as well as independent information about optical distortion. Knowledge 
of the instrument roll angle is important because solar rotation can be transferred 
into measurements of north-south motions; for example, a roll error of 
$0.1^\circ$ would introduce a systematic 3.5 m\,s$^{-1}$ 
northward or southward flow.  Investigations that depend on coalignment of filtergrams 
or with measurements from other instruments can also be affected.
The first objective is to determine the difference in the roll between the
two cameras. Pairs of adjacent (in time) images on the front
and side cameras with otherwise identical settings are 
corrected for distortion using the same parameters employed to make the observables.
The images are then divided into $256 \times 256$-pixel regions, each of
which is high-pass filtered and circularly apodized with a raised cosine
between 0.8*128 pixels and 0.9*128 pixels. For each region the two images
are then cross correlated to determine the shift in x and in y. Finally,
for locations inside the solar image, these shifts as a function of position
on the image are fitted to a model of the shifts with parameters for
the x and y offsets, a scale error and a roll angle.

First this was applied to all 640 pairs of images with one particular
polarization and wavelength setting on 1 July 2012, which gave a mean offset
of $0.08361^\circ$ with a scatter of $0.00061^\circ$ and thus a resulting
error of the mean of $0.00002^\circ$. An estimate of the accuracy of the
number is difficult to obtain, but the rms residual in each direction is
roughly 0.1 pixels and, as shown in the next subsection, the distortion
model is likely uncertain by about the same amount. Since the average
solar radius is $\sim 1900$ pixels, it is reasonable to
expect an uncertainty of the order 0.1/1900 radians or $0.003^\circ$.
Tests performed on a few other days gave average roll offsets that are
slightly higher, about $0.08376^\circ$. A nominal value of $0.0837 ^\circ$
has been adopted for the roll difference between the two cameras.

As in the case of other datasets, the intermediate data series were not saved,
but a rerun of the code on images with current calibration and processing
gives values consistent to the number of significant digits given above.

To investigate the time dependence of the roll difference, the
process is repeated on the daily calibration images taken by HMI,
specifically on the pairs of tuned continuum images obtained with the
two cameras at about 06:00 and 18:00 UT. 

\begin{figure}[!htb]
\centering
\includegraphics[width=0.95\textwidth]{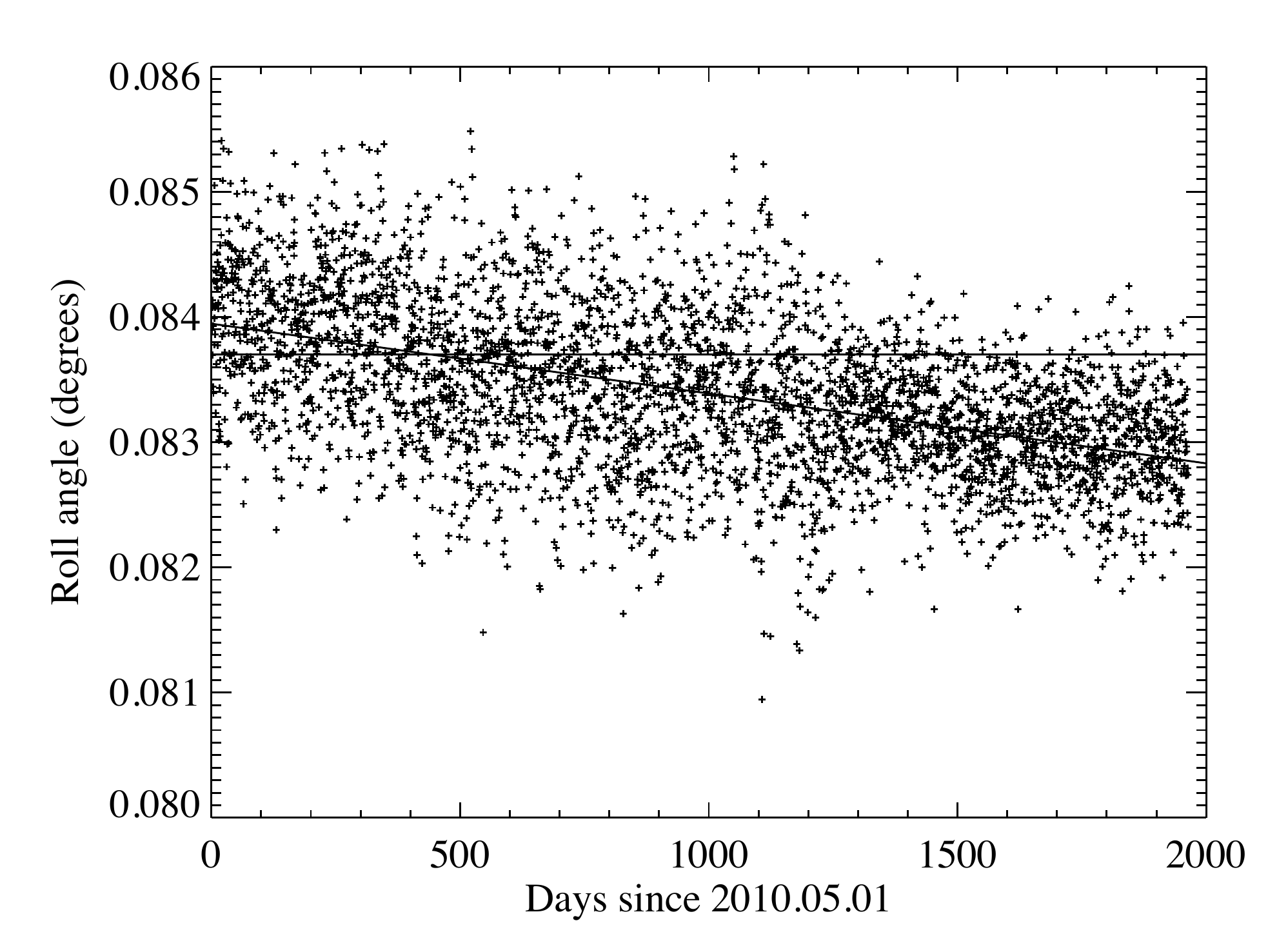}
\caption{
The roll-angle difference between the two HMI cameras determined from the
daily calibration images, shown as a function of time. Also shown is the
adopted nominal value of $0.0837^\circ$ and a linear fit to the data with 
time. The time period covered is 2 May 2010 through 15 September 2015.
}
\label{fig:CameraRoll}
\end{figure}

Figure \ref{fig:CameraRoll} shows that the roll difference drifts slowly with time at a uniform rate of
$-0.00020^\circ$\,

\noindent yr$^{-1} \pm 0.00006^\circ$\,yr$^{-1}$. This corresponds
to an offset of roughly 0.03 pixels at the limb over the five
years, which may be significant when one attempts to combine data from the two
cameras. While the cause of the small drift is not known, three possibly 
relevant improvements to the instrument temperature control scheme were 
made that may have decreased the scatter in the measurement: one on 16 July 2013 
(mission day 1142) to the optical bench, another to the telescope tube on 
25 February 2014 (day 1395), and the third on 16 and 26 June 2014 (days 1517 and 1527) 
to the front window (see \citet{Bush2015} for details.)

The Venus transit of 5\,--\,6 June 2012 provided a unique opportunity to
test the accuracy of the absolute roll angle and parts of the distortion map.
The normal filtergram sequence was run on the front camera, but on the side
camera images were taken in the true continuum in linear polarization.
Images first undistorted in the standard way had a simple model of the solar limb
darkening removed.  Then an area of roughly $100 \times 100$ arcsec around
the expected position of Venus is extracted from each image and the radial
derivative of the intensity is calculated. This is then multiplied by an
apodization function to isolate the limb, cross correlated with a $180 ^\circ$-
rotated version, and the center position of the Venus image is determined.

The Venus positions are then fit, separately for each camera, to the
ephemeris using a model that determines the image offset, image scale,
and roll angle. Considering the accuracy of the spacecraft orbit, the
ephemeris is presumed to be perfect. The small roll-angle values reported
in this section are the p~angle relative to the nominal orientation of the
instrument, i.e. p~angle = 180 - {\sc crota2}, with solar south at the 
top of the CCD.  The original analysis determined absolute roll angles
of $-0.0142 ^\circ$ and $+0.0709 ^\circ$ for the two cameras, with the
difference being $0.0851 ^\circ$.  A reanalysis of the same images processed
with current calibration software gives roll angles of $-0.0140 ^\circ$
and $+0.0712 ^\circ$, corresponding to a roll difference of $0.0852 ^\circ$.
Compared with the value determined by direct comparison reported in the
previous section ($0.0837 ^\circ$), this confirms that the absolute roll values are likely accurate to a few thousandths of a degree.

\begin{figure} [htb]
\centering
\includegraphics[width=0.95\textwidth]{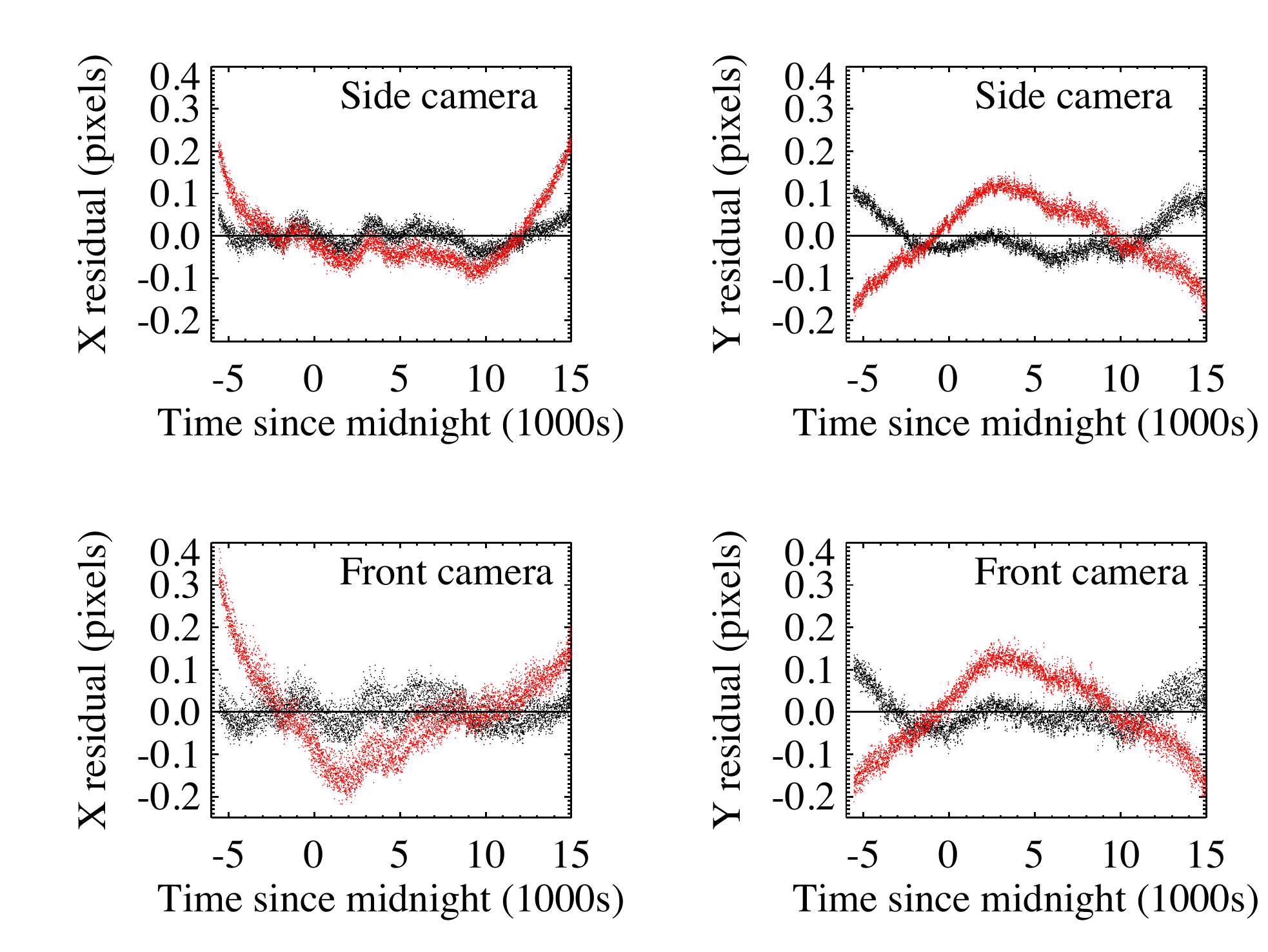}
\caption{Residuals from the fit of the ephemeris data to the measured 
positions of Venus. Red points show the results using raw data, while the 
black points use the distortion-corrected data. The top panels show the x and y 
residuals for Camera 2 (the side camera) and the bottom for Camera 1 
(the front camera). Times are given in thousands of seconds relative to 
0 UT on 6 June 2012. The side camera observed only continuum intensity in 
linear polarization.
}
\label{fig:VenusResiduals}
\end{figure}

Residuals in x and y for each camera are shown in Figure \ref{fig:VenusResiduals}. 
The distortion-corrected data (black points) show substantially reduced 
residuals, but systematic deviations of order 0.1 pixels still remain. 
This is consistent with roll uncertainties of the order $0.003 ^\circ$,
confirming that the discrepancies between the methods are negligible.

A second analysis using the Venus transit was performed to better characterize
the image distortion of the side camera. A PSF estimate (see Section \ref{PSF}) 
for HMI was deconvolved from the images using a Richardson-Lucy algorithm to 
improve the estimates of the Venus-center location.  The results for roll and 
distortion are not significantly different. 

The similarities in the results for the front and side cameras shown in Figure
\ref{fig:VenusResiduals} suggest that the remaining distortion errors arise
in parts of the optical path common to both cameras. The increased scatter
(even sometimes double values) in the residuals for the front camera shown in
the lower panels of Figure \ref{fig:VenusResiduals} were due to filtergrams
taken in different wavelengths, so there is a remaining sensitivity to tuning
at the level of 0.03 pixels. This confirms that the instrumental distortion
model is accurate to about 0.1 pixel, at least for the pixels lying on the
Venus path.

Finally the two sets of numbers must be reconciled. The roll difference from
the direct comparison ($0.0837 ^\circ$) differs from the Venus data ($0.0851
^\circ$)  by less than the expected uncertainty. Furthermore, the
Venus numbers are effectively derived from only a small portion of the solar
images, whereas the direct comparison is effectively an average over the full
disk. Since one of the objectives is to be able to combine full-disk data from
the two cameras, it was decided to use the direct-comparison difference and
the well-determined absolute average p~angle from the Venus measurements. For
the two cameras we obtain the results given in Table \ref{tab:pangles}.

\begin{table} [htb]
\caption{P-angle Table}
\label{tab:pangles}
\begin{tabular}{r l}
\hline
Camera & P angle \\
\hline
Front Camera 1 & $(-0.0142+0.0709)/2 - 0.0837/2 = -0.0135 ^\circ$ \\
Side Camera 2  & $(-0.0142+0.0709)/2 + 0.0837/2 = +0.0702 ^\circ$ \\
\hline
\end{tabular}
\end{table}

Clearly it would be desirable to obtain independent determinations of the
absolute roll. While there have been no other planetary transits since
launch, the next Mercury transits will occur on 9 May 2016 and 11 November
2019. Use of more frequent lunar transits 
presents significant difficulties. Only the ingress can be used due
to the massive thermal perturbations later in the transits. Complications
also arise because the Moon is far from perfectly spherical, so one would have to
use accurate maps of the lunar topography and likely use the lunar mountains
clearly visible in the HMI images to derive an estimate. An attempt to detect
the star Regulus was made, but was not successful.

\subsection{Polarization}\label{sec:Polarization}

This section describes some of the issues with telescope polarization and
corrections made to minimize contamination.  The pipeline module {\sf polcal} 
performs the calibration in several steps.

\begin{itemize}

\item 
First the model described in \citet{SchouPolarization2012} is used to determine
the modulation matrix for the frames. This takes
into account the polarization selector position and
temperature. Because the temperature gradient across the front window is not known, 
it is assumed to be zero.  A single average temperature is used for
all frames.  Given the nature of the model, the I~$\rightarrow$~(Q\,U\,V)
and (Q\,U\,V)~$\rightarrow$~I terms used here are zero and the I\,$\rightarrow$\,I
term is 1. Because the calibration model is only given on a $ 32 \times 32$
grid, so is the modulation matrix at this point.

\item At each pixel a least-squares fit is performed to determine the
demodulation matrix, which relates the observed intensities in the various
frames to I\,Q\,U\,\&\,V.

\item For each pixel in the $4096 \times 4096$ image the demodulation matrix is
linearly interpolated from the $32 \times 32$ grid and applied to the observed
intensities.

\item A telescope polarization correction is applied by subtracting a
small part of I from Q\,U\,\&\,V. The model used and how it was determined
is described in Section \ref{sec:Telpol}.

\item A polarization-PSF correction is made by convolving I with 5x5 kernels and adding
the result to Q\,U\,\&\,V. The model used and how it was determined is
described in Section \ref{sec:PolPSF}.

\end{itemize}

Calibration of the circular polarization measured with the front camera is
briefly described in Section \ref{sec:LoSPolarization}.

The calibration requires the temperatures of some HMI components
that impact the polarization calibration. The temperatures are 
updated once a day for the entire day. 
In NRT mode default temperature values are used.

\subsubsection{Telescope Polarization}\label{sec:Telpol}
The determination of the telescope polarization was performed after launch
because it is notoriously difficult to determine accurately on the
ground and quite straightforward to determine on orbit.

To determine the telescope-polarization term 820 12-minute IQUV averages with
good quality from nine days between 3 May 2010 and 3 September 2010 were used.
The images of (Q,U,V) were first binned to $256 \times 256$ pixels and divided
by the corresponding binned I images; then for each wavelength, polarization,
and pixel the median-in-time over the 820 samples was calculated.  The median
was used to suppress the effect of solar activity.  Because 
activity is still quite visible in the line and because using the continuum is
in any case preferable, the two wavelength positions closest to the continuum
(I0 and I5) were averaged, which also has the effect of canceling the effects
of the orbital velocity variations to lowest order.

\begin{figure} [!ht]
\centering
\includegraphics[width=.85\textwidth]{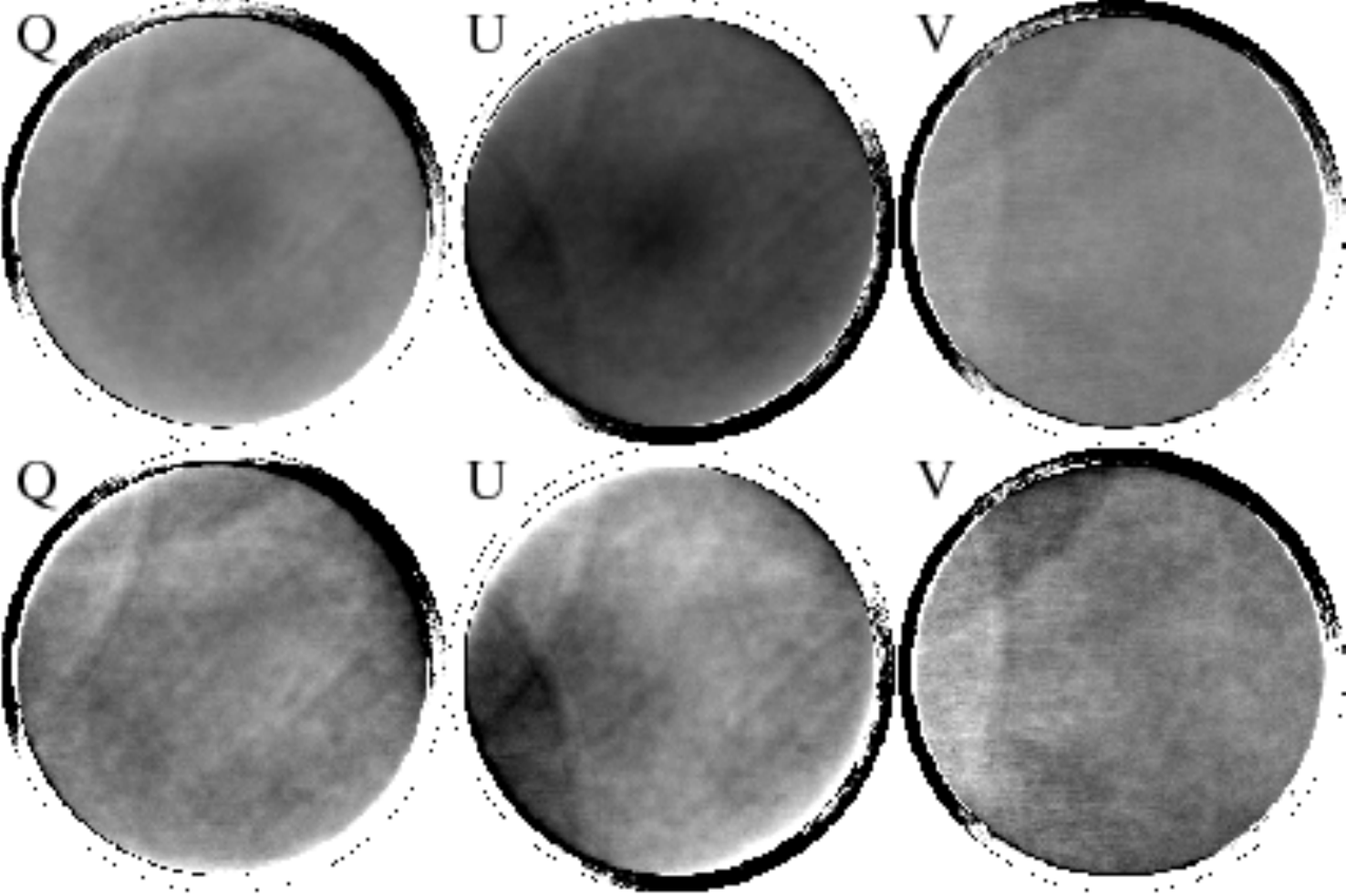}
\caption{The telescope polarization terms, determined as described in the main 
text.  Top row shows the raw images on a scale of $\pm 2.5\cdot 10^{-4}$, while 
the bottom row shows the residuals after subtracting the polynomial fit from Q
and U on a 
scale of $\pm 10^{-4}$.}
\label{fig:telpolarity}
\end{figure}

The resulting images are shown in the top row of Figure
\ref{fig:telpolarity}. The dominant effect is an offset, but there is also
a radial gradient.  A 4th-order polynomial
in the square of the distance to the center of the image models this well. Since
the effect in V is small, no correction is applied to it.  The bottom row in
Figure \ref{fig:telpolarity} shows the residual after the polynomial has been
subtracted.  While the constant part could indeed be due to a small amount
of polarization in some optical component, the cause of the radial variation
is unknown. However, the variation is very
similar for Q and U.  The rms residuals within a distance of 0.85 times the
half width of the image (corresponding roughly to 0.93 times the radius of
the solar image) are 14 ppm, 28 ppm and 18 ppm for Q, U and V, respectively.
The most prominent features in the residuals are the arcs on the left.
The cause of the arcs is unknown, but they may be due to a ghost reflection.

The instrument easily meets the original specification for polarization 
which is 1000 ppm for I\,$\rightarrow$\,Q\,U\,V and 100 ppm for cross-talk among Q\,U\,\&\,V
\citep{SchouPolarization2012}.

While the correction is modeled as if it originates in the
telescope part of the instrument, it is clear
that it actually does not. During commissioning a different set of polarization
settings were used and the constant terms were quite different. As such it is
likely that the root cause lies elsewhere, such as in second-order effects
in the waveplates, in the fold mirror, or in the polarizing beamsplitter;
however, this has not been investigated further.

Given this lack of physical understanding, the term has not been used to
determine the corresponding (Q,U,V)~$\rightarrow$~I terms, which are left
at zero. Given that the terms are of the order $10^{-4}$, the effect on the
intensity term is in any case very small.

Some of the files used for the original calibration are no longer available
and reproducing them with exactly the same calibrations and corrections is 
not practical. To verify that the results are reliable, the analysis was
repeated on the already-calibrated data using the same set of 820 12-minute
averages. Ideally this should result in zero for the calibration terms. In
reality it leads to polynomials that are nearly constant across the disk
with offsets for Q and U of $-1.4 \cdot 10^{-6}$ and $1.1 \cdot 10^{-6}$.
For V (which does not have a calibration applied) the mean correction is
$-1.6 \cdot 10^{-9}$. The cause of the small change is unknown, but 
one of the differences in processing is that bad pixels may have been unfilled 
in the original analysis, and a second is that the polarization-PSF correction 
described below was not applied. In any case the differences are negligible.

\begin{figure} [!htb]
\centering
\includegraphics[width=.80\textwidth,height=0.75\textheight]{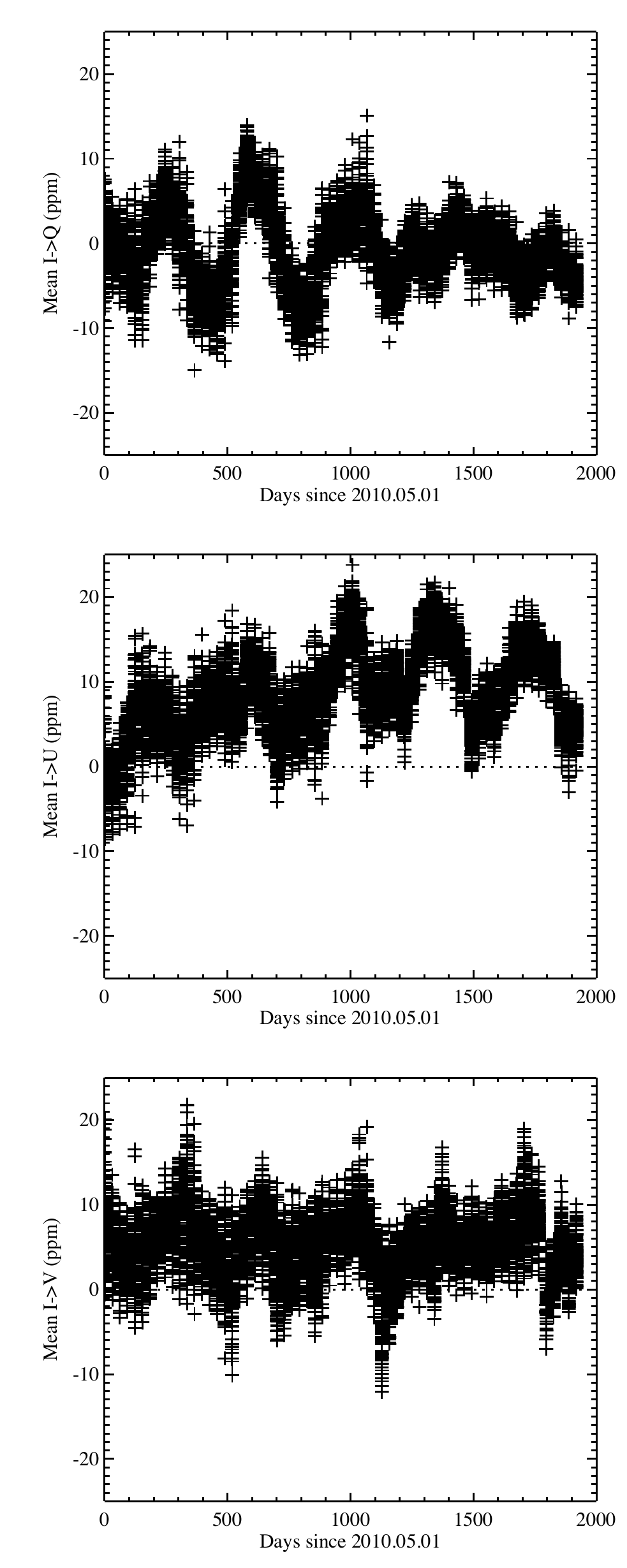}
\caption{Plots of the mean of the telescope polarization terms in parts per million
after the standard correction has been applied as a function of time.  Only results 
for the first day of each month between May 2010 and August 2015 are shown. The means 
are calculated inside a radius of 1741 pixels, corresponding to roughly 0.93R.}
\label{fig:meanbig}
\end{figure}

To determine whether the correction is stable in time
12-minute averages with perfect {\sc quality} for the first day of each
month from May 2010 through August 2015 are analyzed in the same way. 
Figure \ref{fig:meanbig} shows the mean of the telescope polarization terms. The top panel shows 
I~$\rightarrow$~Q, the middle panel I~$\rightarrow$~U, and the bottom panel
I~$\rightarrow$~V, in parts per million.  Some temporal changes are seen,
with a magnitude of up to at most $\sim 2 \cdot 10^{-5}$ with some annual
periodicity. The cause of these changes is unknown, but could be due to
changes in the front window temperature. In any case, the changes are
small compared to the residuals shown in Figure \ref{fig:telpolarity} and the 
correction has been kept fixed in time.

\subsubsection{Polarization-PSF Correction}\label{sec:PolPSF}

\begin{figure} [htb]
\centering 
\includegraphics[width=.65\textwidth]{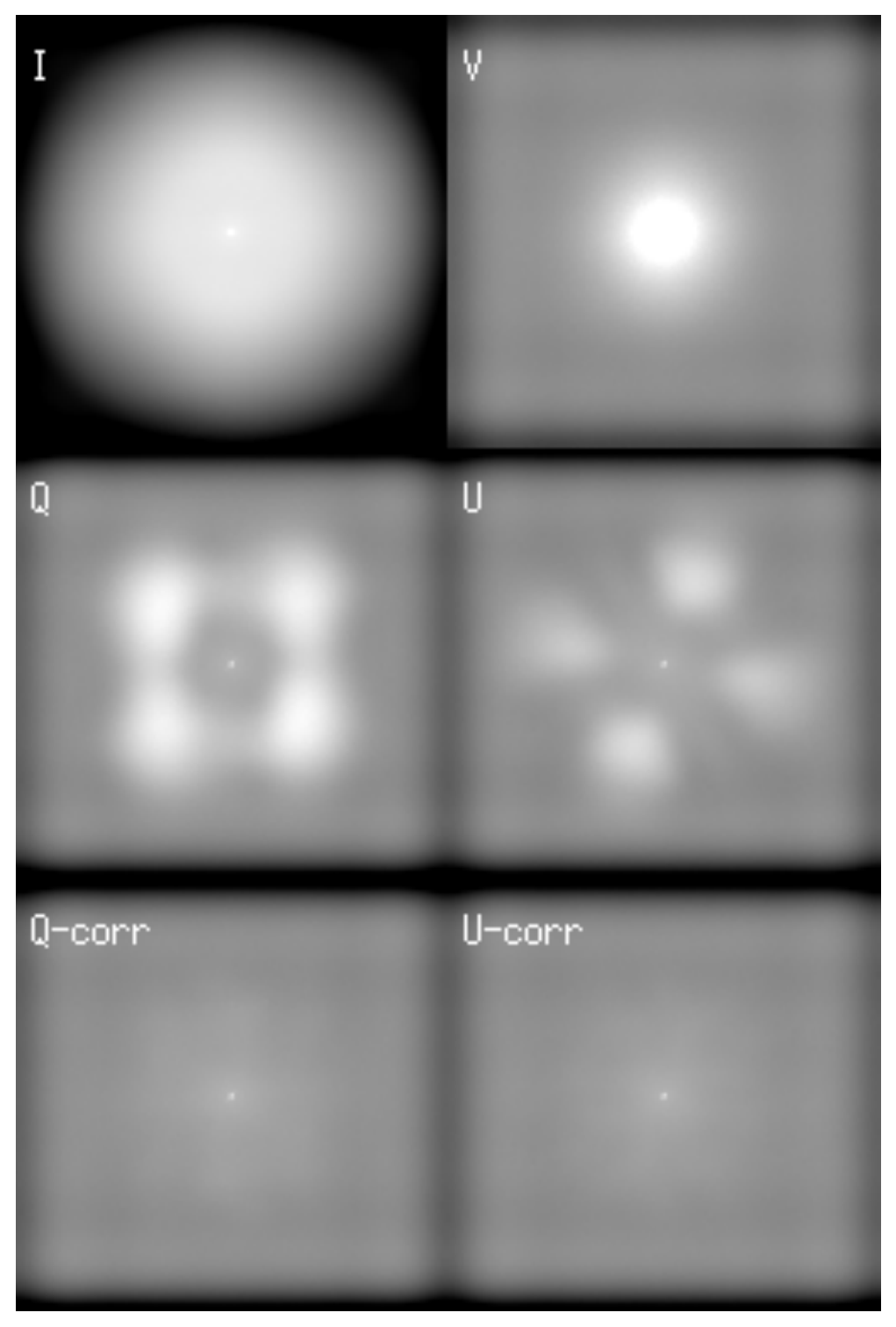}
\caption{Average power spectra of I, Q, U and V with and without correcting
for the polarization-PSF effect.  Intensity I is on a logarithmic gray scale
of $10^5$, while Q, U, and V are on a log scale of 100.  The Q, U and V spectra
are all on the same power scale.  The average is over the samples used for
estimating the correction (excluding 3 August 2010, due to a large sunspot)
and is shown for disk center.  The individual images were mean subtracted
and circularly apodized with a cosine between 0.8 and 0.9 times the half
width of the patch.}
\label{fig:psfpow}
\end{figure}

After the telescope polarization correction is applied, another
artifact becomes apparent, {\em viz.} a granulation-like pattern appearing in Q
and U (but again not in V). An analysis of the average power spectrum of the
I, Q, U, V components (Figure \ref{fig:psfpow}) shows that the pattern
is not simply a leak of I into Q and U, but rather appears to be a filtered
version of I.

While the instrumental cause is unknown, the spectra look similar to what one might expect
if there were a different PSF when observing, for example, I+Q and I-Q. To
correct for the effect, a least-squares minimization procedure (including
a small amount of regularization) is used to find the best 5x5 kernel
that, when convolved with I, best reproduces Q, U, and V. We impose the
additional mathematical constraint that the sum of the kernel is zero in
order to avoid interference with the telescope polarization correction.
This estimate is computed with a subset of 612 samples from the same
datasets used for the telescope polarization.  From these a $512 \times 512$
patch at disk center is extracted and only the two continuum-wavelength
polarizations are used.  As can be seen from the bottom panels of Figure
\ref{fig:psfpow}, this procedure dramatically reduces the contamination.
Larger sized kernels and spatially varying kernels were also evaluated,
but did not substantially improve the results.

Because the datasets as processed for the original analysis are no longer
available, new ones had to be constructed using images computed using the
current calibration to verify the reproducability of the result. 
This was done by convolving I with the adopted kernels and adding it back 
into to the already corrected Q and U images before repeating the analysis. 
The results are very nearly, but not exactly, identical
to the original. The differences are much smaller than the photon noise
and do not cause noticeable differences.

\begin{figure} [!htb]
\centering
\includegraphics[width=.72\textwidth]{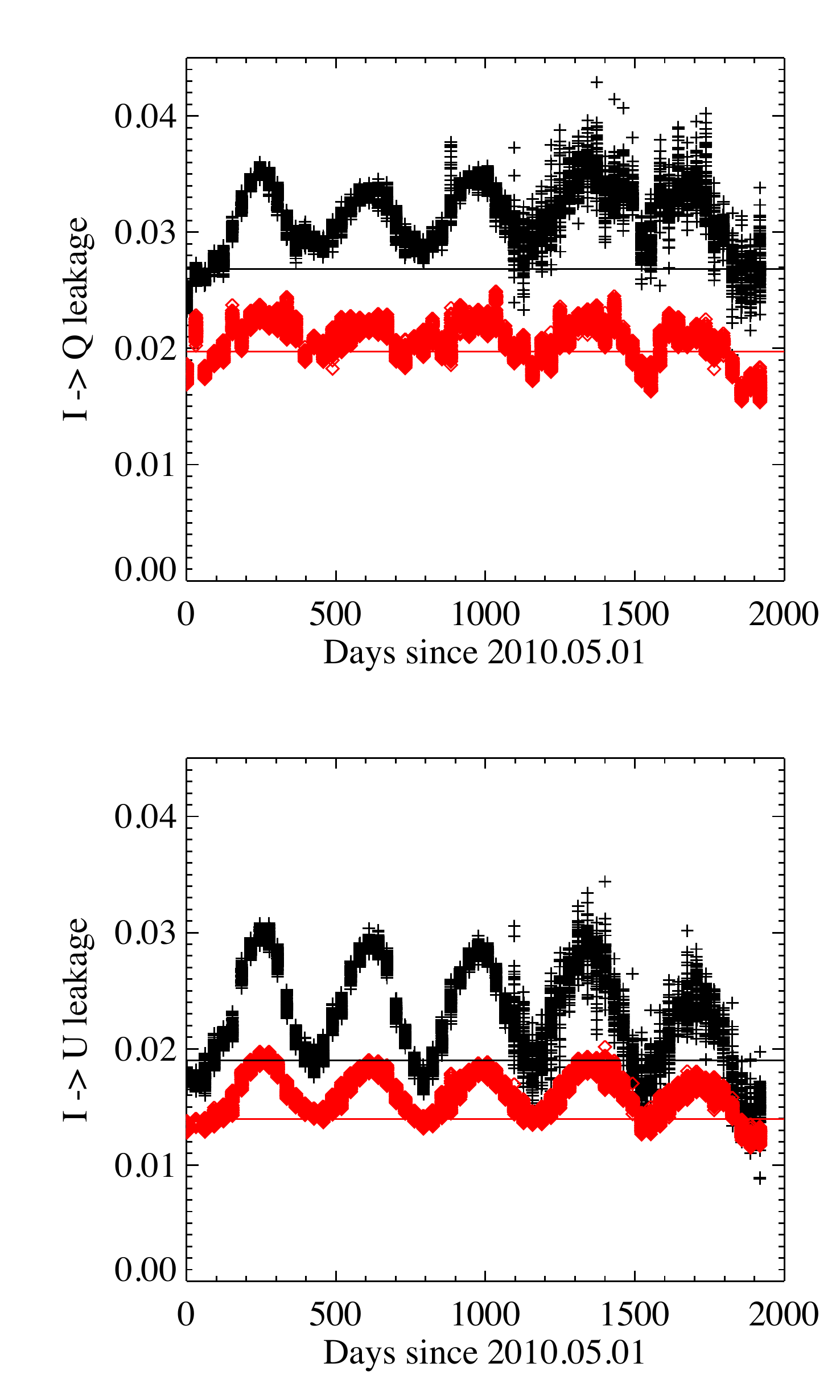}
\caption{The top plot shows parameters for the polarization-PSF leakage from I to Q, while the bottom plot shows the I to U polarization-PSF leakage. Black plusses show the central pixel in the kernel suggesting that about 2 -- 3\% of the intensity signal leaks into the same pixel in Q and U. Because the sum of the kernel is zero, an
equivalent negative signal leaks into other nearby pixels. Red diamonds indicate the norm of the other elements. The smaller telescope polarization term is shown in Figure \ref{fig:meanbig}.}
\label{fig:psfbig}
\end{figure}

The polarization-PSF correction has been estimated over the entire mission. 
Results are shown in Figure \ref{fig:psfbig}.  Here it is, unfortunately, the 
case that there are substantial variations with time.  
The cause of this effect is not understood either. The effect varies with
polarization selector setting and once again the likely causes include
second-order effects in the waveplates, the fold mirror, and the polarizing
beamsplitter.  It is interesting that the short-term temporal variations
increase about the same time that the thermal control scheme for the optical
bench changed.

\subsubsection{LoS Polarization Contamination}
\label{sec:LoSPolarization}

\begin{figure} [!htb]
\centering
\includegraphics[width=.89\textwidth]{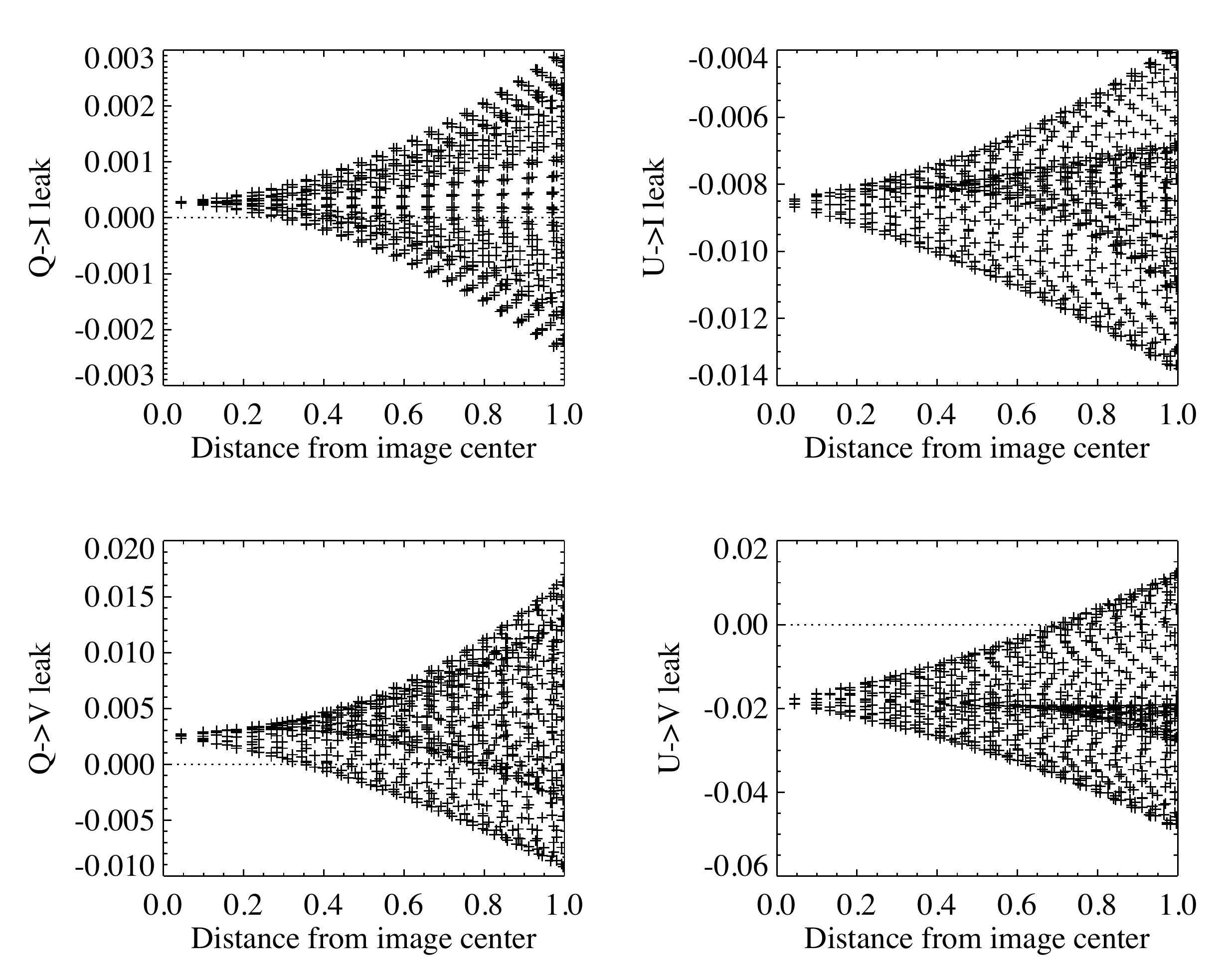}
\caption{Computed leaks from the linear polarization into the intensity and circular
polarization. Points from the entire CCD are plotted as a function of distance 
from the image center. The model was calculated for a polarization selector
temperature of 22.5\,C, representative of the mission. As mentioned in the
main text, the telescope temperature is not used.}
\label{fig:LoSLeakage}
\end{figure}

For the front camera, where only LCP and RCP are observed, it is not possible
to perform a full demodulation. Rather, it is assumed that only I and V are
non-zero, and only these two are inverted for. In reality Q and U are non-zero
and will leak into the resulting I and V. Given the existing polarization
model, it is straightforward to calculate these leaks and the results are
shown in Figure \ref{fig:LoSLeakage}. 
Indeed, among the many possible polarization-selector settings, ones with small
leaks were selected when the framelists were created.
The terms are quite small and much better that the original specification of 5\% for leakage. 

\subsection{Filters}\label{sec:Filters}

The HMI filter system consists of the entrance window, a broad-band blocking
filter, a multi-stage Lyot filter, the last element of which is tunable,
and two tunable Michelson interferometers. The following sub-sections describe
the detection and mitigation of wavelength non-uniformities determined
for the filter elements in phase maps (Section \ref{sec:Phasemaps}), the
correction for interference fringes created by the front window that are
most visible in CalMode (Section \ref{sec:Fringe}), and the I-ripple created
by imperfections in the tunable filter elements that leaks into the look-up
tables (Section \ref{iripple}).

\subsubsection{Phase Maps: Filter-Element Wavelength Non-Uniformity}
\label{sec:Phasemaps}

Ideally the wavelength would be same at each point of a filtergram, but in 
practice the filter elements are not completely uniform and drift with time.
Thus the actual wavelength at each pixel depends on the ray path through the 
filter elements in the instrument. We characterize these imprefections as 
phase maps.

\begin{figure}[!htb]
\centering
\includegraphics[width=\textwidth]{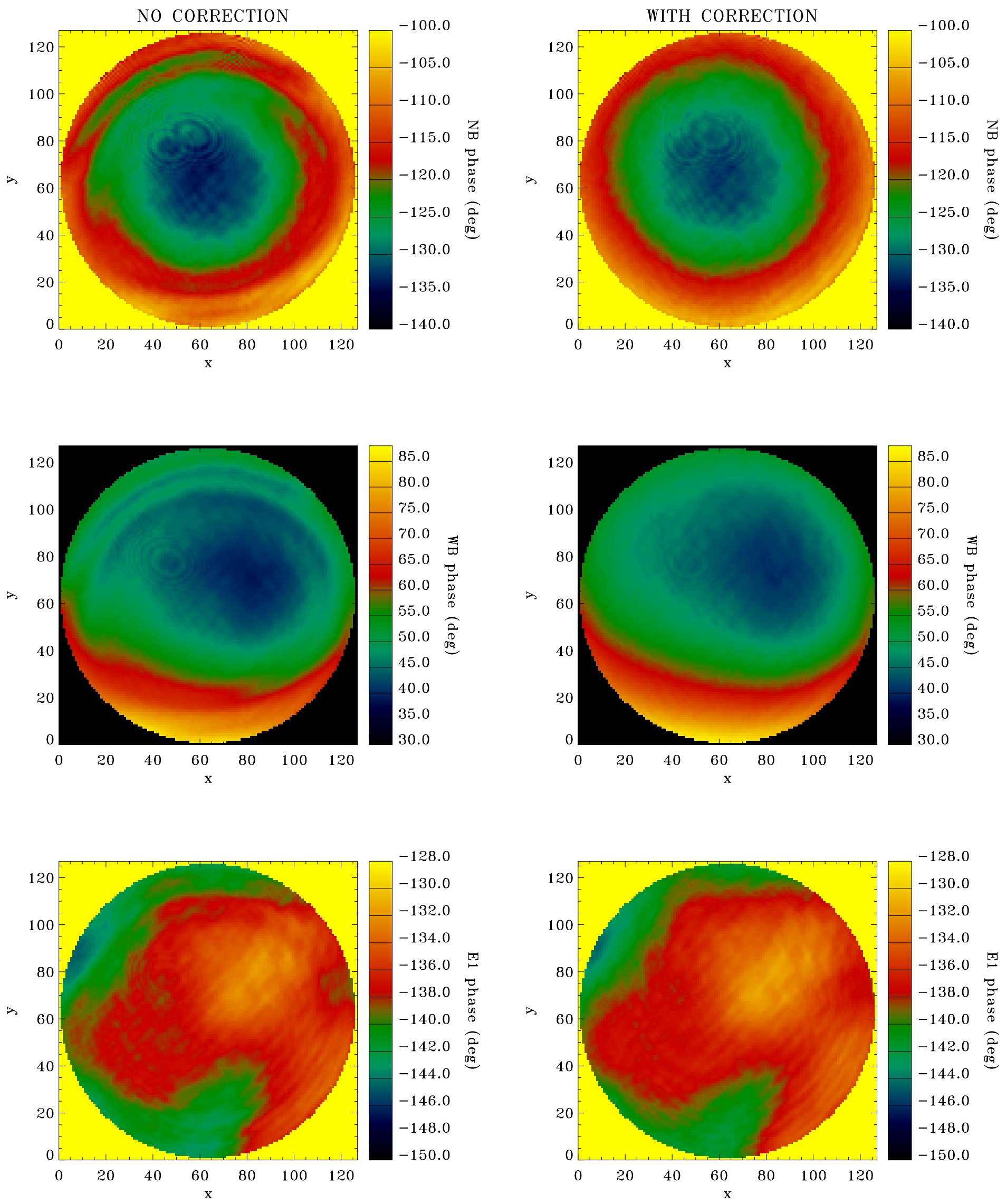}
\caption{Phase maps for the HMI tunable filter elements. From top to bottom the rows show phase maps of the narrow-band Michelson, wide-band Michelson, and E1 Lyot element. The two columns illustrate the 
impact of the CalMode interference fringes on the phase maps. 
Panels on the right show the corrected maps.}
\label{fig3}
\end{figure}

The tunable elements of HMI -- the two Michelsons and the narrowest Lyot element,
E1 -- have a transmittance $T(\lambda)$ that is modeled as:
\begin{equation}
T(\lambda) = \frac{1+ B \, \mathrm{cos}(2 \pi \lambda/\mathrm{FSR} + \Phi + 4 \phi) }{2}
\end{equation}
where FSR is the free spectral range of the element, 
$\phi$ is the phase resulting from the tuning motor position, and
$\Phi$ is the phase due to the properties of the filter. The $\Phi$ for each
element can be determined by tuning each filter separately in what is called
a {\em detune} sequence. Detune sequences are currently obtained every other 
week to make spatially resolved maps of $\Phi$ for all three tunable elements 
for each CCD \citep{Couvidat2012}. These phase maps are used to
determine the effective wavelength at each location and thus to provide a
spatially dependent calibration of the look-up tables used to determine the
observables with the MDI-like method (described in Section \ref{MDI-like}).
Because the variations across the filters have low spatial frequency, phase
maps are computed and stored on a $128 \times 128$ grid and interpolated to
the full resolution of the CCDs when used. Figure \ref{fig3} shows the phase maps for the tunable filter
elements. The Narrow Band Michelson is the most sensitive to tuning phase, 
$\sim 31$~m\,s$^{-1}$ per degree of phase. The CalMode fringe correction is 
described in Section \ref{sec:Fringe}.

The spatially dependent phase-map patterns are relatively constant in time for all 
three filter elements; however, the central wavelengths do drift. 
The narrow-band Michelson drifts at $\sim 6$~m\AA~per year and has a small
annual periodicity. The wide-band Michelson drift is $\sim 30$~m\AA~per
year and is slowing with time. Both contribute about the same amount to
an annual shift in the wavelength zero point that corresponds to a velocity
shift of $\sim 100-200$~m\,s$^{-1}$. The drift is such that the elements remain fairly
well co-tuned for extended intervals. The broader E1 Lyot element drifts
more slowly, $< 7$\,m\AA~per year. The observables look-up tables
are updated approximately once per year to account for the drifts when the
velocity offset exceeds a couple hundred m\,s$^{-1}$.

It was noticed early on that the average phases of the tunable elements
differ slightly between front and side cameras. 
The maximum difference is $\sim +0.55^\circ$ for the narrow-band Michelson,
and it exhibits a very slow increase with time ($\sim 4.3 \cdot 10^{-5}$
degrees per day). The broad-band Michelson and Lyot E1 filter elements have 
smaller phase differences of about $-0.1^\circ$ and $-0.13^\circ$ that
appear stable in time. 
These front/side-camera phase differences likely originate in the polarizer of
the narrow-band Michelson. A small leak of the orthogonal polarization will
be picked up by the polarizing beam-splitter used to separate the two
light paths.

\subsubsection{Removal of CalMode Fringes from the Phase Maps}
\label{sec:Fringe}

The phase maps of the tunable elements are determined using detune sequences made in CalMode. In CalMode calibration lenses are inserted into the HMI light path to switch conjugate planes in the optical path. This puts
an image of the entrance pupil onto the CCD, rather than an image of the
Sun. The purpose of CalMode is to provide uniform disk-average line profiles
all across the CCD field, thus eliminating solar spatial information as well
as the effects of having different angular distributions of rays through
the filter section.

The front window of HMI is constructed of multiple layers of different
glass and glue (with different refractive indices); therefore it acts as
a weak Fabry-Perot interferometer. In CalMode the fringes from the front
window, and to a lesser degree the blocking filter, are imaged onto the CCD.
Consequently the phase maps of the tunable elements also show the interference
fringes. The front-window fringes are not present in the regular ObsMode
images, but because the phase maps are used to derive the look-up tables
for the MDI-like algorithm (Section \ref{lookup}), the CalMode interference
fringes bleed into LoS observables. For helioseismic purposes this is mainly
a cosmetic issue that has minimal scientific impact on the determination of
frequencies. On the other hand, it can adversely impact secondary objectives,
such as determination of surface flows directly from the Doppler shift.

Biweekly detune sequences are used to compute (uncorrected) phase maps.
Each phase map contains five $128 \times 128$ images: one for each of the
three tunable filter elements and two more for the line width and line depth
maps of the Fe\,{\sc i} line fitted together with the phases. The method
described below mostly removes the obvious effects of the fringes present in
the CalMode phase maps, but the user is cautioned to remember that this
does not necessarily mean that the inferred phase values for the CalMode
fringes are actually {\em correct}.

A new correction is computed each time we re-tune the HMI instrument or
change the line or filter calibration (See Table \ref{tab:Caltimes} in
Section \ref{polcor}), since re-tuning requires computing new look-up tables
for the MDI-like algorithm. Finally, it is noteworthy that the CalMode
fringe correction has only been made to look-up tables used to compute
observables collected from 1 October 2012 onwards. Earlier observables
have not be reprocessed, as indicated in the {\sc calver64} keyword (see
\urlurl{jsoc.stanford.edu/jsocwiki/CalibrationVersions} for details.)
For that reason, earlier observables computed using the phase maps still show
the fringe pattern.

To model the fringes we start by assuming that each of the five phase maps can be written as
\begin{equation}
A(x,y,t)=A_0 (x,y)+A_L(x,y) t + A_C (x,y) \mathrm{cos}(\phi(t))
 + A_S (x,y) \mathrm{sin}(\phi(t)) ,
\label{eq_fringe}
\end{equation}
where $A_0$ represents a constant term; $A_L$ is a term describing the overall drift;
$A_C$ and $A_S$ describe the fringes, all arbitrary functions of space; and
$\phi$ describes the phase of the fringes (effectively the glass thickness)
as a function of time. As it turns out, the fringes are most cleanly visible
in line depth, so we start by fitting Eq. \ref{eq_fringe} to that. The
fit is performed using the terms of a singular value decomposition (SVD) 
as the initial guess and alternately fitting $\phi$ and the spatial terms.

Once $\phi$ is determined, Eq. \ref{eq_fringe} is fit for each of the
five variables and an estimate of the corrected Obsmode phase map is 
determined by subtracting the fringe
term $A_C (x,y) cos(\phi(t))+A_S (x,y) sin(\phi(t))$ from the original values.

Clearly, modeling the fringes like this is far from perfect, but the
result is nonetheless that the amplitude of the large-scale fringes is
dramatically reduced.  Unfortunately this leaves behind a number of 
smaller-scale fringes. Repeating the above procedure on the residuals 
(replacing the $A$ terms with equivalent
$B$ terms and $\phi$ with $\phi_1$) to remove these is not nearly as
efficient as for the larger fringes, but does nonetheless improve the
results significantly, so both corrections are applied.

\begin{figure}[!htb]
\centering
\includegraphics[width=0.9\textwidth]{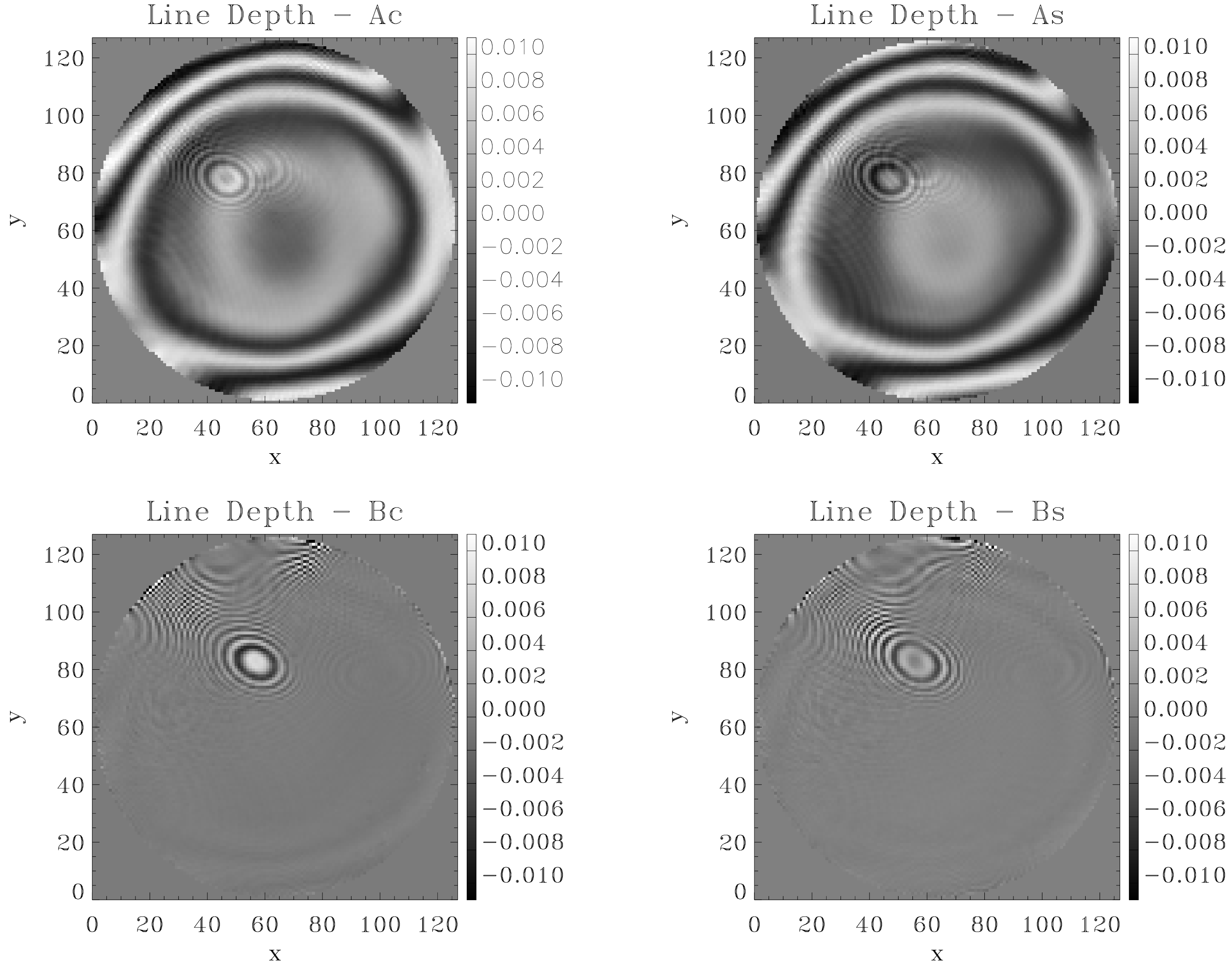}
\caption{The larger- and smaller-scale fringe terms, $A_C, A_S, B_C$ and $ B_S$, determined as described in the text according to Equation \ref{eq_fringe} that are used to remove interference fringes in the phase maps. The values shown are for line depth for Calibration13 (Section \ref{sec:CalibrationProfiles}).}
\label{fig3interference}
\end{figure}

Figure \ref{fig3interference} shows the $A_C$, $A_S$, $B_C$ and $B_S$ terms
for line depth. The A terms capture large-scale concentric
fringes, parts of the oval-shaped medium-scale features above and to the left
of image center, and some of the strong small-scale pattern. The B terms capture 
the stronger small-scale fringes most prominent near the top of the image and 
more of the oval-shaped features.
Figure \ref{fig3} shows the impact of the correction on the tuneable-element 
phase maps: the left panels show the raw maps, while the right panels show the same
phase maps after the fits to the CalMode interference fringes have been removed. 
The CalMode fringe corrections are as much as 25\% the magnitude of the 
of the corrected phase maps, e.g. $\pm 2.5^\circ$ in the narrow-band Michelson, 
but with a much different spatial scale.
While the phase patterns intrinsic to the elements remain, it appears that
the CalMode front-window fringes have been mostly removed.
The intense small-scale fringes near the top of the circle are most obvious in 
the narrow-band Michelson and have been dramatically reduced in amplitude.
The roughly concentric pattern is
significantly reduced in both Michelsons, though it is less obviously
present in the Lyot. Unfortunately, the oval-shaped medium-scale features 
bleed through in both Michelsons, but not as much in the Lyot.
Small-scale fringes not due to the CalMode fitting are also present.
It is interesting to note that $\phi_1 \approx 2.25~\phi$. Given
the thicknesses of the different glass elements (6\,mm, 3\,mm, and 6\,mm in
the order traversed), one might naively have expected a factor of 2.00 or
2.50, the deviation presumably being due to the different thermal expansion
coefficients and/or different changes in the refractive index with temperature.

More elaborate correction schemes were also investigated, but they did
not provide a significant improvement.

\subsubsection{I-ripple Characterization \label{iripple}}

What we call I-ripple is an intensity variation in the HMI output that depends 
on the instrument tuning. I-ripple results from imperfections in the tunable filter 
elements, such as small misalignments in wave-plates. 
It is most apparent when using a uniform and constant light source. 
No correction is currently made for I-ripple when deriving the filter-transmission profiles. 
This introduces a small error in these profiles and therefore in the 
look-up tables used by the MDI-like algorithm. Simulations show that I-ripple produces a systematic zero-point error
in velocity of at most a few tens of meters per second that varies linearly
with Sun-SDO velocity.  For details of the derivation and the results of HMI 
ground calibrations, see \citet{Couvidat2012} and references therein.

For a specific tuning phase $\phi$ of the tunable Lyot element, E1, 
the transmitted intensity can be modeled as:
\begin{equation}
\frac{I(\lambda)}{\bar{I}(\lambda)} = K_0 + [ K_1 \, \mathrm{cos}(\phi/2) + 
	K_2 \, \mathrm{sin}(\phi/2)]^2 \label{eqiripple}
\end{equation}
where $\bar{I}$ is the average intensity over all possible tuning phases,
$\phi$, and $K_0$, $K_1$, and $K_2$ characterize the I-ripple.  Though this
specific equation was originally derived for I-ripple resulting from a
misalignment-like feature in the Lyot half-wave plate, it proved to be
equally good for other sources of I-ripple, such as a combination of
a tilt in the entrance polarizer and quarter-wave plate of a Michelson
interferometer. Therefore this same equation is used to model the I-ripple
of the two Michelson interferometers.

\begin{figure} [!htb]
\centering
\includegraphics[width=0.9\textwidth]{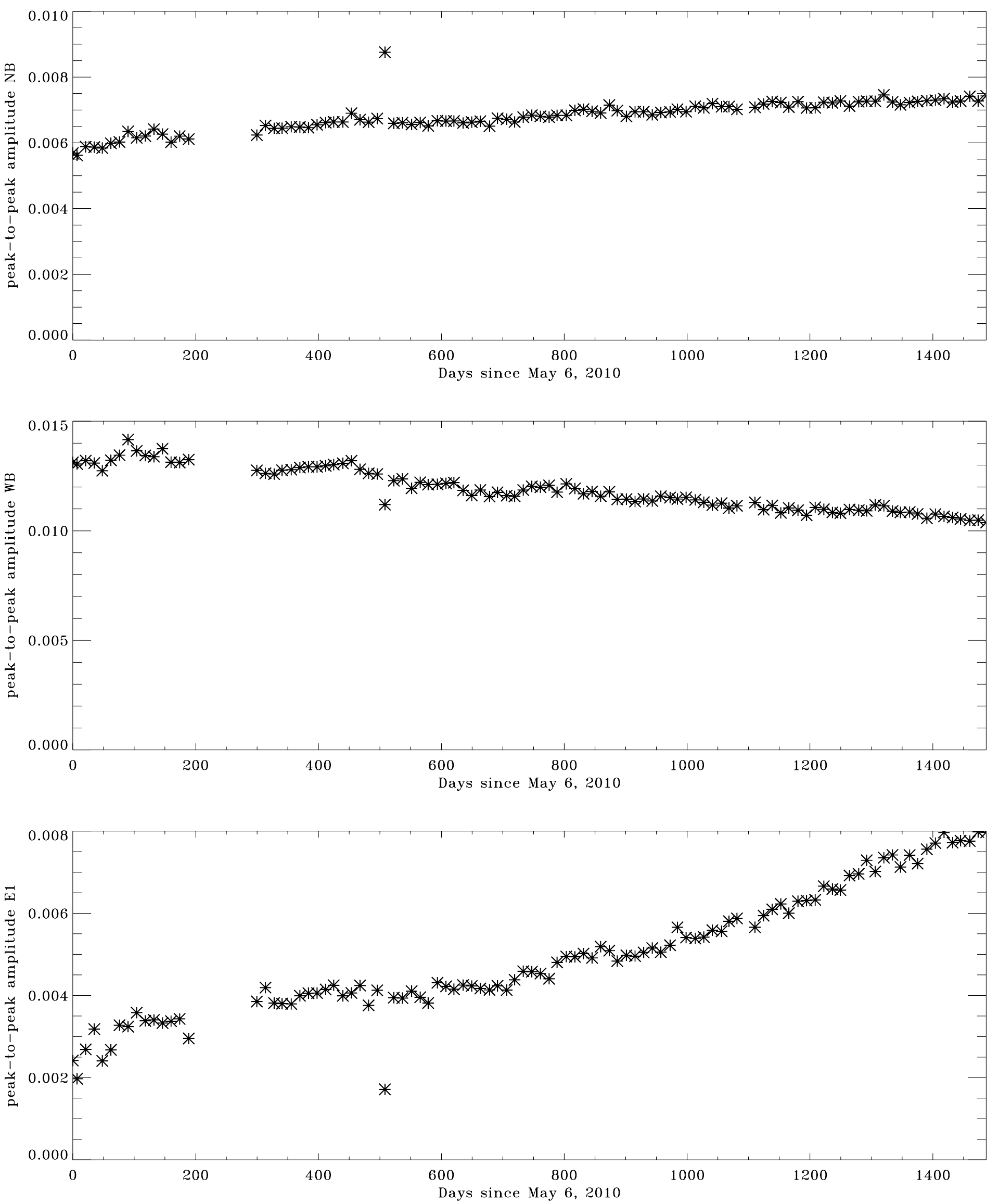}
\caption{Peak-to-peak amplitudes of the tunable-element I-ripple for the first 
$\sim 1500$ days of operation. From top to bottom the panels show the amplitude as a fraction of the transmitted intensity in the Narrow-Band Michelson, the Wide-Band Michelson, and the tunable Lyot element, E1. The measured amplitudes of the I-ripple are consistent with those measured before launch \citep{Couvidat2012}.}
\label{fig4} 
\label{fig:Iripple}
\end{figure}

Figure \ref{fig:Iripple} shows the temporal evolution of the measured I-ripple
(in terms of peak-to-peak amplitude) for each of the three tunable elements
since the beginning of the mission. It is obtained by fitting the I-ripple
of each tunable element to the intensities of a detune sequence using
Equation \ref{eqiripple}. In a detune sequence, the wavelength tuning of the
fiilter elements are varied independently, allowing one to distinguish featues
in the individual elements. The peak-to-peak variation in the transmitted
intensity $I$ is computed as $K_1^2+K_2^2$.

Unfortunately, it is not possible to determine whether the result of each fit
includes only I-ripple effects, or if other time-dependent imperfections in the
tunable elements --- ones that are not included in our transmittance model
--- also bleed into these results. Regardless, Figure \ref{fig:Iripple}
shows that the intensity transmitted by the tunable elements varies with
time. Currently, the I-ripple is not taken into account when computing
the observables. Including this effect when fitting the detune sequences
does improve the goodness of fit, as might be expected when adding
parameters. However, including the I-ripple in the filter 
transmission profile does not significantly improve the issues identified
in the observables. In particular, there was no positive impact on the 
24-hour oscillations detailed in Section \ref{24h}.

\subsection{Computation of Line-of-Sight Observables with the MDI-Like Method}
\label{sec:MDIIntro}
\label{MDI-like}

This section describes the algorithm implemented to compute
the LoS observables. It is called the MDI-like
method because it is based on the technique employed to produce
the SOHO/MDI observables.

MDI was designed so that the FWHM of its filter transmission profiles matches
the FWHM of the Ni\,{\sc i} line, and the four equally spaced wavelength
samples cover an interval equal to twice this FWHM. Consequently with MDI
nearly all of the spectral power in the solar-line shape is captured by the
first Fourier coefficients and the phase derived from the cosine and sine
components is an estimate of the line position.

HMI was not designed this way because the dynamic range
corresponding to twice the FWHM of the Fe\,{\sc i} line is too small to
accommodate the large velocity variations resulting from the SDO orbit.

\subsubsection{The HMI Implementation of the MDI Algorithm \label{HMI-version}}

The HMI algorithm has been described elsewhere ({\it e.g.} 
Eqns. \ref{Eqn:a1} -- \ref{eqcont} here are from 
\citealt{CouvRajaWachter2012}).
This report provides further details about
how it is implemented in the LoS HMI-observables pipeline.  For each of the
$\sim 12$ million illuminated pixels on an HMI image, the MDI-like algorithm starts
by estimating the first and second Fourier coefficients $a_n$ and $b_n$
(with $n=1$ or $n=2$) of the Fe\,{\sc i} line profile $I(\lambda)$, where
$\lambda$ is the wavelength:
\begin{equation}
a_1 = \frac{2}{T} \int_{-\frac{T}{2}}^{+\frac{T}{2}} I(\lambda) \, \mathrm{cos}
	\biggl(2\pi \frac{\lambda}{T}\biggr) \mathrm{d}\lambda 
	\;\;\;\; \mathrm{;} \;\;\;\; 
	b_1 = \frac{2}{T} \int_{-\frac{T}{2}}^{+\frac{T}{2}} I(\lambda) \, 
	\mathrm{sin}\biggl(2\pi \frac{\lambda}{T}\biggr) \mathrm{d}\lambda
\label{Eqn:a1}
\end{equation} 
and 
\begin{equation} 
\
a_2 = \frac{2}{T} \int_{-\frac{T}{2}}^{+\frac{T}{2}} I(\lambda) \, 
	\mathrm{cos}\biggl(4\pi \frac{\lambda}{T}\biggr) \mathrm{d}\lambda 
	\;\;\;\; \mathrm{;} \;\;\;\; 
	b_2 = \frac{2}{T} \int_{-\frac{T}{2}}^{+\frac{T}{2}} I(\lambda) 
	\, \mathrm{sin}\biggl(4\pi \frac{\lambda}{T}\biggr) \mathrm{d}\lambda ,
\label{Eqn:a2}
\end{equation} 
where $T$ is the period of the observation wavelength span. Nominally $T=6
\times 68.8=412.8$\,m\AA\, \mbox{\it i.e.} six times the nominal separation
between two HMI filter transmission profiles. Subsequent discretization
of Equations \ref{Eqn:a1} and \ref{Eqn:a2}, e.g. as in Equation \ref{eq4}, 
requires the assumption that the Fe\,{\sc i} line profile is periodic with period $T$.

We assume that the Fe {\sc i} line has the following Gaussian profile:
\begin{equation}
I(\lambda)=I_c-I_d \, \mathrm{exp}\biggl[-\frac{(\lambda-\lambda_0)^2}{\sigma^2}\biggr]
\label{Eqn:Ilambda}
\end{equation}
where $I_c$ is the continuum intensity, $I_d$ is the line depth,
$\lambda_0$ is the Doppler shift, and $\sigma$ is a measure of the line
width (FWHM$=2 \, \sqrt{\mathrm{log}(2)} \, \sigma$).

The Doppler velocity $v=dv/d\lambda \times \lambda_0$ can be expressed as:
\begin{equation}
v=\frac{\mathrm{d}v}{\mathrm{d}\lambda} \, \frac{T}{2\pi} \, 
	\mathrm{atan}\biggl(\frac{b_1}{a_1}\biggr)
\label{eq1}
\end{equation}
where $dv/d\lambda=299792458.0$ m s$^{-1}/6173.3433$\,\AA\,$=48562.4$ m\,s$^{-1}$\,\AA$^{-1}$.  
The second Fourier coefficients could also be used:
\begin{equation}
v_2=\frac{\mathrm{d}v}{\mathrm{d}\lambda} \, \frac{T}{4\pi} \, 
	\mathrm{atan}\biggl(\frac{b_2}{a_2}\biggr)
\label{Eqn:V2}
\end{equation}

The line depth [$I_d$] estimate is then equal to:
\begin{equation}
I_d = \frac{T}{2 \sigma\sqrt{\pi}} \, \sqrt{a_1^2+b_1^2} \, 
	\mathrm{exp}\biggl(\frac{\pi^2\sigma^2}{T^2}\biggr)
\label{eq2}
\end{equation}
while $\sigma$ is equal to:
\begin{equation}
\sigma = \frac{T}{\pi\sqrt{6}} \, \sqrt{\mathrm{log}\biggl(\frac{a_1^2+b_1^2}{a_2^2+b_2^2}\biggr)}
\label{eq3} 
\end{equation}
However, HMI samples the iron line at only six points, and therefore what we compute 
is a discrete approximation to the Fourier coefficients, rather than the actual 
coefficients. For instance:
\begin{equation}
a_1 \approx \frac{2}{6} \sum_{j=0}^{5} I_j \, \mathrm{cos}\biggl(2\pi \frac{2.5-j}{6}\biggr)
\label{eq4}
\end{equation}
The $b_n$ are determined by a similar formula with cosine replaced by sine. In
the LoS observables code, these $a_n$ and $b_n$ are calculated separately
for the LCP (I+V) and RCP (I$-$V) polarizations.  Applying Equation (\ref{eq1})
returns two velocities: $v_\mathrm{LCP}$ and $v_\mathrm{RCP}$.

Departing from the assumptions made, the actual Fe\,{\sc i} line profile is
not Gaussian ({\it e.g.} see Figure \ref{fig:lineprofiles} in Section \ref{24h}).  
Moreover, the discrete
approximations to $a_n$ and $b_n$ are not accurate due to a reduced number
of sampling points and because the HMI filter transmission profiles are not
$\delta$-functions. Consequently the observables calculated are relative to
the Fe {\sc i} line convolved with the filters.  Therefore, $v_\mathrm{LCP}$
and $v_\mathrm{RCP}$ need to be corrected.

This is the role of look-up tables. They are determined from a realistic model
of the Fe\,{\sc i} line at rest in quiet Sun and from calibrated
HMI filter transmission profiles. Look-up tables are described in Section
\ref{lookup}.

The sensitivity tables vary across the HMI field of view, since each 
CCD pixel samples a different ray bundle in the filters.  
The look-up tables are linearly interpolated at $v_\mathrm{LCP}$
and $v_\mathrm{RCP}$ to derive corrected Doppler velocities $V_\mathrm{LCP}$
and $V_\mathrm{RCP}$. 

Calibration of the HMI filters shows residual errors at the percent
level in their transmittances and free spectral ranges (FSRs), resulting in
imperfect look-up tables.  The SDO orbital velocity is known very accurately
and can be used to somewhat improve these tables. In the HMI pipeline this
additional step is referred to as the polynomial correction (see Section
\ref{polcor}).  As presently implemented this process corrects for the
slow drift in Michelson phases and FSRs, but leaves a residual variation in
observables at the SDO orbital period.

Since the Sun has magnetic fields and the observations are made in LCP and RCP
polarizations, the actual velocity computed will be roughly the centroid of the
unsplit and one or the other of the Zeeman-split components with the relative
strength depending on the direction of the field.  It is convenient that the
splitting of the centroids is a good measure of the LoS component of the 
magnetic field.

Finally, the resulting $V_\mathrm{LCP}$ and $V_\mathrm{RCP}$ velocities are
combined to produce a Doppler-velocity estimate: 
\begin{equation} V = \frac{V_\mathrm{LCP}+V_\mathrm{RCP}}{2} \end{equation}
while the LoS magnetic flux density $B$ is estimated as:
\begin{equation}
B = (V_\mathrm{LCP}-V_\mathrm{RCP}) \; K_m
\label{Eqn:BeqVdiff}
\end{equation}
where $K_m = 1.0/(2.0 \times 4.67 \cdot 10^{-5} \, \lambda_0 \, g_L \,
c)=0.231$ G m$^{-1}$\,s, $g_L=2.5$ is the
Land\'e g-factor, and $c$ is the speed of light \citep[see][]{Norton2006}.

An estimate of the continuum intensity $I_c$ is obtained by reconstructing the 
solar line from the Doppler-shift, line-width, and line-depth estimates:
\begin{equation}
I_c \approx \frac{1}{6} \, \sum_{j=0}^5 \biggl[I_j + I_d \, 
	\mathrm{exp}\biggl(-\frac{(\lambda_j-\lambda_0)^2}{\sigma^2}\biggr) \, \biggr]
\label{eqcont}
\end{equation}
where $\lambda_0$, $I_d$, and $\sigma$ are values retrieved using Equations
(\ref{eq1}), (\ref{eq2}), and (\ref{eq3}), and $\lambda_j$ are the nominal
wavelengths corresponding to each filter profile.

The intensity computations are actually implemented slightly differently in the HMI pipeline.
Tests on synthetic Gaussian lines using proper HMI filter transmittances
show that the theoretical algorithm overestimates the line width of Gaussian
lines by $\sim 20\%$ for a line with $I_d=0.62$ and $\sigma=0.0613$~\AA\
\citep[values in][]{Norton2006}.  Conversely, the line depth is underestimated
by $\sim 33\%$.  The continuum intensity seems only slightly underestimated
(by $\sim 1\%$). We surmise that these errors in the parameters of synthetic
Gaussian lines arise because the number of wavelength samples is small and
the filters are not $\delta$-functions. Unlike velocity shifts (and therefore
magnetic-field strength), the line width and line depth are not corrected
by look-up tables.
In the current implementation, the line depth $I_d$
and line width $\sigma$ returned by Equations (\ref{eq2}) and (\ref{eq3}) are 
multiplied by $K_2 = 6/5$ and $K_1 = 5/6$, respectively, so that both
values are expected to be closer to the actual ones.
The integral of a Gaussian is proportional to $I_d\cdot\sigma$; therefore
the continuum intensity remains unchanged.

Finally, when computing $I_d$ and $I_c$ with Equations
(\ref{eq2}) and (\ref{eqcont}), we find that the observed $\sigma$ 
is sometimes spurious in the presence of locally strong magnetic 
field, especially for pixels away from solar disk center. For that reason the 
observed $\sigma$ derived from Equation (\ref{eq3}) 
is not used for computing $I_d$ and $I_c$. Instead a nominal line width is determined 
from a fifth-order polynomial fit to an azimuthal average around the solar disk 
center of an HMI line-width map obtained during a period of low solar activity 
using Equation \ref{eq3} and corrected by $K_1$,
as a function of center-to-limb distance.  

The line width, line depth, and continuum intensity are computed separately
for both LCP and RCP and the average is saved in the observables data series.

\subsubsection{Observables Look-Up Tables \label{lookup}}

The MDI-like algorithm computes a discrete estimate of the
first and second Fourier coefficients of the Fe\,{\sc i} line profile from
the six measured wavelengths of the observables sequence. Under ideal
circumstances the phases of these two Fourier components would vary linearly
with velocity, but in reality this is only an approximation. Therefore,
a correction to the phase-derived raw velocities is made.
That is the role of the look-up tables.

It is possible to compensate for real conditions by simulating an observation
with a model of the solar line convolved with the actual filter transmission
function. The line parameters used for this process are described in Section
\ref{sec:CalibrationProfiles} and the maps of individual components of the
tunable filter are described in Section \ref{sec:Filters}.  The simulated
intensity signals are computed for the six tuning positions of the filters,
the Fourier coefficients are calculated, and the velocity is determined
as an expected instrumental response to each input sample velocity. Since
the filters vary slowly across the field, the sensitivity table is computed
on a $128 \times 128$ grid that covers the active portion of the CCD for a
range of 821 input velocities in steps of 24 m\,s$^{-1}$; thus the maximum
input velocity range is $\pm 9840$ m\,s$^{-1}$. This range accommodates the
signal contributions due to the $\sim 3500$ m\,s$^{-1}$ SDO orbit, $\sim
2000$ m\,s$^{-1}$ solar rotation, $\sim 400$ m\,s$^{-1}$ supergranulation,
and $\sim 1000$ m\,s$^{-1}$ from granulation and p-mode signals (at HMI
resolution). That leaves at least $\sim 3400$ m\,s$^{-1}$ to account for
Zeeman splitting, which introduces an equivalent 2.16 m\,s$^{-1}$\,G$^{-1}$
splitting from the zero-field line center (see Equation \ref{Eqn:BeqVdiff}).

The resulting look-up tables have dimensions $1642 \times 128 \times
128$, where the first dimension is the number of test velocities times
two, because the look-up tables store both the first and second Fourier
velocities, and the second and third dimensions are the x and y grid
locations. The tables are stored in several different DRMS series: {\sf
hmi.lookup}, {\sf hmi.lookup\_corrected}, {\sf hmi.lookup\_expanded}, and
{\sf hmi.lookup\_corrected\_expanded} where ``corrected'' refers to the fact
that the phase maps have been corrected for the CalMode interference fringe
pattern (Section \ref{sec:Fringe}) and ``expanded'' refers to look-up tables
computed for a larger off-limb radius.

\begin{figure} [!htb]
\centering
\includegraphics[width=0.8\textwidth]{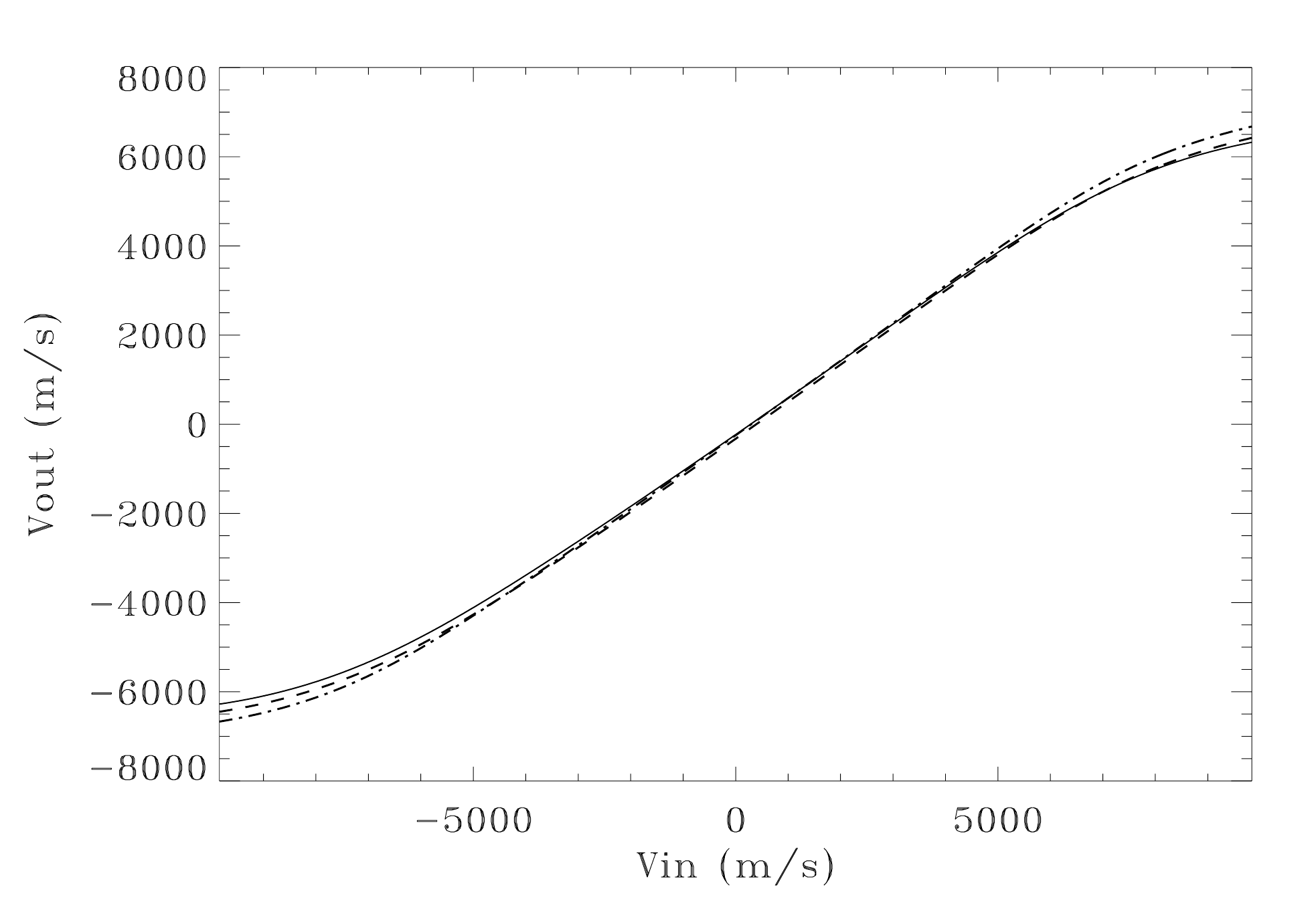}
\caption{Sensitivity table for the first Fourier coefficients determined for
the central CCD pixel on the front camera. The x axis shows the input solar
velocity and the y axis shows the output velocity computed with the MDI-like
algorithm. The lines show sensitivity curves for three of the six calibration
tables used during the HMI mission. The solid line shows the values from the
beginning of the mission determined with Calibration version 11. The dotted-line
shows the sensitivity calculated using Calibration version 12 starting in early 
2012, and dashed line shows the result after the retune in 2015 using version 13. The inverse of this table is used to determine the solar inputvelocity from the raw {\em output} velocity determined with the MDI algorithm.}
\label{fig10}
\label{fig:Sensitivity}
\end{figure}

When the correction is made, a 1642-element sensitivity table of output
velocity values is interpolated for each CCD pixel from the look-up tables
saved on the coarser $128 \times 128$ grid.  The inverse of the sensitivity
table is used to interpolate the input solar velocity as a function of the raw
output velocity determined by the MDI algorithm. As the input velocity shifts
the line away from the center of the tuning range, the sensitivity decreases.

Figure \ref{fig:Sensitivity} shows example sensitivity tables for
the pixel at CCD center for the front camera for the three different
HMI line-profile calibrations over the course of the mission (Section
\ref{sec:CalibrationProfiles}.  Tables are computed for each camera. At
disk center the front- and side-camera sensitivities differ by at most 4
m\,s$^{-1}$ over the entire velocity range. Each time HMI is retuned to
compensate for the wavelength drift in the Michelson interferometers, new
sensitivity tables are produced.

\subsubsection{Line Profile Tweaking and Calibration Changes}
\label{sec:CalibrationProfiles}

\begin{figure} [!htb]
\centering
\includegraphics[width=\textwidth]{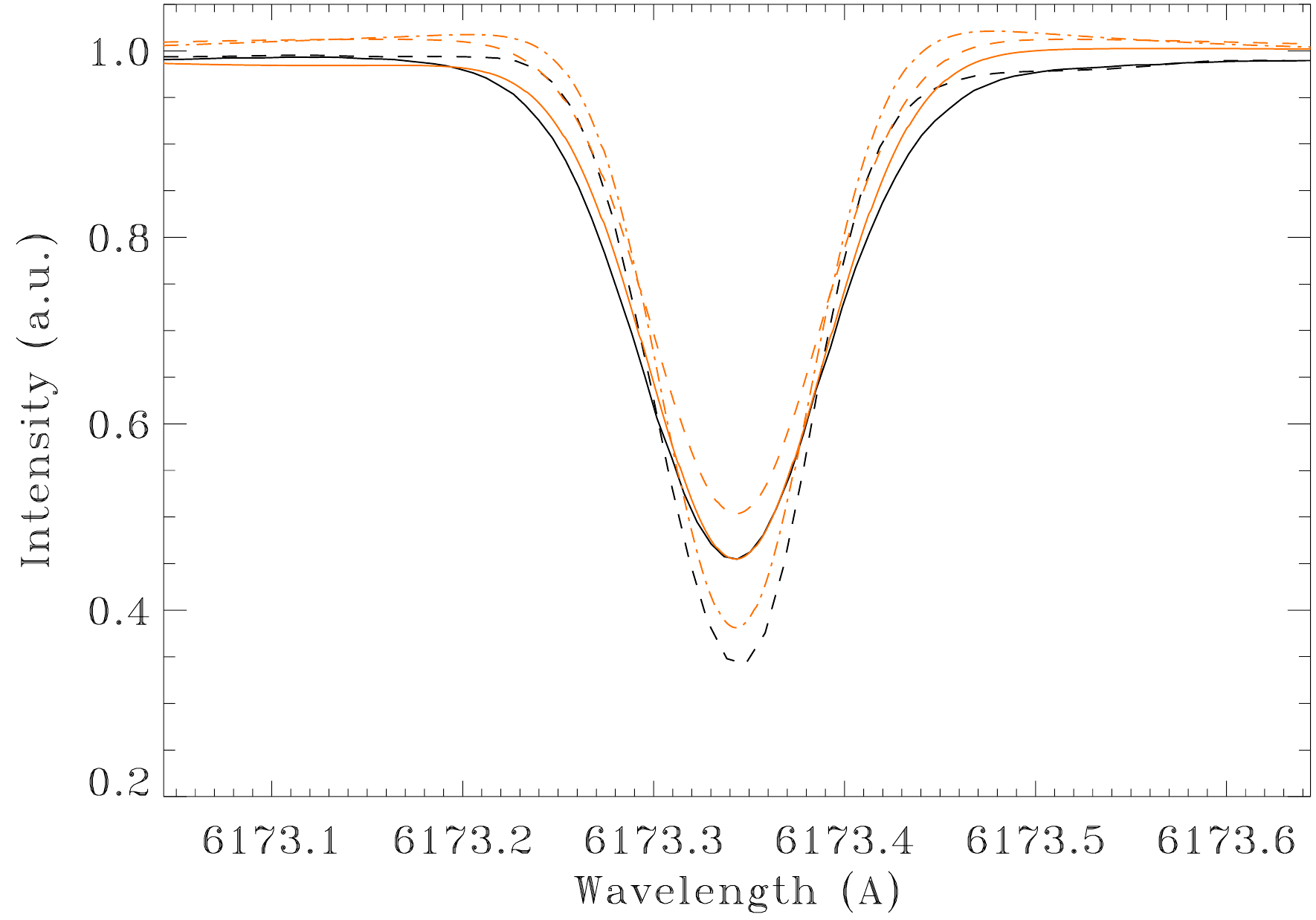}
\caption{Different Fe\,I line profiles used for the calibration of HMI. Three synthetic profiles are shown in red and two observed line profiles (in black). The solid black is from the NSO Solar Atlas and the dashed black is from Mount Wilson Observatory. The solid red is the profile used for HMI calibrations from May 2010 - January 2012, red dashed is January 2012 - January 2014, and red dot-dashed is after January 2014. For ease of comparison each curve has been divided by the nearby continuum intensity for that instrument.}
\label{fig6}
\end{figure}

To produce the look-up tables for the MDI-like algorithm as described in Section \ref{lookup}, we need a model of the Fe\,{\sc i} line profile (at rest and in the quiet Sun). Figure \ref{fig6} shows, in black, two observed Fe\,{\sc i} line profiles: one from the NSO atlas \citep{NSOAtlas1984} derived from the Fourier Transform Spectrometer at
the McMath-Pierce Solar Facility at Kitt Peak and the other from the solar telescopes at Mount Wilson Observatory provided by Roger Ulrich (private communication, 2011).

The two observed line profiles are rather different in terms of depth,
width, and asymmetry. Differences arise because instruments have
different point spread functions (PSFs) and quite different wavelength resolutions,
not to mention targeting, spatial resolution, and other effects. For that reason
it is difficult to determine a {\em correct} line profile and consequently we take 
a somewhat empirical approach. 
The detune sequences regularly taken by the HMI and used to estimate the filter
and Fe\,{\sc i} profiles are sensitive to a combination of different
quantities. The fitting code used to estimate the best parameters for the
filters and the solar lines also has to deal with partly degenerate quantities
(for instance, the filter contrast and the Fe\,{\sc i} line depth). Finally,
some of the quantities characterizing the HMI filters can only be measured
in a lab because they require access to a large wavelength range
and are not known with a high precision. For instance, the filter-element
free spectral ranges (FSR) were measured several times prior to the
SDO launch and yielded inconsistent results. For all of these reasons, it
is not possible to precisely measure the Fe\,{\sc i} line profile using detune
sequences. To accommodate the wavelength drift in the Michelson interferometers
and other long-term changes in HMI, we have had to re-tune the instrument
several times and to change our estimate of the Fe\,{\sc i} line profile to
produce better look-up tables. Three different calibrations have been used
since the SDO launch in which the solar line profile used to produce the look-up
tables has been slightly modified. 
Figure \ref{fig6} shows, in red, the three line profiles used so far.

In the code producing the look-up tables for the MDI-like algorithm, the 
Fe\,{\sc i} line is approximated by a Voigt profile \citep{TepperGarcia2006} and two Gaussians 
(to simulate the line asymmetry):
\begin{eqnarray}
I & = & Ig- dg \, \mathrm{exp}(-l^2) \biggl( 1-\frac{a}{\sqrt{\pi} \, l^2} 
	\, \biggl[(4 l^2+3)(l^2+1) \, \mathrm{exp}(-l^2)-\frac{2 l^2+3}{l^2} 
	\, \mathrm{sinh}(l^2)\biggr]\biggr) \nonumber \\
	& - & A \, \mathrm{exp}(-(\lambda+B)^2/C^2) \nonumber \\
	& + & D \, \mathrm{exp}(-(\lambda-E)^2/F^2)
\end{eqnarray}
where $l=\lambda/wg$ and for $|l| \le 26.5$. 

Coefficients for the three calibration intervals are given in Table
\ref{tab:CalibrationCoefficients}.  These are sometimes referred to
Calibration Versions 11, 12, and 13.  The first calibration was used from 1
May 2010 until 18 January 2012 at 18:15 UT, the second from 18 January 2012
at 18:15 UT to 15 January 2014 at 19:18 UT, and the third has been used ever
since.  In each case, we used the following FSRs for the filter elements:
$168.9$ m\AA\ for the NB Michelson, $336.85$ m\AA\ for the WB Michelson,
$695$ m\AA\ for the Lyot element E1, $1417$ m\AA\ for E2, $2779$ m\AA\
for E3, $5682$ m\AA\ for E4, and $11,354$ m\AA\ for E5.

\begin{table} [htb]
\begin{center}
\caption{HMI Calibration Coefficients by Time Interval}
\label{tab:CalibrationCoefficients}
\begin{tabular}{c|r|l l l l l l l l l l}
 \multicolumn{2}{c|}{Calibration \& Start Date} & Ig & dg & wg & A & B & C & D & E & F & a \\
\hline
11 & 2010.05.01\_00:00 & 1.0 & 0.5625 & 0.06415 & ~0.015  & 0.225 & 0.2  & 0.004 & 0.15  & 0.22 & ~0.03 \\ 
12 & 2012.01.18\_18:15 & 1.0 & 0.53   & 0.0615  & -0.01   & 0.225 & 0.2  & 0.015 & 0.10  & 0.25 & ~0.03 \\
13 & 2014.01.15\_19:18 & 1.0 & 0.58   & 0.058   & -0.0074 & 0.2   & 0.13 & 0.021 & 0.05  & 0.18 & -0.09 \\
\hline
\end{tabular}
\end{center}
\end{table}

\subsection{Polynomial Velocity Correction \label{polcor}}

\begin{figure}[!htb]
\centering
\includegraphics[width=0.95\textwidth]{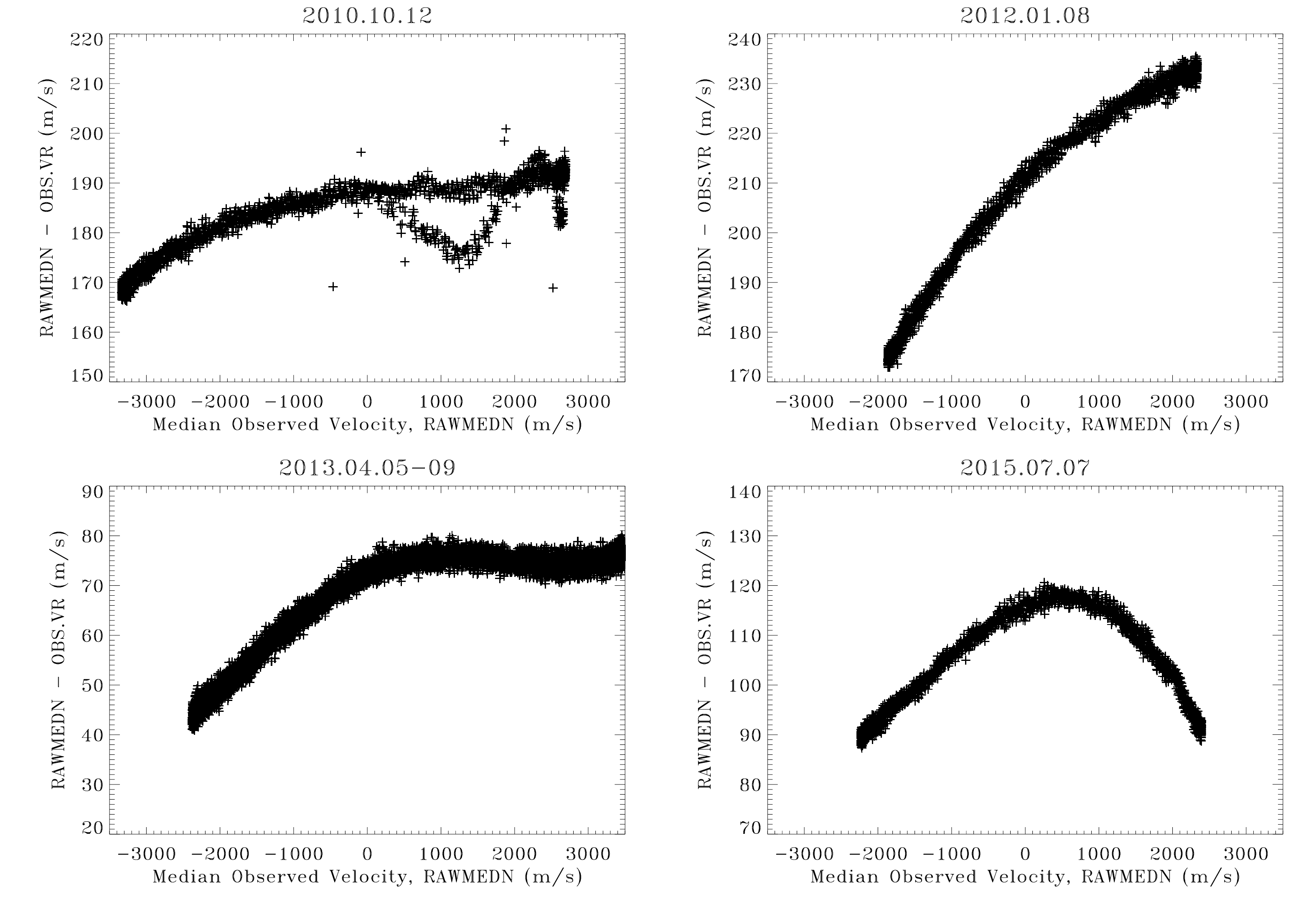}
\caption{Median Doppler velocity returned by the line-of-sight algorithm prior
to applying the polynomial velocity correction for several intervals during the
course of the SDO misison. The panels show the difference between the median
velocity observed in the inner 99\% of the disk, {\sc rawmedn}, and the known
Sun-SDO radial velocity, {\sc obs\_vr}, as a function of {\sc rawmedn}. The
three panels clockwise from the upper left show different 24-hour intervals; the lower-left 
panel shows five consecutive days. Note the difference in the observed range of
{\sc rawmedn} at different times of year. The upper-left panel shows
observations on 12 October 2010 during which the SDO spacecraft completed a roll maneuver.
The roll starts at $\sim$18~UT and takes place mostly during an interval
of positive {\sc rawmedn}, after local noon. During non-roll times the
difference between {\sc rawmedn} and {\sc obs\_vr} for a given {\sc rawmedn}
lies within a very narrow band of about 5~m\,s$^{-1}$. During roll maneuvers
the median velocity varies systematically, as is better illustrated in
Figure \ref{fig:RollSensitivity}. The lower left panel shows the
difference over a five-day interval in April 2013. While slightly broader
than the one-day plots, this panel demonstrates the slow evolution of the
daily velocity pattern.} 
\label{fig:RollDayVelocityDifferences}
\label{fig13} 
\end{figure} 

\begin{figure}[!htb]
\label{fig5}
\label{fig:DailyPolynomialCoefficients}
\centering
\includegraphics[width=\textwidth]{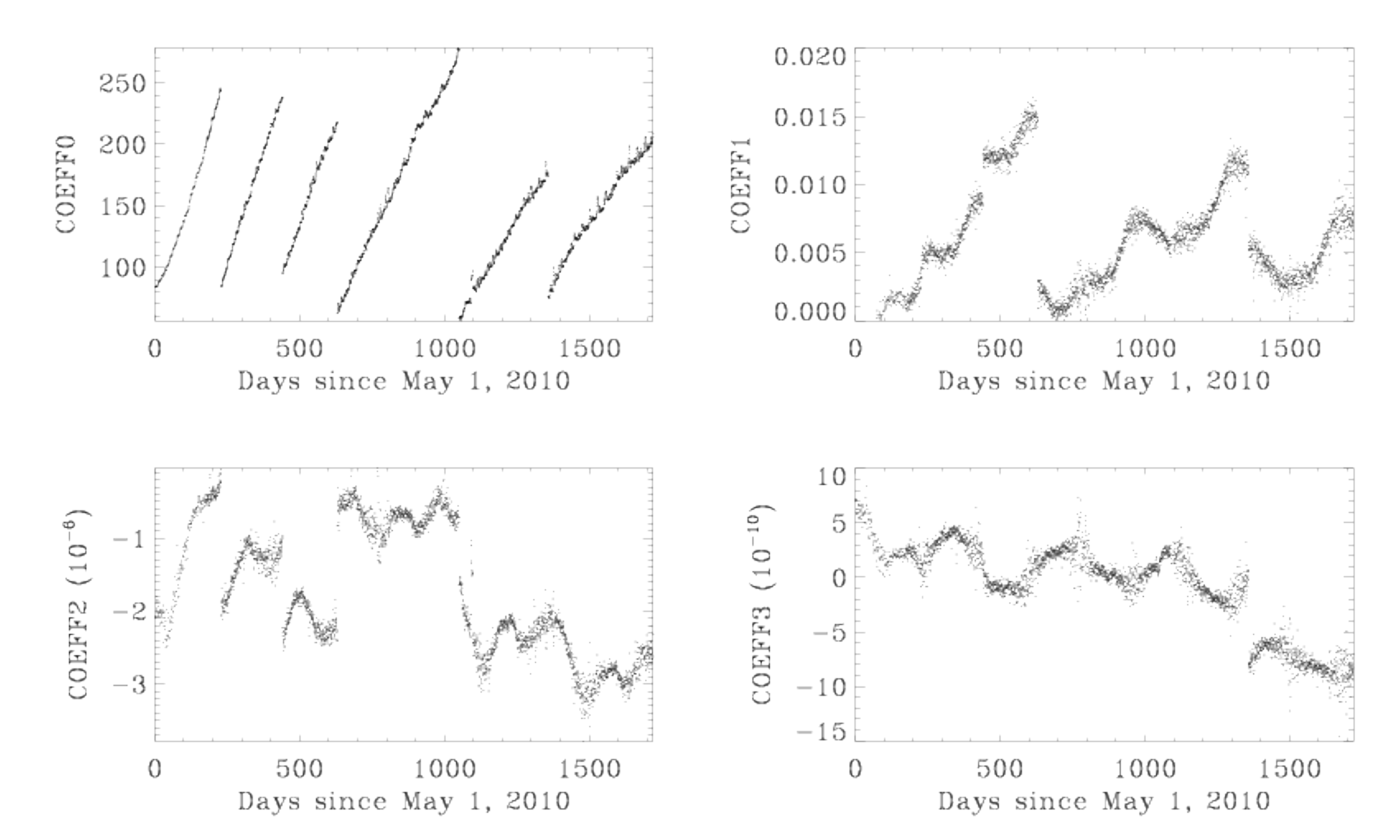}
\caption{Daily polynomial coefficients $C_0, C_1, C_2,$ and $C_3$ used to 
correct the computed Doppler velocities as a function of time. In Equation 
\ref{eqn:DopplerDaily}, $C_0$ is the constant term in the measured daily offset 
between the median disk velocity, {\sc rawmedn}, and the known Sun-SDO velocity, 
{\sc obs\_vr}.}
\label{fig:CoeffDrift}
\end{figure}
  
After calibrating the HMI Doppler signal as described above, it was clear that
an artifact proportional to the SDO orbit velocity remained.
A procedure was developed to use the accurately known spacecraft velocity
to make a post-calibration correction to the $V_\mathrm{LCP}$ and $V_\mathrm{RCP}$
used to compute the Doppler velocity and LoS magnetic field. This was done by
computing a third-order polynomial fit to a 24-hour segment of the difference
between the uncorrected full-disk median Doppler velocity {\sc rawmedn} and the 
known Sun-SDO radial velocity {\sc obs\_vr} as a function of {\sc rawmedn}. 
\begin{equation}\label{eqn:DopplerDaily}
\mbox{\sc rawmedn - obs\_vr} = C_0 + C_1 \cdot \mbox{\sc rawmedn} + 
	C_2 \cdot \mbox{\sc rawmedn}^2 + C_3 \cdot \mbox{\sc rawmedn}^3
\end{equation}
The expectation was that this would account for the slow phase drift of the
Michelsons, a possible error in the Narrow-Band Michelson FSR, and a non-linear
contribution likely due to errors in the look-up tables. It had been noticed
that the median, which would be expected to be nearly a constant offset
from {\sc obs\_vr}, varied through the day in a slowly changing systematic way.

Figure \ref{fig:RollDayVelocityDifferences} shows the difference between
{\sc rawmedn} and {\sc obs\_vr} plotted as a function of {\sc rawmedn} for
four days or intervals at different seasons throughout the mission. On one
of the days, 12 October 2010, a spacecraft roll maneuver took place (upper
left panel). See
Section \ref{RollCalibration} for a discusion of the effects of instrument roll. 
Otherwise, the daily velocity difference generally follows a smooth curve that can be 
well fit using Equation \ref{eqn:DopplerDaily}. The residual variability, which includes
solar signals, is less than 5~m\,s$^{-1}$ for a given {\sc rawmedn}; however, the
daily variation in the difference is 15~--~60~m\,s$^{-1}$. The overall offset
from zero drifts with time and is determined by the tuning of the instrument filters. 
Note that the range in {\sc rawmedn} and the range in the velocity difference are not
the same for each day and depend on the orbit of the spacecraft and Earth.
The correction
determined by the fit is applied to the measured $V_\mathrm{LCP}$ and 
$V_\mathrm{RCP}$ velocity at each pixel in the image
and the corrected median velocity is given in the keyword {\sc datamedn}.

Figure \ref{fig:CoeffDrift} shows the time variation of the four polynomial
coefficients of the fit since 1 May 2010. $C_0$ represents the offset between
{\sc rawmedn} and {\sc obs\_vr}, and it increases with time primarily because of the
wavelength drift in the two Michelson interferometers. The discontinuous
jumps in $C_0$ are due to the retuning of the instrument. Jumps in the
other coefficients reflect those and other changes to the calibration
procedures. 
Coefficient $C_1$ (upper right) shows sharp discontinuities when the
calibration changes (see Table \ref{tab:Caltimes}), suggesting a 
temporal dependence of the instrument FSR that is decreasing with time.
The polynomial coefficients are computed for times separated by 12 hours 
and recorded in the DRMS data series {\sf hmi.coefficients}.
The analysis is run only on the nrt data in {\sf hmi.V\_45s\_nrt}
rather than on the definitive data. Note that because the NRT data are
not permanently archived, reproducing the fits exactly is impractical.

\begin{table}
\caption{HMI Tuning and Recalibration Times}
\label{tab:Caltimes}
\begin{tabular}{l|r|l|l}
\hline
Date & Filtergram & Tuning & Calibration \\
 & Serial \# {\sc fsn} & & \\
\hline
2010.04.30\_22:24 &  4609085 & Initial Tuning & Calibration 11 \\
2010.12.13\_19:45 & 15049882 & Retuning & \\
2011.07.13\_18:35 & 24812717 & Retuning & \\
2012.01.18\_18:15 & 33508227 & Retuning & Calibration 12 \\
2013.03.14\_06:42 & 52877107 & Retuning & \\
2014.01.15\_19:13 & 67002794 & Retuning & Calibration 13 \\
2015.04.08\_18:51 & 87614596 & Retuning & \\
2016.04.27\_18:56 & 105328287 & Retuning & \\
\hline
\end{tabular}
\end{table}

The correction is applied to the $V_\mathrm{LCP}$ and $V_\mathrm{RCP}$
returned by the MDI-like method by
first interpolating the nearest sets of daily coefficients before and after the
{\sc t\_obs} of each computation and then computing a correction for each circular 
polarization, {\em e.g.}:
\begin{equation}\label{eqn:PolDopplerDaily}
 v_\mathrm{LCP} = V_\mathrm{LCP}  - ( C_0^\prime + C_1^\prime \cdot V_\mathrm{LCP} + 
	C_2^\prime \cdot V_\mathrm{LCP}^2 + C_3^\prime \cdot V_\mathrm{LCP}^3 ),
\end{equation}
where $C_n^\prime$ are the interpolated polynomial coefficients.
The resulting $v_\mathrm{LCP}$ and $v_\mathrm{LCP}$ are used to compute the corrected
Doppler velocity and LoS magnetic field products.

The polynomial correction suffers from three main shortcomings. First, it is
computed for the daily velocity range covered by {\sc obs\_vr}, which is quite
limited compared to the full velocity range of the solar signal. Consequently,
for velocities outside the range of {\sc obs\_vr}, the polynomial correction
amounts to an extrapolation and this can be problemmatic for a cubic fit.
Some of the 12-hr and 24-hr variations seen in strong LoS magnetic fields may
be due to the side effects of this extrapolation. Second, the correction is
computed for a spatial average over the CCD, and therefore local differences
are not taken into account. In particular the correction seems to be less
accurate toward the solar limb. Finally, the correction coefficients are
computed each 12 hours to capture the slow change in shape with time of
the plots in Figure \ref{fig:RollDayVelocityDifferences}. However, because
the likely errors in the FSRs and the phase maps are changing very slowly
in time, the present procedure probably introduces some unnecessary noise
in the LoS Doppler and magnetic field products. The HMI team continues to
explore better calibration procedures.

This correction, which may in part correct for an error in the FSR, is not
applied to the  Stokes observable or down-stream vector magnetic field 
inversion. That is consistent with the presence of a 24-hr systematic variation 
seen even in weak fields in the vector-field products discussed in Section \ref{24h}.

\subsubsection{Daily Variation in LoS Doppler-Velocity Data}
\label{sec:DailySolarRotation}

\begin{figure} [!htb]
\centering
\includegraphics[width=0.8\textwidth]{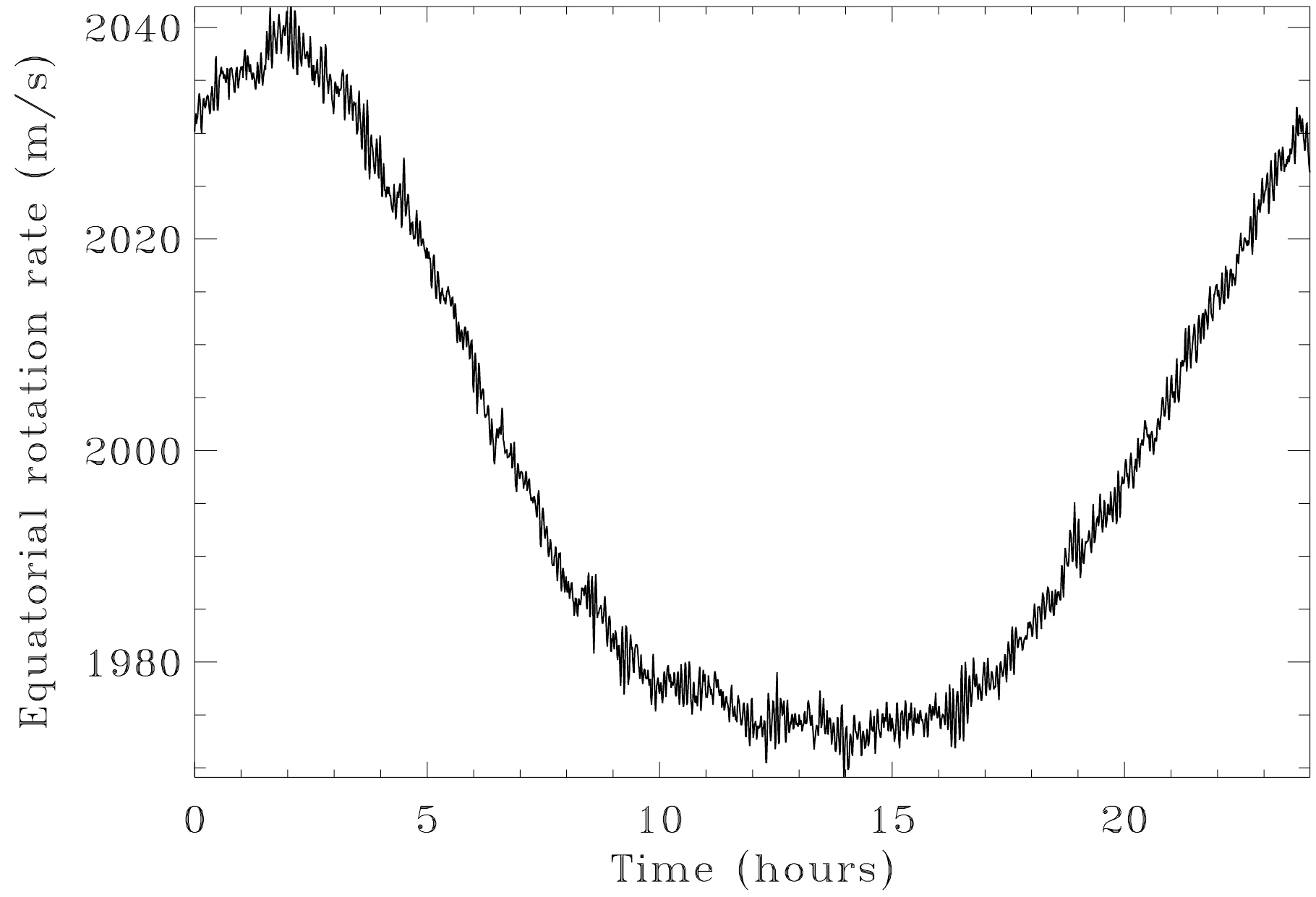}
\caption{Daily variation in the computed solar equatorial rotation velocity 
on 5 January 2015. SDO attained maximum sunward velocity of 2249.81~m\,s$^{-1}$ at 13:46 UT.}
\label{fig11}
\label{fig:SolarRotation}
\end{figure}

Even after applying the polynomial velocity correction, a residual error
remains in HMI Dopplergrams.  After removing the three components of the
SDO velocity projected onto each pixel, what remains can be described as
an offset from zero plus a nearly linear scale error that changes slowly
with time of year. The mean value has a 12-hr period with peak-to-peak
variation of $\sim$20\,m\,s$^{-1}$. We calculate a fit to the residual
solar velocity pattern using three terms: solar differential rotation using
the traditional form
$ V_\mathrm{rot}(\lambda) = V_\mathrm{eq} - B \cdot \mathrm{sin}^2(\lambda) - C \cdot \mathrm{sin}^4(\lambda) $
with $\lambda$= latitude, but with B == C; a term equivalent to a rotation
in the N-S direction; and radially symmetric limb shift (also known as a convective blue
shift). The equatorial rotation rate $V_\mathrm{eq}$ from these fits has nearly the same
daily shape as {\sc obs\_vr} with a typical amplitude of $\sim 1.5$\% of
{\sc obs\_vr}.  The differential rotation terms B==C, often show a mix of 12-hr
and 24-hr periods with peak-to-peak values of $\sim 40$~m\,s$^{-1}$ about
the average of $\sim 300$~m\,s$^{-1}$.  An example of the fit equatorial
rotation velocity is shown in Figure \ref{fig:SolarRotation} for January 5,
2015. That day, the peak-to-peak variation in the equatorial rotation rate
was $\sim 65$~m\,s$^{-1}$.  The N-S ``rotation" term, which should be zero if the
roll angle is correct, shows only small variations about a (much too large) average 
value of $\sim 60$~m\,s$^{-1}$. These values are about the same when the fit is
restricted to the inner half the disk area, showing that the issue is not
just with values near the limb.

These numbers suggest that the reported velocity has both scale errors
and errors across the field that are not corrected, and possibly even
exaggerated, by the polynomial correction coefficients. If the polynomial
correction is not applied, the equatorial rotation error amplitude is 
significantly reduced, but the mean amplitude is larger.

A scale error, such as a 1\% error in the FSR, will produce a similar signal
in the fit rotation rate. An uncorrected FSR error will introduce both 24
and 12-hour periodicity into the vector field inversions. But it is not that
simple. The N-S error suggests an error in one or more of the filter phase maps.
The HMI team is continuing to study these issues.

\section{Performance, Error Estimates, and Impact on Observables}

The following sections describe some of the remaining issues and uncertainties
in the HMI observables.  The information is intended to provide investigators
with a better understanding of how best to use and interpret measurements,
as well as how to avoid being mislead by systematic and other features in
the observations. As much as possible we have tried to be quantitative.

\subsection{Sensitivity of Doppler Velocity to Roll } 
\label{RollCalibration}

\begin{figure} [!htb]
\label{fig15}
\centering
\includegraphics[width=0.90\textwidth]{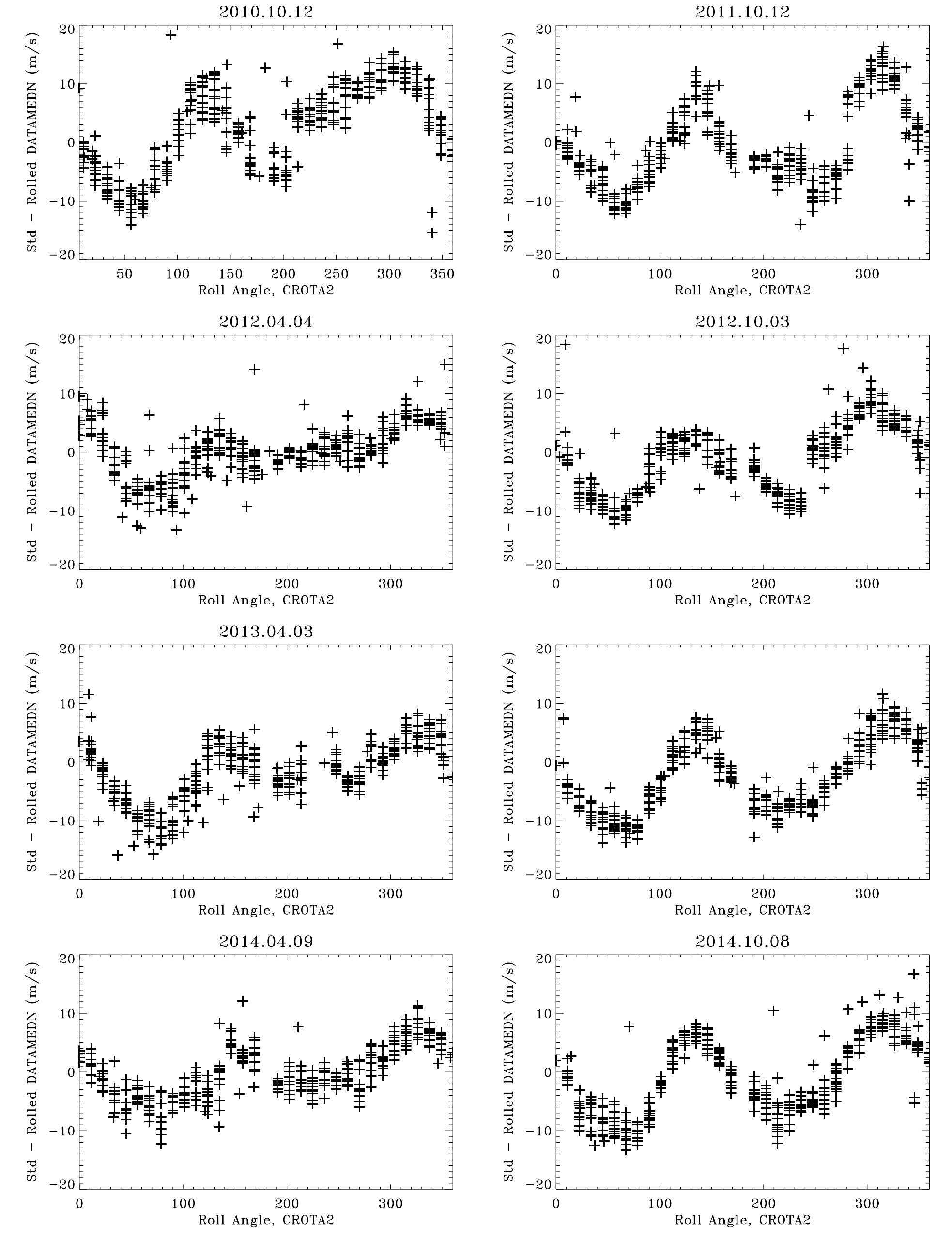}
\caption{The dependence on roll angle of the corrected full-disk median
velocity determined during several roll maneuvers.  Plotted as a function of
roll angle is the difference between the corrected full-disk median velocity,
{\sc datamedn}, measured during the spacecraft roll and the velocity measured
at the same {\sc obs\_vr} when the spacecraft was not rolled.  
Note that the nominal {\sc crota2} for HMI is $\approx 179.93^\circ$.  The panels show
results for eight SDO rolls from October 2010 to October 2014. There is a
systematic dependence on roll angle at the $\pm 10$~m\,s$^{-1}$ level. Curves
in October and April are a little different, owing to the difference in
the range of {\sc obs\_vr}, as shown in Figure \ref{fig13}.}
\label{fig:RollSensitivity}
\end{figure}

The measured value of the solar velocity should not depend 
on the orientation of the instrument.  At each roll angle, light from a
specific location on the Sun is projected onto a different CCD pixel and
the angular distribution of the light rays is different through different
parts of the filters. The senstivity of the HMI velocity measurement to the orientation of the
solar image passing through the instrument can be determined by rotating
the instrument. 

Roll maneuvers are performed in April and October of each year, during which
the SDO spacecraft is rotated 360 degrees around the Sun-spacecraft axis. The
roll calibrations take approximately six hours because the spacecraft pauses
for $\sim 7$ minutes after each $11.25^\circ$ step in roll. Data collected
during the pauses are used for measuring the oblateness of the Sun \citep[{\it
e.g.},][]{Kuhn2012}. 

Changing the roll angle, reported in the keyword {\sc crota2}, also 
helps characterize the spatial dependence of the HMI instrument and the angular sensitivity of the filter system. 
Light rays from a location on the Sun travel through the HMI instrument 
in such a way that they sample nearly the entire filter. However, the
incidence angle of the rays at each point in the filter depends systematically 
on solar image location, as does the wavelength of the spectral line due to solar
rotation velocity. The angle-wavelength relationship in the filter changes when 
the instrument rotates. If the change is not small compared with solar rotation, 
then the angular dependence may be of concern.

The upper left panel in Figure \ref{fig:RollDayVelocityDifferences} shows
the difference between the uncorrected median velocity, {\sc rawmedn},
and the known Sun-spacecraft velocity on a day when there was a six-hour
roll maneuver. It is easy to see the deviation from the usual polynomial
trend in the signal in the right half of the October 2010 panel of Figure
\ref{fig:RollDayVelocityDifferences}, when the roll was underway. 

The sensitivity to roll angle is revealed more clearly in Figure
\ref{fig:RollSensitivity}. For each roll maneuver we plot as a function of
roll angle the difference between the polynomial-corrected median velocity
measured when the spacecraft was and was not rolled. To do that the two
measurements of {\sc datamedn} made on the same day at times when the
Sun-spacecraft velocity was closest to being the same are compared.

If the HMI calibration were perfect, the median corrected velocity would
not depend on roll angle and would show only random variations of a few
m\,s$^{-1}$ due to instrumental noise and solar signals, including global
solar oscillations. Figure \ref{fig:RollSensitivity} shows that this
is not the case. There is a systematic variation of the median velocity
with roll angle that has a $180^\circ$ periodicity with an amplitude of
5\,--\,10~m\,s$^{-1}$, about 0.25\,--\,0.5\% of solar
rotation. The shape of the curve appears a little different in April than
in October. During October the rolls begin at about 18 UT, when the {\sc
obs\_vr} values are just becoming positive. During April the rolls begin
at 6 UT, a little before SDO local midnight, just before the value of {\sc
obs\_vr} becomes negative.  Since the LoS observables algorithm returns
a slightly different Doppler velocity depending on roll angle, there must
be a small wavelength sensitivity to roll angle in the filters in the HMI
optical path. This provides a rough estimate of the large-scale systematic
error in the determination of the Doppler velocity.

\subsection{24h Variations in Observables \label{24h}}

Temporal variations with a 24-hour period are found in every HMI observable.
The LoS velocity is discussed in Section \ref{polcor}. Effects in
the inverted vector magnetograms and line-of-sight magnetograms are
discussed by \citet{Liu2012} and \citet{Hoeksema2014}.

The major cause of this variation is the Doppler shift of the spectral
line relative to the nominal position. The geosynchronous orbit of SDO
has a daily change of velocity relative to the Sun of up to $\pm\,3500$
m\,s$^{-1}$ on top of an annual variation of several hundred m\,s$^{-1}$ due
to Earth's orbital motion. Solar rotation together with other additional motions
further complicate this issue. The SDO orbit also indirectly impacts the
observables in other ways. For instance, the relative position of the HMI side (vector) 
camera with respect to the Earth during the course of a day produces a 
change in the amount of heat received. This change in 
thermal environment impacts the Level-1 filtergrams, and consequently 
the observables, in subtle ways.

\citet{Hoeksema2014} showed an example of the 24-hour oscillations in the
HMI vector magnetograms of a simple and stable active region. NOAA AR \#11084
was tracked for four days, from 1 -- 4 July 2010. For each magnetogram,
two groups of pixels were selected with relative intensity in the 
ranges 0.0 -- 0.35 and 0.65 -- 0.75 relative to the nominal continuum.
The two groups correspond to pixels in the sunspot umbra and penumbra.
They also selected a quiet-Sun area of $40 \times 30$ pixels. The mean field
strength in each of the three intensity groups exhibits periodic 24-hour
variations. Based on analysis of 20 stable sunspots, \citet{Hoeksema2014}
find that the velocity sensitivity of the daily variation increases with 
field strength; a vector-field LoS-component measurement of 2500~G is expected to
change by $\pm 48$~G for a $\pm 3$\,km\,s$^{-1}$ change in {\sc obs\_vr}.
The amplitude of the daily field-strength variation is less than 5\% in a
strong-field region. The random uncertainty in the HMI vector-field total 
magnitude varies across the disc, but is $\sim 100$~G for a typical pixel 
in a 720s vector magnetogram. Note that because the HMI analysis assumes a filling factor of one, we
generally do not distinguish between magnetic field strength and flux density.

\begin{figure}[!htb]
\centering
\includegraphics[width=0.95\textwidth]{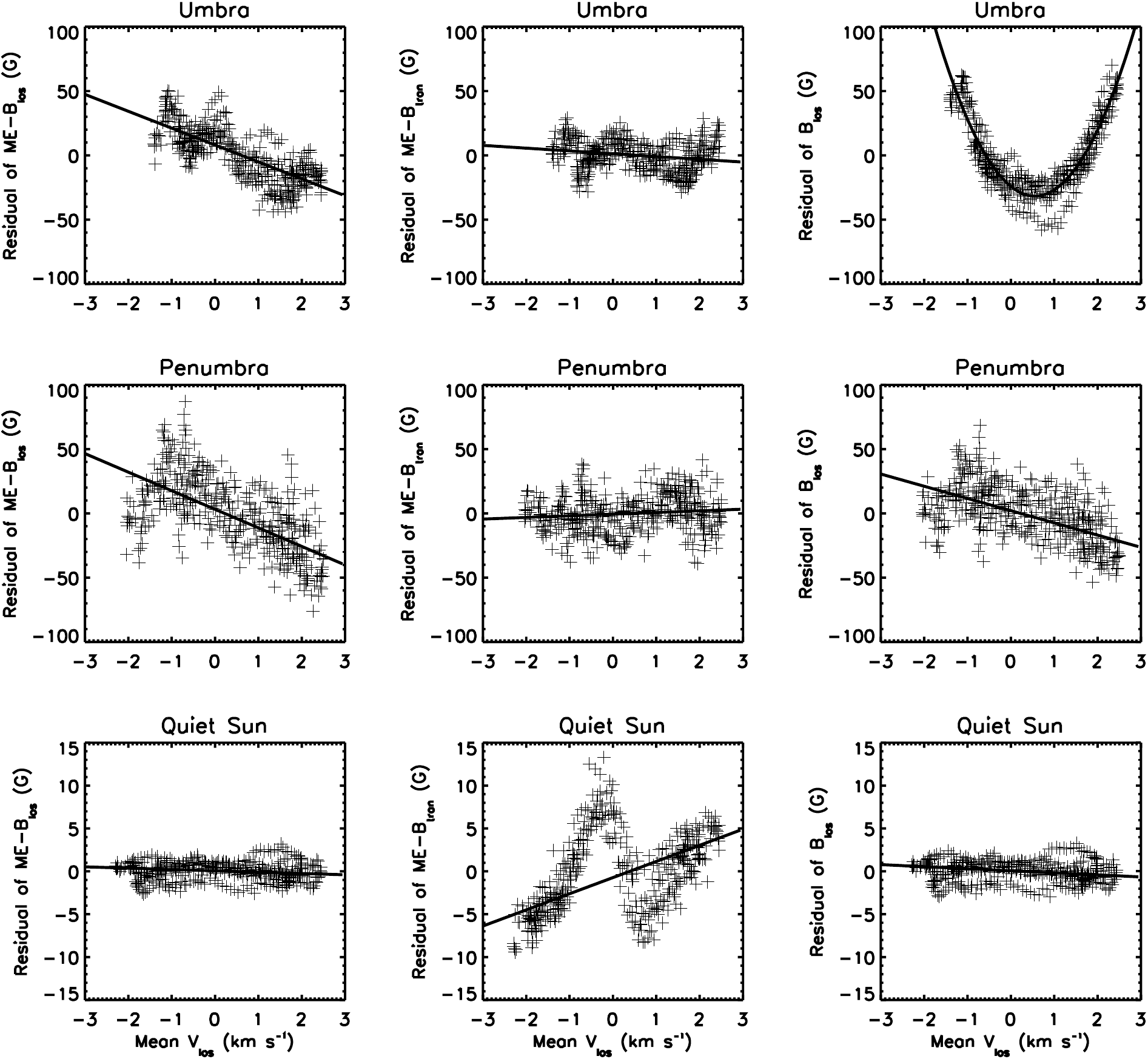}
\caption{Scatter plots show how the measured magnetic field depends on velocity
for regions of different field strength. The panels show the magnetic field
residual, which is the difference between the measured field and a four-day
polynomial fit (see text for details) versus velocity. From top to bottom the
rows show umbra ($\approx 2500$\,G), penumbra ($\approx 1300$\,G), and a quiet
Sun region ($\approx 100$\,G). Pixels are actually selected by intensity (see
text). Panels from left to right show the line-of-sight component of the full
vector magnetic field, the transverse component of the vector field, and the
line-of-sight field derived using MDI-like algorithm. All filtergrams are from
the side camera. The curve in the top right panel is a second-order polynomial
fit to the data; solid straight lines in the other panels are linear fits. 
Note the difference in scale in the quiet-Sun panels in the bottom row.}
\label{variationvsvelocity}
\end{figure}

To further explore the source of the 24-hour oscillation, we consider separately
the line-of-sight and transverse components of the vector field computed
using the Milne-Eddington inversion, hereafter called ME-B$_{los}$ and
ME-B$_{tran}$, respectively. The ME inversion depends on 
a fit to all of the Stokes components; however, the transverse field is most sensitive
to the Q and U components. 
Following \citet{Hoeksema2014}, we first
fitted the four-day time series of the field components (both ME-B$_{los}$
and ME-B$_{tran}$) in each intensity group with a third-order polynomial. The
difference between the field strength and the polynomial fit is the residual,
which eliminates most long-term evolutionary trends in the time series,
but retains sensitivity to variations with a period less than two days.
Shown in left and middle columns of Figure \ref{variationvsvelocity} are
residuals of the LoS component of the vector field (ME-B$_{los}$, left column)
and the transverse field component (ME-B$_{tran}$, middle column) for
umbra (top), penumbra (middle), and quiet Sun (bottom). For comparison,
we plot in the right column the residuals of the LoS field derived using
the MDI-like algorithm (B$_{los}$ hereafter). The curve in the top-right
panel is a second-order polynomial fit; the solid straight lines in the 
other panels are linear fits.

Velocity-dependent variations are clearly seen in the LoS field of both
ME-B$_{los}$ and B$_{los}$ in the sunspot (both umbra and penumbra). There
is no significant variation in the transverse field component ME-B$_{tran}$
in the sunspot. The unique quadratic form of the residuals of B$_{los}$
in the umbra (Figure \ref{variationvsvelocity}, top right panel) indicates
that in strong-field regions the determination of the LoS magnetic field
using the MDI-like algorithm depends on the magnitude of the velocity,
but not on the sign. Strong magnetic field, together with orbital
velocity, may shift either the left or right circular polarization
(LCP or RCP) components too far from one of the HMI sampling
positions. 
This may be one of the reasons behind the velocity-dependent
variation in strong field that is significant in the LoS component of the
field. This will be further discussed in Section \ref{sec:LoSErrors}.

There is no velocity-dependent variation obviously visible in either
ME-B$_{los}$ or B$_{los}$ in the quiet Sun region (bottom left and right
panels of Figure \ref{variationvsvelocity}). Surprisingly, the transverse
component shows a relatively strong and complex variation in quiet-Sun
regions near zero velocity (ME-B$_{tran}$, bottom middle panel of Fig.
\ref{variationvsvelocity}).  Note that the field-strength scale for the
quiet-Sun is much less than for the umbra and penumbra. The specific cause
of this variation is still under investigation.

The rest of this section describes efforts made to characterize and mitigate
the 24-hour variations in the HMI LoS magnetic-field-strength values. Unfortunately the
source of the fluctuations and a successful mitigation method remain problematic.

To investigate the effects of the limited HMI spectral resolution,
spectropolarimetric observations were collected using the Interferometric Bidimensional Spectropolarimeter (IBIS) 
instrument at the National Solar Observatory's Dunn Solar Telescope. The sunspot in NOAA AR \#10960 was observed for seven hours on 8 June 2007 near disk center (S07\,W17), and the full Stokes profiles of the Fe\,{\sc i} line were measured at 23 wavelengths. These Stokes profiles are averaged over the first hour of observations and interpolated onto a fine-wavelength grid
\citep[see][]{CouvRajaWachter2012}. The increased number of wavelengths allows better modeling of the sensitivity
of the lower spectral-resolution HMI filters. A Stokes-I profile from a quiet-Sun region is used as the reference line profile when computing look-up tables for the MDI-like algorithm applied to IBIS data. The measured high-spectral-resolution LCP and RCP profiles for a pixel inside the  sunspot umbra are shifted in wavelength to simulate various Doppler velocities and magnetic field strengths. The simulated Doppler velocities span the range from $-2016$ to $+1860$ m\,s$^{-1}$. That range is typical for the daily variation in SDO {\sc obs\_vr} values. At each velocity and field strength, the MDI-like algorithm is applied to simulated HMI values for LCP and RCP derived by convolution with HMI filter transmission profiles.

\begin{figure}[!htb]
\centering
\includegraphics[width=0.7\textwidth]{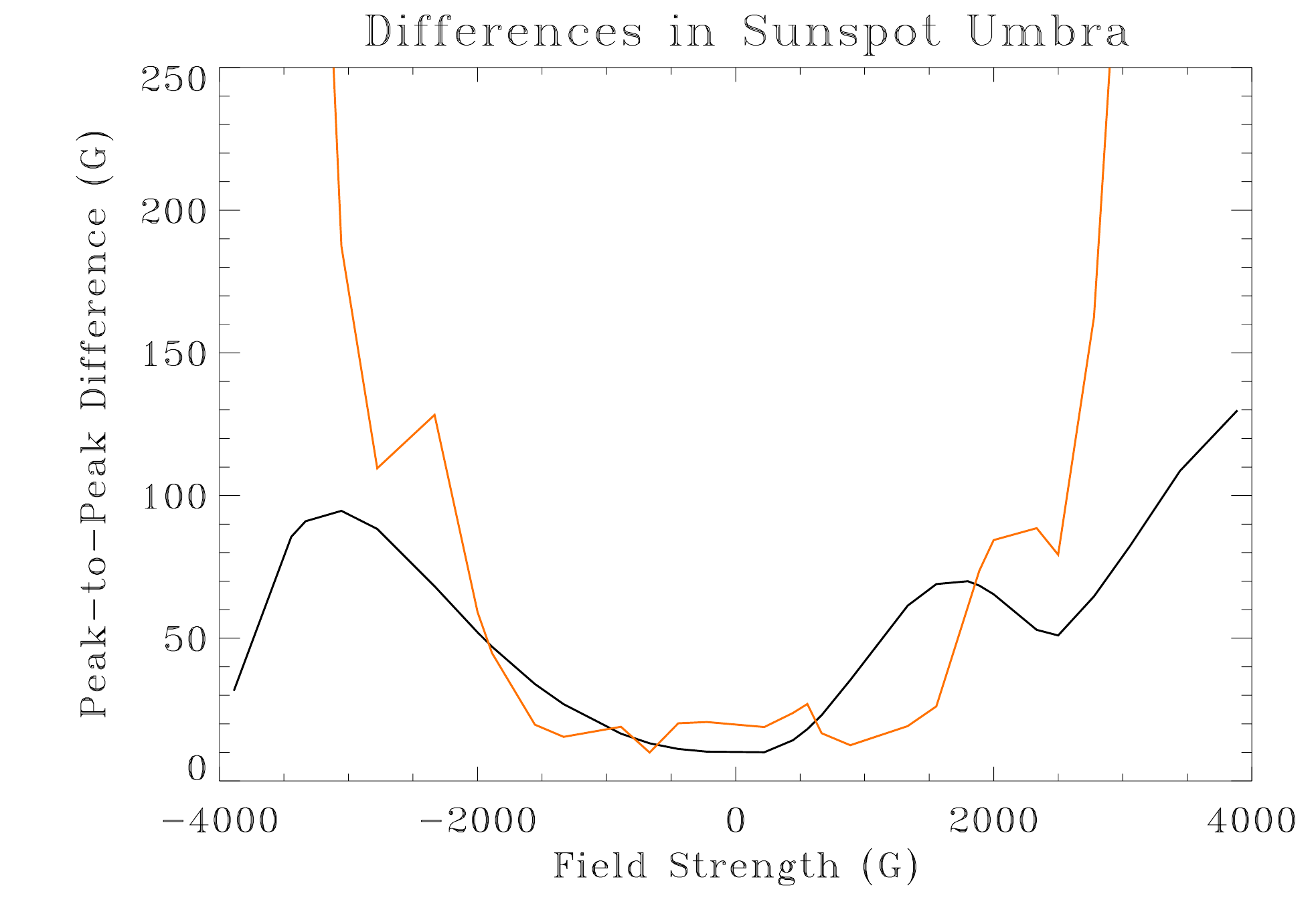}
\caption{Computed peak-to-peak difference in the LoS magnetic-field strength
returned by the observables algorithm as a function of model field strength,
for the Doppler velocity ranging from $-2016$ to $+1860$ m\,s$^{-1}$). 
That range is typical of the daily Sun-SDO orbital velocity.
The black curve shows the field-strength variation for the MDI-like algorithm. 
The red curve is for a least-squares fit using a Voigt profile.
}
\label{fig:field-velocity}
\end{figure}

The black curve in Figure \ref{fig:field-velocity} shows the dependence of the
daily amplitude variation on field strength. The curve is not symmetric with
field polarity, but the daily variation remains below 30\,G for field $<
1000$\,G and $< 75$\,G for field strengths below 2250\,G. 
In practice the
specific pattern depends on the {\sc obs\_vr} range, and it is complicated by 
the position-dependent photospheric velocity. 
Even for a given field strength at a given location on the solar surface,
the daily peak-to-peak variation will vary during the year with the orbital
velocity of SDO. The systematic daily variation is much larger than the random per-pixel 
photon noise in the HMI LoS magnetic field measurement: $\sim 7$~G for the 45s 
and $\sim 5$~G for 720s.

\begin{figure}[!htb]
\centering
\includegraphics[width=0.5\textwidth]{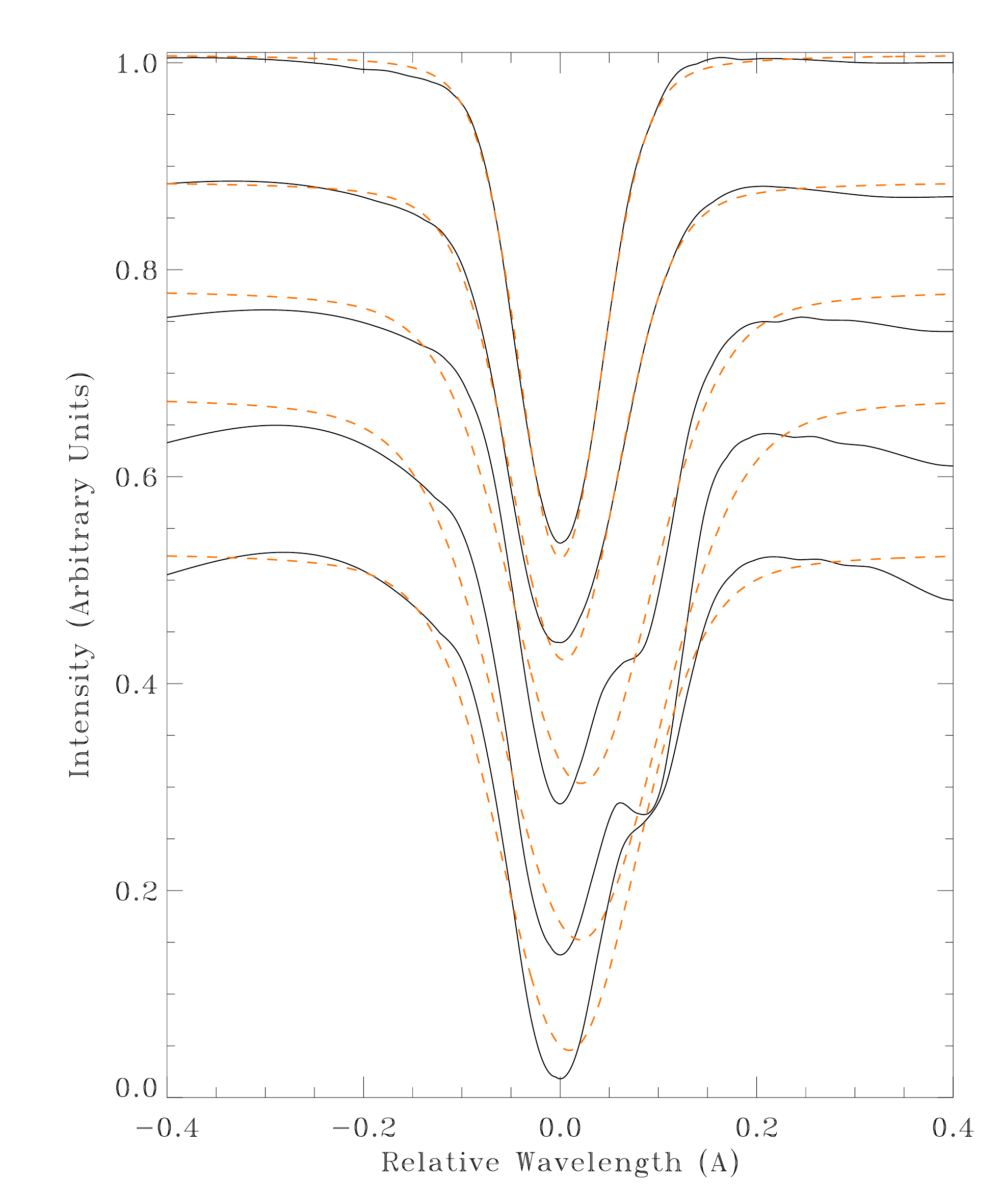}
\caption{Changes in the high-resolution observed LCP Fe\,{\sc i} line profile at 
various locations across the sunspot from quiet Sun (top black curve) to the 
darkest part of the sunspot umbra (bottom black curve) corresponding to 
different field strengths. The curves have been divided by the continuum value and offset for clarity. Red lines 
show the least-squares fit to the observed line profile with a Voigt profile. The complex LCP profiles are what is expected in a sunspot where there is a mixture of $\sigma$ and $\pi$ polarization components coming from a region with inclined field and less than 100\% filling factor. HMI makes filtergrams at six wavelengths spaced by 76~m\AA ~centered on the unshifted wavelength.}
\label{fig:lineprofiles}
\end{figure}

To assess the performance of the MDI-like method for determining the LoS field, 
the red curve in Figure \ref{fig:field-velocity}
compares the daily peak-to-peak variation determined for a least-squares regression 
using a Voigt-profile fit to
the same synthetic HMI line profiles. The curve is a little more symmetric
and the daily range is marginally smaller than the MDI-like method for
LoS field strengths between 1000 and 1800\,G, but overall the least-squares
fit does not reduce the daily peak-to-peak difference. The potential advantage
of the least-squares fit is that it adapts to changes in the Fe\,{\sc i} line width
and line depth in the presence of strong field. However, this does not seem
to be especially beneficial for reducing the daily peak-to-peak variation.
The main issue with the Fe\,{\sc i} line profile in strong fields is not that
the line width or line depth changes, but rather that the shape of the line is
quite different from a Voigt or Gaussian profile. Stronger fields alter this
shape more than weaker fields, as can be seen in the measured and fitted line
profiles in Figure \ref{fig:lineprofiles}.

\begin{figure}[!htb]
\centering
\includegraphics[width=0.75\textwidth]{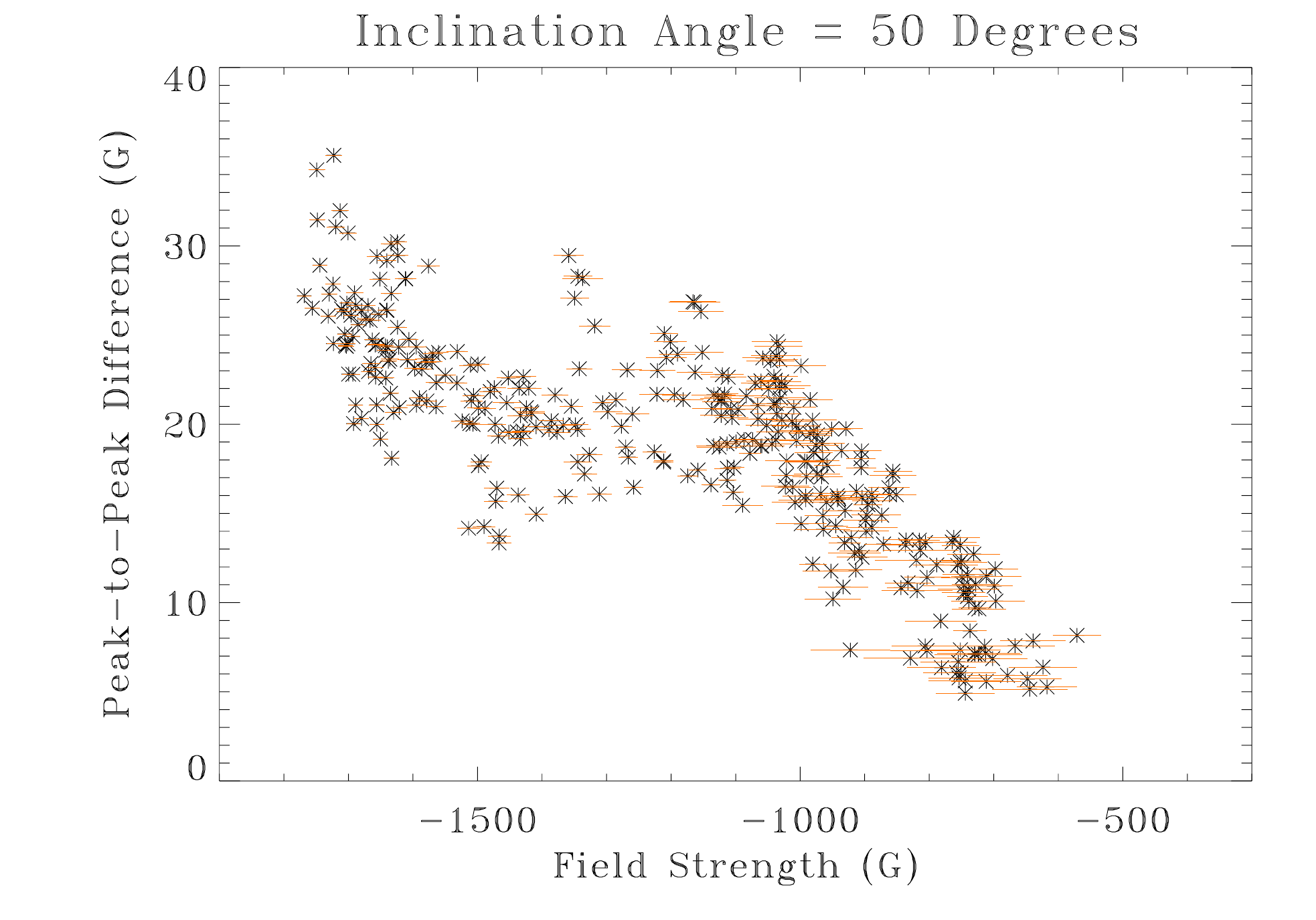}
\caption{Peak-to-peak difference in the LoS magnetic field strength returned
by the MDI-like algorithm as a function of observed total vector-field magnitude for 
the Doppler velocity range $-2016$ to $+1860$ m\,s$^{-1}$. All data points shown have an inclination angle between 49.5$^o$ and 50.5$^o$; this inclination range selects a reasonable number of points with a range of field values typical of the penumbra. Field strength is derived by applying a Milne-Eddington inversion to the high-spectral-resolution IBIS Stokes profiles. Error bars are estimated from the inversion algorithm. Peak-to-peak
variation is determined by applying the MDI-like method to Doppler-shifted
lower spectral-resolution values computed by convolving the HMI filter
profiles with the IBIS observations.}
\label{fig:fielderrors}
\end{figure}

As another test, we consider all of the pixels having a given field inclination
and calculate the peak-to-peak difference in the field strength returned
by the MDI-like algorithm for various field strengths. The vector magnetic
field magnitude in the same sunspot is determined by performing a Milne-Eddington
(ME) inversion \citep{Rajaguru2010} on the Stokes profiles averaged over
the first hour of IBIS observations and interpolated onto a fine-wavelength
grid \citep[see][]{CouvRajaWachter2012}. Figure \ref{fig:fielderrors} shows
the peak-to-peak difference in the LoS magnetic field strength returned by
the MDI-like algorithm as a function of total field magnitude for simulated
Doppler velocities ranging from $-2016$ m s$^{-1}$ to $+1860$ m s$^{-1}$.
The points plotted here all have an inclination angle of magnetic field to
the line of sight between 49.5 and 50.5 degrees. This inclination value was
selected simply to provide a reasonable number of points over a reasonable
range of field magnitudes. The points tend to cluster near the umbra-penumbra boundary.
Error bars for field strength are estimated by the
ME inversion algorithm \citep{Rajaguru2010}.  The figure shows a systematic
dependence of the peak-to-peak difference on field strength. This dependence
is probably not linear. The increase in the daily variation from about 10\,G
to 30\,G for total field strengths from about 500 -- 1800\,G is consistent
with the results shown in Figure \ref{fig:field-velocity} when correction
is made for the projection from total field strength to LoS field strength
at 50$^{\circ}$ inclination.

Several possible approaches to reduce the amplitude of the 24-hour variation 
have been explored. Unfortunately, none result in a signficantly better outcome. 
For example, current implementation of the MDI-like algorithm 
ignores the I-ripple of the tunable filter elements (see Section \ref{iripple}). 
Better estimates of the phase and amplitude of the I-ripple were tested, as
were more realistic Fe\,{\sc i} line profiles.  Neither change resulted
in a lower amplitude for the 24-hour variations. Nor did an attempt
to implement a spatially dependent polynomial correction (Section \ref{polcor}). 
Certain HMI filter characteristics, such as the exact free spectral 
range (FSR) of the tunable elements, are sufficiently uncertain that adjusting
them may be part of the solution. The instrument team continues to
investigate the cause of the variation and to determine a way to mitigate it.

 \subsection{Errors with the LoS Algorithm} \label{sec:LoSErrors}

As mentioned in Section \ref{24h}, the MDI-like algorithm may 
produce significant errors in the presence of strong fields.  The shape of
the Fe {\sc i} line in a strong and inclined field differs significantly 
from a Voigt profile and from the synthetic profile used to produce the 
look-up tables, while the wavelength shift resulting from the Zeeman effect
may push the LCP, RCP, or both components partly or totally outside the 
dynamic range of HMI in places where there is high velocity.

\begin{figure}[!htb]
\centering
\includegraphics[width=0.62\textwidth]{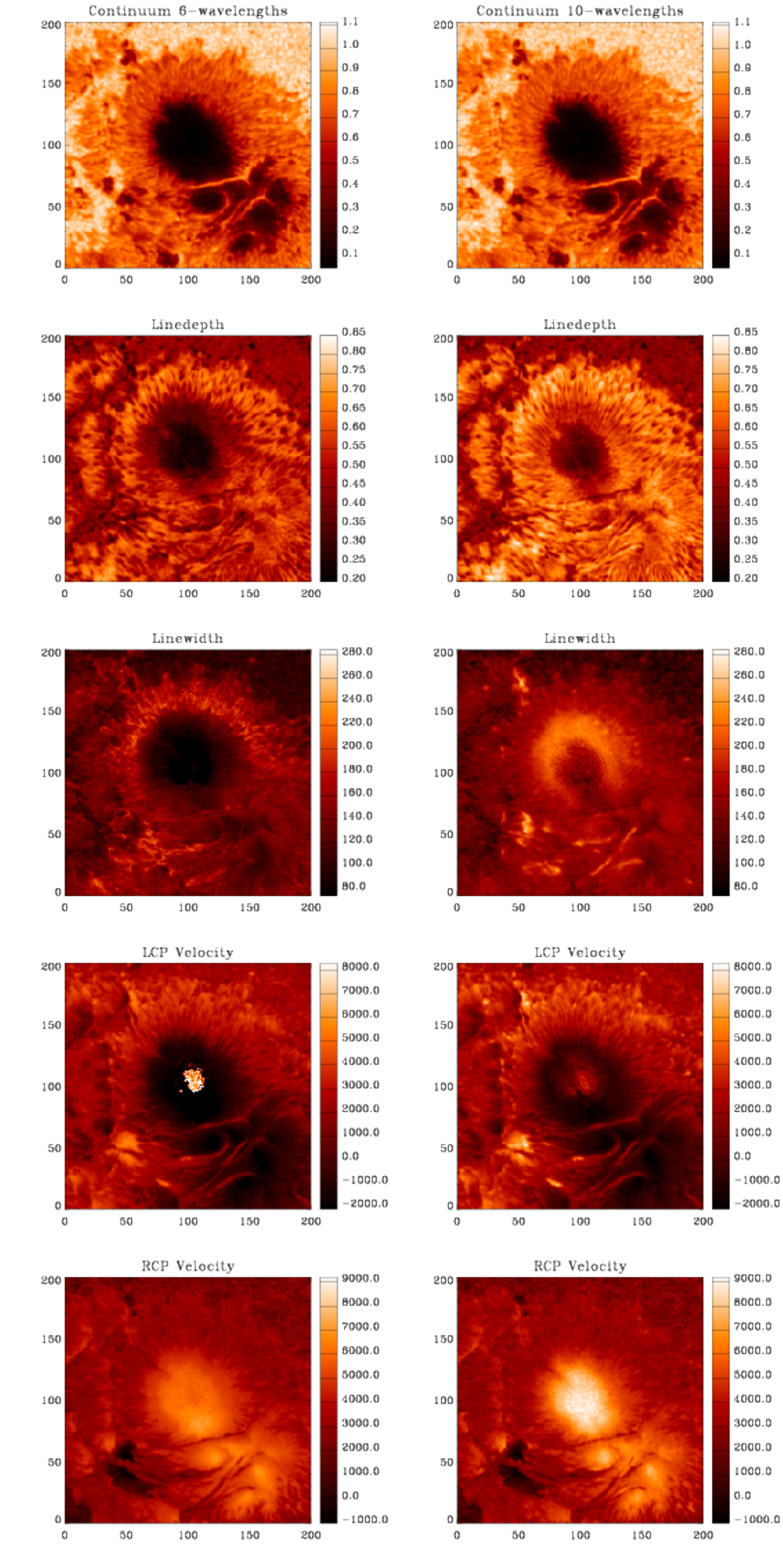}
\caption{LoS observables returned by the MDI-like algorithm based on observations 
at six and ten wavelengths in different columns. Results are derived from a 
ten-wavelength sequence taken on 24 October 2014 for NOAA AR \#12192. 
The color bars for each quantity in the left and right columns are the same.}
\label{fig17}
\label{fig:tenpointsequence}
\end{figure}

One way of evaluating such errors is to take a 10-wavelength observables
sequence, rather than the usual six.  On 24 October 2014, such a special
observing sequence was run for about an hour. The cadence on the front
camera was 75 seconds, rather than 45 seconds, and ten equally spaced
wavelengths were taken. There was a large sunspot that day, NOAA 12192.
Figure \ref{fig17} shows a comparison of the LoS observables quantities
returned by the MDI-like algorithm when using six or ten wavelengths.
The six-wavelength plots show the standard observables
obtained just prior to the run of the special sequence.  While the continuum
intensity appears fairly robust, the line width is especially sensitive to
the change in the number of wavelengths.  With six wavelengths, the computed
Fe {\sc i} line width decreases as we move from quiet Sun toward the sunspot
umbra. This is the reverse of what is expected in the presence of a magnetic
field, and of what the 10-wavelength sequence returns.

The panels showing the individual LCP and RCP velocities highlight the
differences. The LCP velocities derived
with six wavelengths show saturation inside the sunspot umbra, due to the
wavelength shift produced by the Zeeman effect (the LCP profile is partly
outside the dynamic range of HMI with only six wavelengths).  Moreover,
the RCP velocities with six wavelengths are underestimated compared to 10
wavelengths. This results from the RCP profile lying partly outside the
dynamic range of only six wavelengths: the MDI-like algorithm, based on a
discrete estimate of the Fourier coefficients, assumes that the line profile
is periodic. If the profile is truncated, then the condition of periodicity
produces a line profile whose shape is significantly different from a quiet-Sun
profile, resulting in large errors in the estimate of the velocity shift.

\begin{figure}[!htb]
\centering
\includegraphics[width=0.8\textwidth]{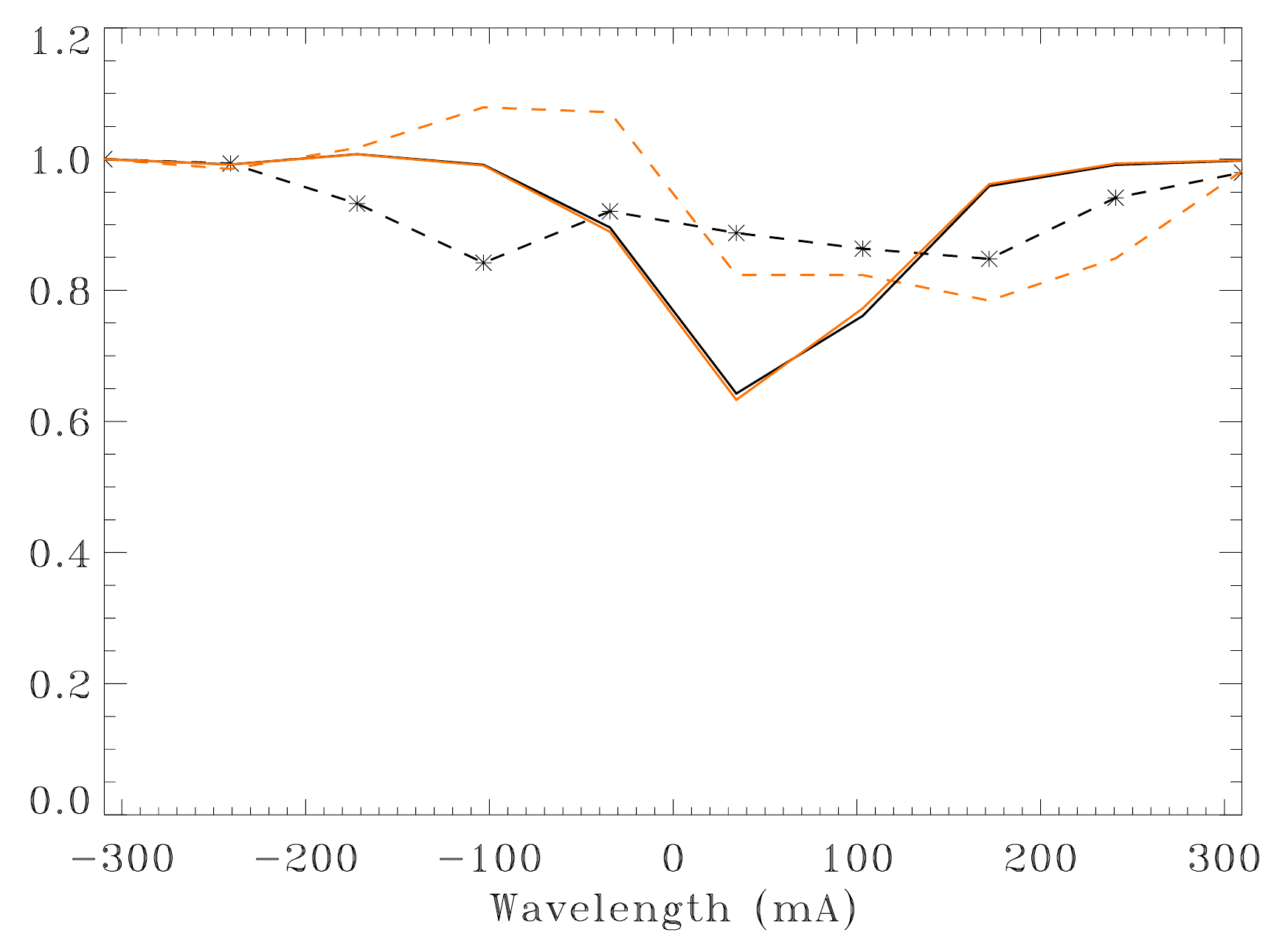}
\caption{LCP (black lines) and RCP (red lines) components of the Fe {\sc i} 
line profile in a pixel within a sunspot umbra (dashed lines) and in a quiet-Sun 
pixel (solid lines). Results are from a ten-wavelength observables sequence taken on 24 October 2014.
Intensity units are normalized to the value of the intensity at -310~m\AA.}
\label{fig18}
\label{fig:tenpointRCPLCP}
\end{figure}

The impact of a magnetic field on the line profile is also evident in Figure
\ref{fig:tenpointRCPLCP}, which displays the LCP and RCP components obtained during 
the 10-wavelength sequence.  This figure shows values for a pixel in a sunspot umbra (dashed
lines) and in the quiet Sun (solid lines).  The magnetic field in the sunspot
shifts the LCP (black lines) and RCP (red lines) components, but they remain
within the dynamic range of HMI (even with only six wavelengths).  However,
the line shapes are significantly distorted compared to the quiet-Sun shape
used to produce the look-up tables: both LCP and RCP components are shallower
and wider, and the LCP profile does not have a clearly
defined minimum. A fit of the profile by a Gaussian function finds a minimum
at a lower wavelength than the location of the actual minimum. This results
in an underestimation of the actual LoS magnetic-field strength in the umbra.

Therefore, in the presence of a strong field the MDI-like algorithm will
underestimate the LoS field magnitude, and may result
in saturation if the Doppler shift resulting from motions (Sun-SDO radial
velocity, solar rotation, acoustic waves, convection, etc.) combined with
the Zeeman effect send the one or both of the LCP and RCP profiles outside
the range of the instrument \citep[see also][]{Liu2012}. Even in the absence of saturation, the fact
that the Fe {\sc i} line profile is different from the one used to produce
the look-up tables will result in errors in the observables determination
and contributes to the 24-hour oscillations.

\subsection{Magnetic-Field Error with Stokes-Vector Inversion}

Inversion of the full Stokes vector produces
better estimates of the magnetic field than the LoS strength returned by
the MDI-like algorithm. For one thing, the Very Fast Inversion of the Stokes
Vector (VFISV) code \citep{Borrero2011,Centeno2014} fits the width and
depth of the Fe {\sc i} line profile, thus taking into account the broadening
and shallowness of this profile in presence of a magnetic field. However, the inversion has two significant limitations: the fixed six-point  spectral-line sampling and the simple model used by the inversion code. 
If the magnetic field is strong (e.g. in an umbra), the split of the line may be 
larger than HMI's spectral range, particularly where the velocity is large, and
the intenstity profiles may be very shallow. This will reduce the reliability of 
the inversion. In addition the Milne-Eddington model used in VFISV
assumes that none of the physical quantities in the atmosphere vary with 
depth, except the source function. In places where strong gradients in velocity or
magnetic field exist, the quality of the fit will degrade as the gradients
become more significant.

\begin{figure} [!htb]
\centering
\includegraphics[width=0.9\textwidth]{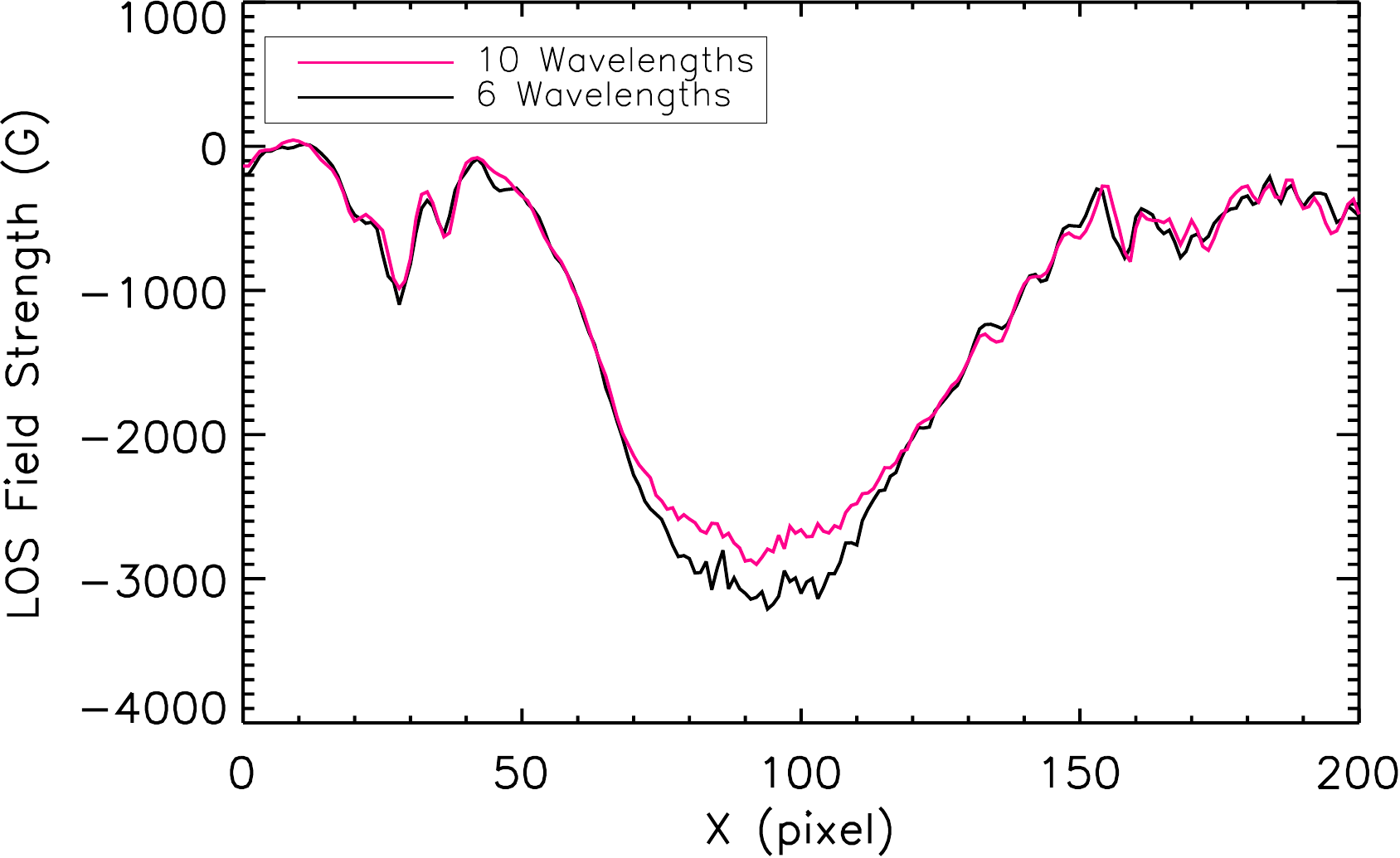}
\caption{Comparison of inversion results of VFISV for the calculated 
line-of-sight component of the vector magnetic field --- cosine of the 
inclination angle times the field strength --- derived from six- and ten-wavelength 
observable sequences. We show a cut across the umbra of the sunspot in 
NOAA AR \#12192 on 24 October 2014. A 75-second-cadence observables sequence 
with ten wavelengths rather than the usual six was run for an hour that day.} 
\label{fig16}
\label{fig:VFISVsixten}
\end{figure}

To better quantify these
uncertainties, we use the results of the special ten-wavelength observables sequence
of 24 October 2014 described in the previous section.  Full Stokes profiles
were produced for each time step in the interval covered by this sequence,
and VFISV was run to invert the vector magnetic field.  The large sunspot
in NOAA AR \#12192 allowed for a comparison of inverted field strengths in the presence of
strong field. To make comparison easier with the front-camera observables,
we looked at the LoS component for the field strength returned by VFISV 
(field strength multiplied by the cosine of the inclination angle). 
Six-wavelength results
we obtained by looking at standard VFISV inversions just prior to the start
of the special sequence run. As can be seen in Figure \ref{fig:VFISVsixten},
the LoS field strength inside the umbra of the sunspot is overestimated (in
absolute value) using six wavelengths compared to ten. This plot
can be used to provide a rough estimate of systematic errors we might have in
the inverted field strength from VFISV with the standard observables sequence.

\subsection{Temperature Dependence of CCD Gain}

\begin{figure} [!htb]
\centering
\includegraphics[width=0.95\textwidth]{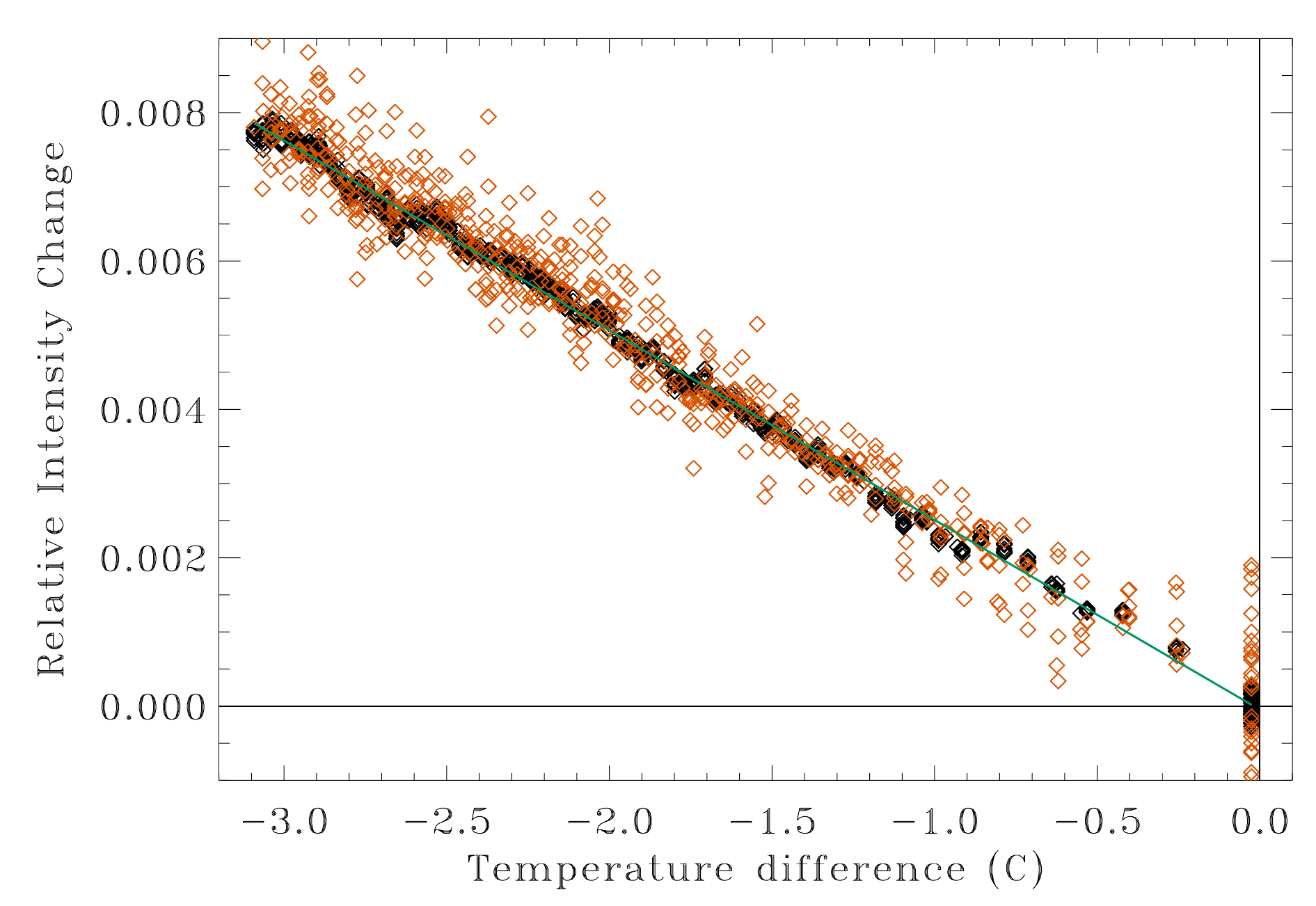}
\caption{Dependence of the HMI front-camera CCD gain on temperature difference. The plot shows the difference in the corrected median intensity as a function of detector-temperature difference for pairs of filtergrams with identical filter tuning and the same Sun-SDO velocity on 28 September 2011. The filtergrams represented by darker black diamonds had higher overall intensities due to the tuning of the Lyot filter, and therefore less noise. The trend line is the result of a linear regression. See text for details.}
\label{fig19}
\label{fig:CCDGain}
\end{figure}

The gain of the HMI CCDs varies with CCD temperature, as shown in
Figure \ref{fig:CCDGain}. 

This figure is derived from detune sequences taken on 28 September 2011. About
10 hours of detunes were run in Obsmode on the HMI front camera. For each Level-1 filtergram the median intensity value of the on-disk pixels, {\sc datamedn}, is corrected for the known non-linearity of the CCD and for changes in Sun-SDO distance. For each particular Sun-SDO radial velocity, {\sc obs\_vr}, there are generally two or more observations made with the same filter tuning ({\sc fid}) at different times of the day at two different CCD temperatures. Figure \ref{fig:CCDGain} plots the fractional change in {\sc datamedn} as a function of the difference in temperature for pairs of observations made with the same Sun-SDO velocity and same filter tuning. Low {\sc fid} values (6000 to 6008) with high transmittance due to the tuning of the E1 Lyot element
are represented by black diamonds, while lower-intensity higher {\sc fid} values are shown as red diamonds. The trend line is the result of a linear regression. The slope is $\approx -0.0025$, so {\sc datamedn} decreases by about $0.25\%$ when the CCD temperature increases by one Kelvin. The CCD temperature variations on this date are typical, with peak-to-peak variation
of less than $\approx 3$\,K during the day. The mean daily CCD temperature varies
by a degree or two over the course of a year.

This sensitivity of the CCD gain to temperature is not negligible for the
measures of intensity. However, since the determination of the velocity and 
magnetic observables depends on intensity differences, the effect is small.  
The HMI team plans to include a correction for this effect in the observables 
pipelines. Complexities in dealing with the temperature records have delayed the 
implementation of this correction.

\subsection{Correcting the  HMI Point Spread Function \label{PSF}}

In a real optical system not all of the light from a point in the object 
ends up at one point in the image, rather it is spread out according
to what is known as the point spread function (PSF). 
The final image is the convolution of the original image with the PSF
(noting that the PSF is often spatially variable). The PSF 
includes contributions from both diffraction and
imperfect optics. Deviations from the diffraction-limited case are often 
divided into two categories - those from large-scale wavefront errors, 
which result in a reduced sensitivity to small-scale variations, and 
those due to scattering from dust and scratches that result in large-angle 
scattering.  The latter is often referred to as stray light and may also 
include such things as ghost images.

It is convenient to describe the PSF in terms of its Fourier transform,
known as the optical transfer function (OTF) and the magnitude of that,
known as the modulation transfer function (MTF). Though not true generally,
because all of the OTFs considered in our analysis are real (the PSFs are 
symmetric), the MTF and OTF are the same.
For an ideal diffraction-limited telescope the OTF is given by \citep{bracewell95}:
\begin{equation}
\label{eqn:MTF4}
     OTF_{ideal}(\rho) = \frac{2}{\pi}  \left [
        acos(\rho^{\prime}) -\rho^{\prime}\sqrt{(1-\rho^{\prime2})}\right ]
\,;\:\: \mathrm{ where} \:\:
     \rho^\prime = \frac{f \lambda} {PD}\rho \approx 1.82\rho \;.
\end{equation}
Here $\rho^\prime$ is the spatial frequency normalized by the optical Nyquist 
frequency, $\rho$ the spatial frequency in pixels$^{-1}$,
$D$ the diameter of the telescope (140mm),
$f$ the effective focal length (4953 mm), 
$\lambda$ the wavelength (6173 \AA), 
and P the pixel size (12 microns).
The ideal PSF is then

\begin{figure} [!htb]
\centerline{\hspace*{0.015\textwidth}
 \includegraphics[width=.95\textwidth]{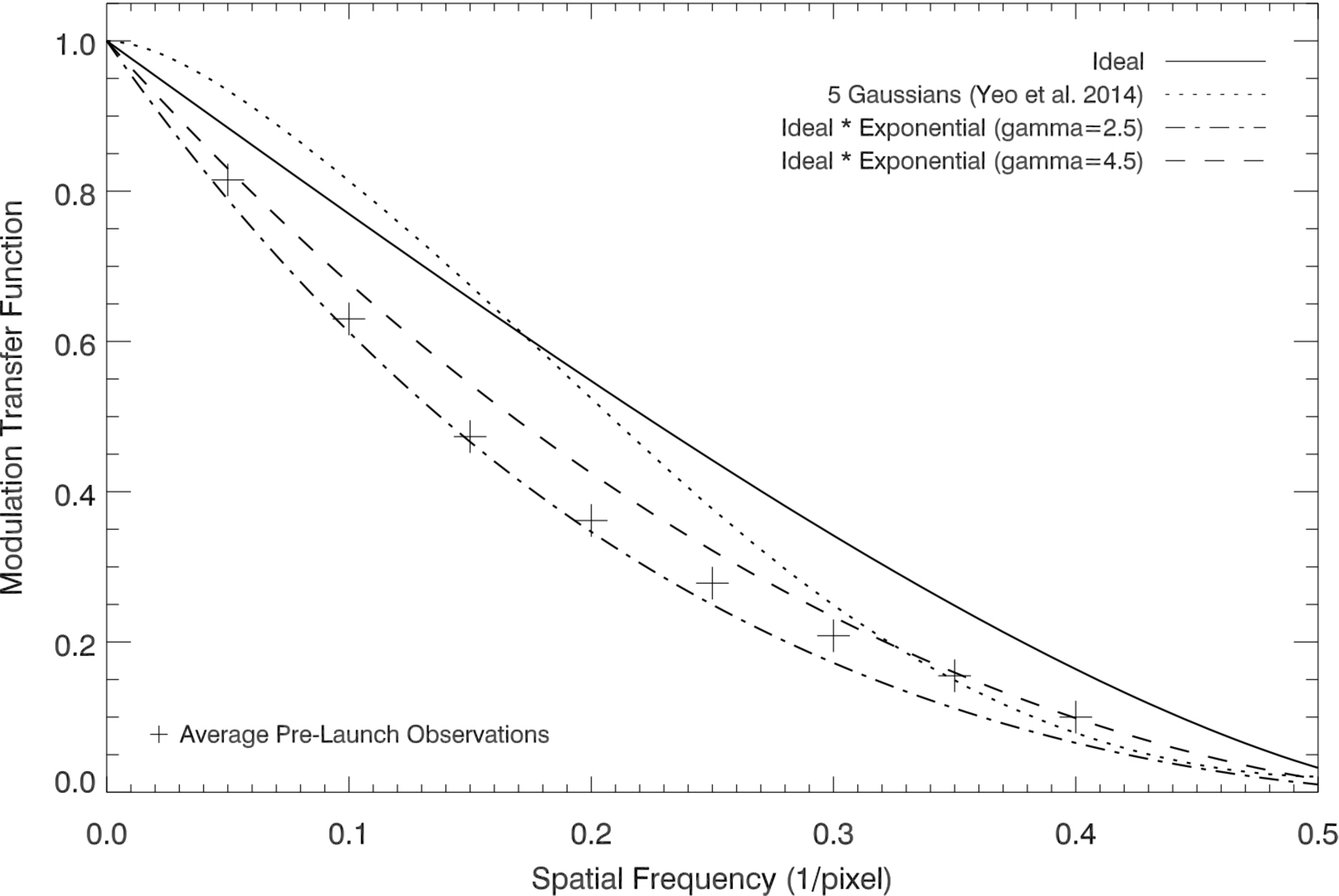}}
\caption{The ideal MTF is shown as a function of spatial frequency. Also
plotted are two MTFs that are the ideal OTF multiplied by a simple
exponential with $\gamma = 2.5$ and $\gamma = 4.5$. The \citet{Yeo2014}
MTF is overplotted, as is an MTF determined by the ideal multiplied by an
exponential function (Equation \ref{eqn:MTF7}). The symbols represent the
average of three of the ground based curves observed during instrument
calibration, as reported in \citet{Wachter2012}. }
\label{fig:fig2}
\label{fig:MTFPlot}
\end{figure}

\begin{equation}
     \label{eqn:MTF2}
        PSF_{ideal} ( r ) = \left( \frac {2J_1(r^{\prime})}
             {r^\prime} \right)^2 
		\,;\:\:\mathrm{where}\:\:
     \label{eqn:MTF3}
        r^\prime = \frac{\pi PD}{f \lambda} r .
\end{equation}
Here r is the radius in pixels and $J$ is a Bessel function. 
Note that the OTF for HMI does not go to zero by the pixel Nyquist frequency, because
HMI undersamples by a factor of $\sim 1.1$, resulting in some aliasing at
the highest spatial frequencies.

To model the real optical performance a variety of approaches have been taken.
Some of these, such as modeling the PSF (or equivalently the MTF) as a sum
of simple functions, such as Gaussians \citep{Yeo2014}, exponentials, etc.,
are inherently unphysical, because they indicate sensitivity above the optical
Nyquist frequency and do not properly describe the ideal diffraction-limited
case \citep{Wedemeyer-Bohm2008}. Having said that, the use of Gaussians is
convenient and is adequate for some purposes. 

The form of the PSF derived here is an Airy function convolved with a
Lorentzian. The parameters are bound by observational ground-based testing
of the instrument conducted prior to launch \citep{Wachter2012}, by full-disk
data used to evaluate the off-limb behavior of the scattered light, as well
as by data obtained during the Venus transit. The PSF correction has been
programmed in both {\sc C} and {\sc cuda C} and runs within the JSOC environment using
either a CPU or GPU. A single full-disk intensity image can be deconvolved
in less than one second. In contrast, \citet{Yeo2014} model the HMI PSF as
the sum of five Gaussians that are fit to only the Venus-transit data, and 
the correction routine runs much more slowly.
This new PSF has already been used by \citet{Hathaway2015} to forward-model 
solar-convection spectra and by \citet{Krucker2015} to investate footpoints 
of off-limb solar flares.

The PSF model has two components. The first accounts for the large-scale
wave-front errors and the second describes the long-distance scattering.
For the large-scale errors we parameterize the PSF measured before launch 
by \cite{Wachter2012} as:
\begin{equation}
\label{eqn:MTF7}
     OTF(\rho) = OTF_{ideal}(\rho) \times exp ( - \pi \rho^\prime / \gamma ),
\end{equation}
with an adjustable parameter $\gamma$, as shown in Figure \ref{fig:MTFPlot}.
The best value for the exponential of Eq. \ref{eqn:MTF7} is $\gamma = 4.5$,
as determined by fitting the Venus-transit data from 5 June 2012 measured
with the side camera. The result agrees well with the pre-launch measurements,
shown by the plus symbols in the figure.
\clearpage
The shadow of Venus is too small to effectively measure the long-distance scattering.
Lunar-eclipse observations from 7 October 2010 show that HMI continuum intensity
filtergrams have a light level 0.34\% of the disk-center continuum intensity
at positions 200 pixels onto the lunar disk; the scattered light falls 
off roughly exponentially with distance.  This motivated adding an additional term to the PSF. As noted above, simply adding such an exponential term to the PSF leads to an unphysical solution. However, the error is quite small, so for the purpose of testing the idea we model the PSF as:
\begin{equation}
\label{eqn:MTF8}
     PSF(r) = {\mathscr{F}}(OTF) +  c\times exp ( \frac{-\pi r}{\xi r_{\mathrm{max}}} ) ,
\end{equation}
where $c = 2 \times 10^{-9}$, $\xi = 0.7$, $r_{\mathrm{max}}$=2048, and $r$ is in pixels. 
If we only considered light
scattered from less than 10\,$^{\prime\prime}$ away, as \citet{Wedemeyer-Bohm2008}
did for SOT, then the additional term would not be necessary. The lunar
eclipses are not appropriate for fitting other components of the MTF because
of the temperature perturbation to the HMI instrument. A more accurate analysis
of scattered light is underway.

\begin{figure}[!htb]
\centerline{\hspace*{0.015\textwidth}
\includegraphics[width=0.99\textwidth,clip=]{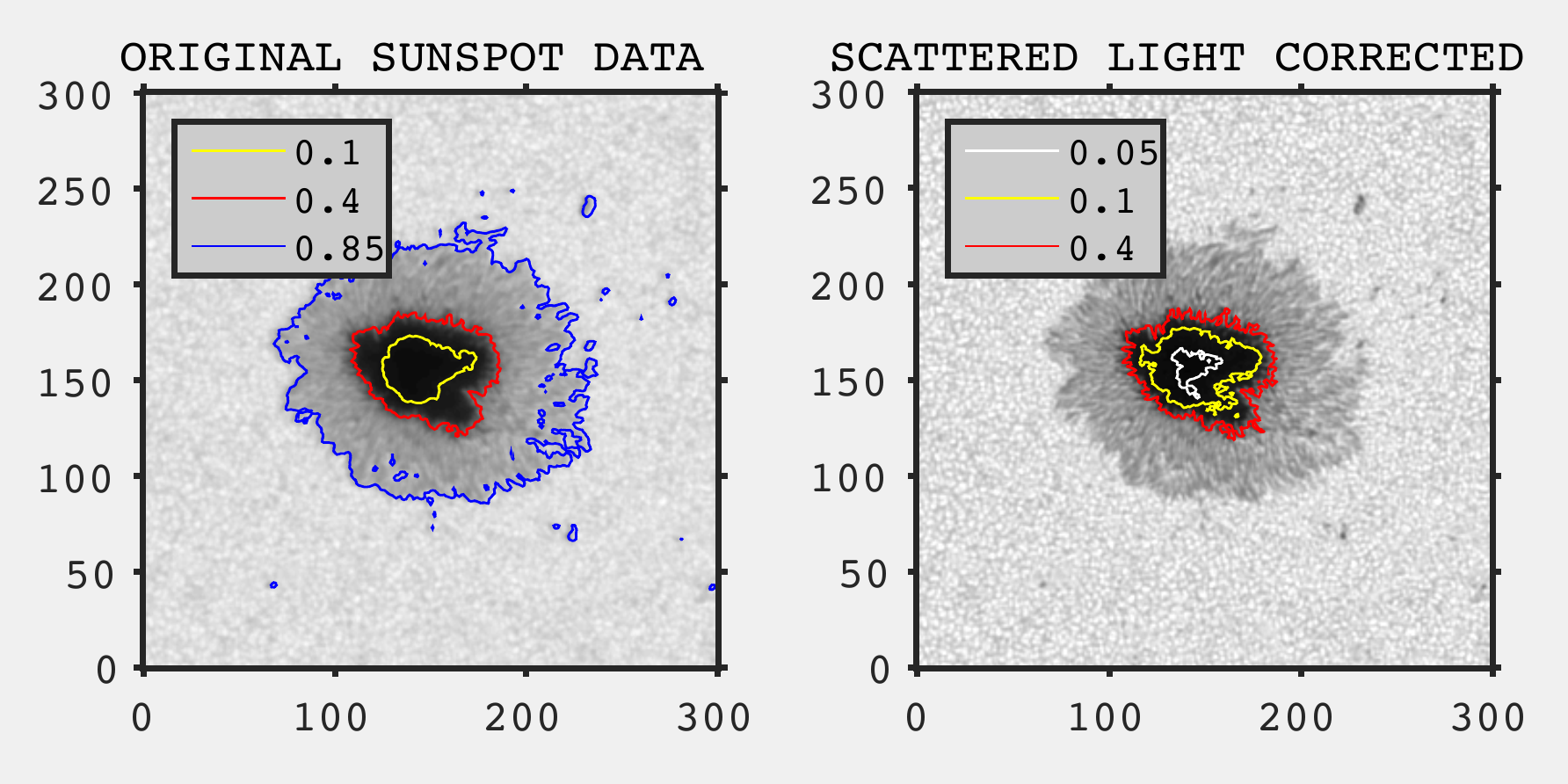} }
\caption{A cropped filtergram image containing a sunspot from the HMI
side camera taken on 18 November 2013 is shown at left. The corresponding
deconvolved image, with scattered light removed, is shown at right. The
axes are labeled in 0.5" pixels. Colored contours indicate the fraction of 
the quiet-Sun continuum intensity. The dark core of the sunspot changes from 
5.5\% of the nearby quiet-Sun continuum intensity in the original image to 3.3\% 
in the deconvolved image. The inferred minimum temperature changes from 3370~K 
in the original to 3140~K in the deconvolved.  The granulation contrast doubles; 
the standard deviation of the intensity in the quiet-Sun region is 3.7\% of the 
average in the original vs. 7.2\% in the deconvolved image.}
\label{fig:figobsv}
\end{figure}

The result is normalized so that the integral of the final PSF is unity.
Deconvolution is carried out using a Richardson-Lucy algorithm on a graphics
processing unit. The recovered images are compared to the originals to
determine the increase in the granular intensity contrast and the decrease
in minimum umbral intensity. Figure \ref{fig:figobsv} shows a large, isolated 
sunspot before and after applying the scattered-light correction.

\section{Summary and Conclusion}

The HMI instrument was designed to make continuous full-disk observations
of the photosphere at a rapid cadence in order to better understand solar
variability and the causes of space weather. During the first five years
of the mission the instrument has taken $4096 \times 4096$ narrow-band
filtergrams about every 3.75 seconds on each of two cameras while tuned to
one of six wavelengths and four polarizations. The investigation, instrument,
pipeline analysis of the individual filtergrams, and the in-flight performance
of the instrument are described elsewhere.

This report describes the way the project computes what are called the
line-of-sight and Stokes-vector observables from HMI filtergrams. The LoS
observables computed using 45-s sequences of filtergrams from the front
camera are Doppler velocity, LoS magnetic field, line width, line depth,
and continuum intensity. The side-camera sequence takes 135\,s and from it
are computed the four Stokes parameters, [I\,Q\,U\,V], at each wavelength
averaged every 720\,s. The vector observables pipeline stops with the Stokes parameters
because deriving the full vector field is model dependent \citep[for details 
about the higher-level vector products,
see][]{Hoeksema2014}. Averaged LoS observables are also computed independently 
using the side-camera Stokes-I and Stokes-V images on the same 720-s cadence. The 
observables for the entire mission are available from \urlurl{jsoc.stanford.edu}.

Observables are generated on a strict cadence for times equally spaced for an observer situated exactly 1 AU from the Sun.
They are computed from co-aligned, flat-fielded filtergrams corrected
for gaps and cosmic rays and interpolated to a common time, size, center,
and roll angle to correct for temporal and spatial variations and to address
a number of instrumental optical, spectral, and polarization effects,
as described in the middle parts of Section 2. Section \ref{MDI-like}
describes the MDI-like algorithm used to compute the complete set of LoS
observables at each pixel in the final image. To the extent that changing
performance and better knowledge of the instrument alters the calibration,
adjustments are made to the parameters of the calculation and look-up tables. Information provided in keywords allows the user to determine what calibration has been applied to any particular instance of an observable.

The noise levels in the HMI observables are better than the original specifications.
There remain some systematic errors associated with the effects of the SDO orbit,
specifically the impact of the large daily orbital-velocity changes relative
to the Sun. This is one of the largest unresolved calibration issues for
HMI data. An {\em ad hoc} correction is made to each LoS velocity measurement;
the correction is extrapolated from a daily polynomial fit to the 15 -- 60\,m\,s$^{-1}$ 
difference between the measured full-disk median Sun-SDO velocity and the known value
(Section \ref{polcor}). Work is currently underway to improve this correction, in particular to
reduce the dependence on extrapolation for line shifts outside the range of
the daily orbit variation near disk center. See, for example, recent work by
\citet{Schuck2016}.

The HMI magnetic-field measurements are similarly affected. The daily variations
in the LoS and vector field components have been characterized, but no fix is
currently available. In umbral and penumbral field regions, the line-of-sight
component of the vector field varies by about $\pm 50$\,G, or 2 - 5\% through the
day, while the transverse component is relatively insensitive to velocity.
In contrast, in weak field regions, there is a curious $\pm 7$\,G variation
in the transverse component observed near 0 relative velocity, but little
sensitivity in the line-of-sight component. It is important to note that all
of these variations are much lower than the $\sim 100$\,G uncertainty in the 720s vector field total magnitude, but they are systematic errors 
that vary smoothly with time rather than being random. 
The HMI LoS magnetic field determined with the MDI-like algorithm has a different 
sensitivity to velocity ($\sim \pm 35$\,G during a day) and
saturates at significantly lower field intensity. It also has a much
lower per-pixel random noise, 5~G for 720s and 7~G for 45s magnetogram. Periodic magnetic-field variations have been noticed by others. For example \citet{Smirnova2013} analyzed MDI-like LoS measurements in six regions and 
concluded that periodic variations were important in strong-field regions 
(above 2000~G). They found inconsistent results in the amplitude and period 
of systematic variations.

Other instrument issues have been explored to varying degrees. 

The understanding of instrument polarization has a number of outstanding
issues. We do not understand the origin of the telescope polarization and
the polarization effects described in Section \ref{sec:IQUV}. Though we have
found a way to correct for both effects, some observational confirmation
of the source may be desirable, particularly if we want to observe with
different polarization-selector settings. A remaining source of uncertainty is the impact of the unknown 
temperature gradient across the front window.  This means that the amount 
of depolarization is not known \citep{SchouPolarization2012}. 
Furthermore, the temperature gradient certainly evolves in time, albeit slowly. 
One approach is to measure changes
in the observed field strength as the front window heater temperature is
changed (while keeping the images in focus). An attempt at this was made
during commissioning in March 2010, but was unsuccessful due to evolution of
the target spot. Another option may be to measure how the PSF changes with
polarization and temperature, for example by performing a phase diversity
analysis at each setting. Either of these two methods requires running the
front window at several different temperatures and ensuring that everything is
stable at each. This would likely result in much of the data for some number of
days being of variable and questionable quality. Alternatively, it may be possible
to do as was done on the ground, that is to directly measure the birefringence
as a function of position on the front window in CalMode. The problem here
is that the net polarization of integrated sunlight is quite small and that
measurements at different values will likely be needed. The advantage of
this method is that it may be possible to perform at a single temperature.

Section 3 describes an exploratory investigation of observing 10 wavelengths
with HMI. This improves reliability and sensitivity in the strongest
magnetic field regions, but reduces the cadence of the
osbservations and thus the signal to noise. It may also be possible to improve 
the characterization of
the instrument filter transmission profiles enough to improve the vector
field inversion. The HMI team has also considered combining the filtergrams
from the two cameras. That would make the collection of data for the vector
observables more efficient and reduce photon noise, particularly in the linearly
polarized filtergrams. Now that the instrument is much better characterized,
this is a real possibility.

Because the Doppler and magnetic observables are computed from filtergram
differences, they are to first order insensitive to gain variations;
however, the intensity variables are. Modest long-term transmission
changes due to filter degradation are compensated for by lengthening the
exposure time; otherwise, the instrument sensitivity is remakarbly
stable \citep{Bush2015}. Relatively less attention has been given to quantitative 
analysis of the observed intensities and spectral line parameters, so beyond the
characterization of the CCD-gain dependence on temperature, no corrections have 
been made to the intensity variables.

Correction for stray light using the instrument point spread function is
another potentially promising improvement to the analysis, particularly
in high-contrast sunspot regions (Section \ref{PSF}). Though the PSF was
determined on the ground \citep{Wachter2012}, the scattered light analysis
described in Section \ref{PSF} needs to be refined and a physically realistic
model used. The core of the PSF, {\em i.e.} the part resulting from large-scale
wavefront errors, also needs to be determined. The traditional method uses
a phase-diversity technique to analyze images of the same scene (such as the
Sun) taken at multiple focus settings. Such focus sequences are taken on a
regular basis and crude attempts at a phase-diversity analysis have been
made. Unfortunately the results are not entirely consistent and stable,
so this needs to be pursued further.

The HMI data volume is large, but once an appropriate PSF correction scheme
has been determined, applying it to regions of interest should be quite
possible. If GPU computing were to be implemented, the correction could be
made to more of the incoming data. It has not yet been determined how this
would affect the observables.

As shown in Section \ref{sec:Venus} residuals in the fits to the position of
Venus are of order 0.1 pixels, indicating that the distortion model is not
as perfect as one might wish. As described in \citet{Wachter2012}, the model
used is based on offsetting the telescope relative to the stimulus telescope
during ground testing.  An obvious possibility is to use images taken during the
spacecraft offpoints, which are already used for determining the flat field
\citep{Wachter2009}. Another existing source of distortion data comes from
the roll maneuvers. Not only do they provide distortion data near the limb,
they may also be used, as was done by \cite{Korzennik2004}, to estimate the
distortion by cross-correlating images to determine the motion of features,
such as supergranules.

Finally, we note that continuous and timely production of well-calibrated
observables on a regular cadence requires constant monitoring of the instrument
for effects that impact transmission-filter profiles, CCD flat fields, CCD
dark currents, etc. Regular calibration sequences are taken on orbit to allow
this monitoring. The on-orbit calibration and performance as it affects the
generation of calibrated Level-1 filtergrams is described in \citet{Bush2015}.

\begin{acks}

We thank the many team members who have contributed to the success of the
SDO mission and particularly to the HMI instrument. We particularly
thank Jeneen Sommers and Hao Thai for their consistent efforts to
actually run the HMI analysis pipeline. We thank Bj\"{o}rn L\"{o}ptien for
verifying some of the equations. NSO/Kitt Peak FTS data used here were produced by NSF/NOAO.
This work was supported by NASA Contract NAS5-02139 (HMI) to Stanford University.  
\end{acks}

\appendix

\section{Data Management at the SDO Joint Science Operations Center}
\label{sec:DRMSApp}

In order to explain some of the terminology used in describing the data
products, it may be helpful to outline the system of data management used
by the SDO JSOC. The {\em JSOC} is the SDO HMI and AIA Joint Science Operations
Center. The JSOC has two components, the JSOC Science Data Processing
(SDP) located at Stanford University and the JSOC Instrument Operations Center
located at Lockheed Martin Solar and Astrophysics Laboratory in Palo Alto, CA.
When JSOC is used in this paper it refers to the JSOC-SDP.

The JSOC software system for data storage and access is called the Data
Record Management System (DRMS). DRMS is based on (a) the use of a relational
database for metadata, and (b) the virtualization of the storage of bulk data,
also managed with a relational database, via a Storage Unit Management System
(SUMS).

A {\sc data series} corresponds to a database table, in which each column
represents a {\sc keyword} and each row corresponds to a particular
datum described by the values of the {\sc prime keys}. A row in the table
is referred to as a data {\sc record}. Most keyword-value pairs for a data
series correspond to the information that would be stored in a FITS header
for a file associated with a particular record. Some, however, provide
information about the processing history and some about the associated
file-based data in SUMS. Data components for a record that are too large
to store in the DRMS database, called {\sc segments}, are stored as files
in directories in SUMS. Each record that has {\sc segments} of associated
data in SUMS has a pointer, {\sc sunum}, used by the SUMS system to retrieve
and access the file data. SUMS manages the physical location of the files
associated with a {\sc sunum}. The location may be either a file-system path,
a tape file identifier, or even a reference to a SUMS locator in a different
DRMS installation at another site.

Each {\sc record} in a series is identified by a unique record number and usually
by a set of {\sc prime keys} as well, i.e. a set of keywords whose combined
set of values defines a record. In a sense the set of prime-key values is the
record's name. Upon export the prime-key values are often used to generate
a file name. If new versions of a particular record are created, the record
with the highest record number is the current record. Older versions are
often maintained when there might be some historical purpose for them and
they can be accessed by special queries.

It is possible for keywords (including segments) in a data series to
be dynamically or statically linked to those in another series, so that
changes to metadata values automatically flow through derivative series
when appropriate. Whether the link is dynamic or static depends on whether
it points to the prime keys or the record number. This also makes it possible
to create and manipulate data series with subsets or supersets of the bulk
data in other series without actually having to write additional data.

For many users the most common type of SDO {\sc data series} is a set of
images for a span of time. The data come in {\sc records} that contain data
for a single time step specified by a prime keyword, e.g. {\sc t\_rec}. The {\sc
records} consist of metadata and pointers to data files. The metadata are
accessed as keyname-value pairs that describe the data, and the {\sc segment}
pointers indicate the location of the bulky data files stored in SUMS. When
data are exported out of the DRMS system, the series is usually converted to
a set of Rice-compressed FITS files where keyword names are mapped to valid FITS keyword
names and the file data in SUMS are written as FITS data arrays. Thus to an
outside user an exported {\sc data series record} typically looks like a simple
FITS file with static file and keyword values.

The processing at the JSOC is done using a set of programs running in a
semi-automated processing pipeline that manages the data from from raw
spacecraft-telemetry files to completed high-level data products. Data are
provided to external users by several web-based tools, some of which are
linked into the often-used IDL SolarSoft system and other commonly used
solar data access methods. Some processing on request is accomplished by
automatic processing of export requests through the JSOC pipeline.

While this system may seem complex, it does allow nearly automated processing
of more than 100,000 image files per day into higher level products and access
to several hundred million data records and the associated 9,000 Terabyes
of data with support from only a small staff. Data are delivered to external users at more
than 40 million bytes per second, every second every day. The commonly
used collection of data (the working set) is online, as are all of the metadata, 
and it is accessible to all who need it as soon as it is processed. The software
used is also available online. JSOC data may be found directly through the
JSOC site at \urlurl{jsoc.stanford.edu} or via links from the NASA SDO site
at \urlurl{sdo.gsfc.nasa.gov}.

As applied to the HMI Observables as listed in Table 1, it can be seen that
each data series corresponds to all ``images" of a particular observable
at a particular interpolation cadence. Each of the observables series has two
prime keys: {\sc t\_rec}, the target time for the interpolation, which is
at a fixed cadence at 1 AU, and {\sc camera}, describing which of the
two HMI cameras the filtergrams contributing to the observable were drawn.
Each of the observables series, except for {\sc hmi.S\_720s}, has associated 
with it a single segment, representing the corresponding image, be it a 
Dopplergram, magnetogram, continuum photogram, or whatever. The Stokes-parameter 
series has 24 segments, each corresponding to the 2-dimensional 
image of one of the four Stokes parameters at one of the six wavelength
positions.

For HMI, the standard processing does not stop at the observables, 
which are the HMI principal data products. The HMI team also produces higher
level products such as vector magnetograms, sub-surface flow maps, etc.

\newcommand{\adv}{    {\it Adv. Space Res.}}
\newcommand{\annG}{   {\it Ann. Geophys.}}
\newcommand{\aap}{    {\it Astron. Astrophys.}}
\newcommand{\aaps}{   {\it Astron. Astrophys. Suppl.}}
\newcommand{\aapr}{   {\it Astron. Astrophys. Rev.}}
\newcommand{\ag}{     {\it Ann. Geophys.}}
\newcommand{\aj}{     {\it Astron. J.}}
\newcommand{\apj}{    {\it Astrophys. J.}}
\newcommand{\apjl}{   {\it Astrophys. J. Lett.}}
\newcommand{\apss}{   {\it Astrophys. Space Sci.}}
\newcommand{\cjaa}{   {\it Chin. J. Astron. Astrophys.}}
\newcommand{\gafd}{   {\it Geophys. Astrophys. Fluid Dyn.}}
\newcommand{\grl}{    {\it Geophys. Res. Lett.}}
\newcommand{\ijga}{   {\it Int. J. Geomagn. Aeron.}}
\newcommand{\jastp}{  {\it J. Atmos. Solar-Terr. Phys.}}
\newcommand{\jgr}{    {\it J. Geophys. Res.}}
\newcommand{\mnras}{  {\it Mon. Not. Roy. Astron. Soc.}}
\newcommand{\nat}{    {\it Nature}}
\newcommand{\pasp}{   {\it Pub. Astron. Soc. Pac.}}
\newcommand{\pasj}{   {\it Pub. Astron. Soc. Japan}}
\newcommand{\pre}{    {\it Phys. Rev. E}}
\newcommand{\solphys}{{\it Solar Phys.}}
\newcommand{\sovast}{ {\it Soviet  Astron.}}
\newcommand{\ssr}{    {\it Space Sci. Rev.}}

\bibliographystyle{spr-mp-sola}
\bibliography{test}

\end{article}
\end{document}